\documentclass[11pt,a4paper]{article}
\usepackage{jheppub}
\pdfoutput=1

\usepackage[utf8]{inputenc}
\usepackage{adjustbox}
\usepackage{multirow}
\usepackage{MnSymbol}
\usepackage{amsmath}
\usepackage{amsfonts}
\usepackage{extarrows}
\usepackage{xcolor}
\usepackage{tikz}
\usetikzlibrary{arrows}
\usetikzlibrary{shapes}
\usetikzlibrary{fit}
\usepackage[english]{babel}
\usepackage{graphicx}
\usepackage{subcaption}
\usepackage{array}

\newcommand{\co}{\mathbb C}

\newcommand{\be}{\begin{equation}} 
\newcommand{\ee}{\end{equation}}
\newcommand\addvmargin[1]{
  \node[fit=(current bounding box),inner ysep=#1,inner xsep=0]{};
}
\def\c3d#1{\mathcal{C}^{3d}(#1)}

\def\node#1#2{\overset{#1}{\underset{#2}{\circ}}}
\def\cnode#1#2#3{\overset{#2}{{\color{#1}\underset{#3}{\circ}}}}
\def\upnode#1{\overset{\scriptstyle#1}{\overset{\displaystyle\circ}{\scriptstyle\vert}}}
\def\cupnode#1#2{\overset{{\color{#1}\overset{\scriptstyle#2}{\displaystyle\circ}}}{\scriptstyle\vert}}
\def\topnode#1#2{\overset{#1}{\overset{{\displaystyle\circ}\rlap{\,\,$\scriptstyle#2$}}{\scriptstyle\vert}}}
\def\ctopnode#1#2#3{\overset{#2}{\overset{{\color{#1}{\displaystyle\circ}\rlap{\,\,$\scriptstyle#3$}}}{\scriptstyle\vert}}}

\def\flav#1{\overset{\scriptstyle#1}{\overset{\square}{\scriptstyle\vert}}}

\def\orbit#1#2{\overline{{\rm #1}_{#2}}}
\def\ba{\begin{equation} \begin{aligned}} \def\ea{\end{aligned}\end{equation}}

\tikzset{
    7brane/.style={circle,draw=black, fill=white, inner sep=0pt,minimum size=5pt},
    7braneBig/.style={circle,draw=black, fill=white, inner sep=0pt,minimum size=8.2pt},
    toric/.style={circle,draw=black, fill=black, inner sep=0pt,minimum size=5pt},
    dot/.style={circle,draw=black, fill=black, inner sep=0pt,minimum size=1pt},
}

\newcolumntype{C}{ >{\centering\arraybackslash} m{3cm} }

\newtheorem{summary}{Summary}
\newtheorem{conjecture}{Conjecture}

\preprint{Imperial/TP/18/AH/10}

\title{Tropical Geometry and Five Dimensional Higgs Branches at Infinite Coupling}
\author[a]{Santiago Cabrera,}
\author[a]{Amihay Hanany}
\author[b,c]{and Futoshi Yagi}
\affiliation[a]{Theoretical Physics, The Blackett Laboratory, Imperial College London\\London SW7 2AZ, United Kingdom}
\affiliation[b]{School of Mathematics, Southwest Jiaotong University,\\ west zone, high-tech district, Chengdu, Sichuan 611756, China}
\affiliation[c]{Department of Physics, Technion - Israel Institute of Technology, Haifa 32000, Israel}
\emailAdd{santiago.cabrera13@imperial.ac.uk}
\emailAdd{a.hanany@imperial.ac.uk}
\emailAdd{futoshi\_yagi@swjtu.edu.cn}

\abstract{Superconformal five dimensional theories have a rich structure of phases and brane webs play a crucial role in studying their properties. This paper is devoted to the study of a three parameter family of SQCD theories, given by the number of colors $N_c$ for an $SU(N_c)$ gauge theory, number of fundamental flavors $N_f$, and the Chern Simons level $k$. The study of their infinite coupling Higgs branch is a long standing problem and reveals a rich pattern of moduli spaces, depending on the 3 values in a critical way. For a generic choice of the parameters we find a surprising number of 3 different components, with intersections that are closures of height 2 nilpotent orbits of the flavor symmetry. This is in contrast to previous studies where except for one case ($N_c=2, N_f=2$), the parameters were restricted to the cases of Higgs branches that have only one component. The new feature is achieved thanks to a concept in tropical geometry which is called \emph{stable intersection} and 
 allows for a computation of the Higgs branch to almost all the cases which were previously unknown for this three parameter family apart form certain small number of exceptional theories with low rank gauge group. 
 A crucial feature in the construction of the Higgs branch is the notion of \emph{dressed monopole operators}.}

\begin{document}

\maketitle
\flushbottom

\section{Introduction and Summary}
Brane webs of five branes in string theory are well known to capture the dynamics of five dimensional gauge theories and their UV fixed points \cite{Aharony:1997ju,Aharony:1997bh}. Many of the studies which followed were focused on the Coulomb branch of the 5d theories, and this paper is devoted to study ${\cal H}_\infty$ \cite{Cremonesi:2015lsa,Ferlito:2017xdq}, the Higgs branch (or branches) which arise as the gauge coupling is tuned to infinity. As is usual in brane systems of the type of \cite{Hanany:1996ie}, the most useful way of studying the Higgs branch is to introduce $(p,q)$ seven branes as in \cite{DeWolfe:1999hj}, with a possibility to end multiple 5-branes on a single 7-brane as in \cite{Benini:2009gi}. With the seven branes attached at ends of five branes, the Higgs branch ${\cal H}_\infty$ is realized as the gauge coupling is tuned to infinity and the five brane webs are maximally divided into sub webs that move freely between seven branes, thus providing the Higgs branch moduli, one (quaternionic) modulus per each such sub web. While such a picture, already identified in \cite{Aharony:1997bh}, correctly captures the dimension of the Higgs branch, it was not sufficient until now to determine the Higgs branch in full detail. This paper is devoted to fix this problem and to correctly identify the Higgs branch ${\cal H}_\infty$ from the brane realization.

Higgs branches, being hyperK\"ahler, are typically constructed by providing combinatorial data, as encoded in a quiver gauge theory \cite{nakajima1994}. There is a corresponding Lagrangian and for a hyperK\"ahler quotient one uses the F and D term equations to construct the Higgs branch as the space of gauge invariant vacuum equations. If in addition one sets the FI terms to zero, the Higgs branch becomes a hyperK\"ahler cone, or a symplectic singularity \cite{Beauville:2000}. HyperK\"ahler quotient is the traditional way of computing Higgs branches at finite coupling in any dimension, and in particular in five dimensions. In three dimensions, there is an alternative way of constructing a symplectic singularity by taking the same combinatorial data in the form of a quiver (more precisely a graph) and construct the Coulomb branch as the space of dressed monopole operators, as in \cite{Cremonesi:2013lqa} (this notion has been further developed throughout the mathematical literature \cite{Nakajima:2015txa,Braverman:2016wma,Nakajima:2017bdt}). Hence we are faced with two ways of constructing symplectic singularities and one wonders which way is more suitable for the purpose of solving the problem of finding ${\cal H}_\infty$.

Incidentally, there are other ways of constructing symplectic singularities like the space of dressed instanton operators as described in \cite{Cremonesi:2015lsa,Ferlito:2017xdq}. This is a very interesting direction of research, but will not be pursued in the present paper.

It should be stressed that the construction of the moduli space as a space of dressed monopole operators is not restricted to 3d. All the features which are required for such a construction exist in all higher dimensions 4, 5, and 6. The monopole operators are still localized in 3 dimensions, but contribute to the chiral ring just as they do in 3 dimensions. The topological symmetry is still conserved by the Bianchi identity, but due to boundary conditions remains associated with a 1 form current.
This is consistent with \cite{Cordova:2016emh} that finds that there are no higher form currents in 5 dimensional SCFTs. The combinatorial data is hence not an exclusive feature of 3d, but rather of any dimension that includes 3! It is a feature of a co-dimension 3 object. This is further supported by the brane picture of type \cite{Hanany:1996ie} in which a Higgs branch modulus is characterized by a D$p$ brane in between two D$(p+2)$ branes where the boundary conditions remove a vector multiplet, but leave a compact scalar which, together with one of the 3 real scalars, exponentiate to form monopole operators in the construction of the moduli space.

As a result, it makes more sense to call the construction of such a moduli space a \emph{space of dressed monopole operators} rather than a \emph{3d Coulomb branch}, since the former name is more general and does not commit to a particular dimension -- it can be in 3, 4, 5, or 6 dimensions. Furthermore, the latter name commits to 3d and may lead to confusion as coming from some sort of compactification and 3d mirror symmetry \cite{Intriligator:1996ex}. With this comment in mind, below we use the names \emph{3d Coulomb branch} or \emph{space of dressed monopole operators} interchangeably.

To proceed, we recall that the Higgs branch is given by sub webs of five branes which are ending on seven branes, and notice that five branes ending on seven branes carry magnetic charges and are naturally associated with magnetic monopole solutions. This property favors the construction of the Higgs branch at infinite coupling as the space of dressed monopole operators and we proceed by taking this direction. One first needs the combinatorial data in the form of a quiver. Once this is given, one can use a host of techniques which are developed in papers that follow \cite{Cremonesi:2013lqa} and are still under intensive study.

To be more specific we now introduce the class of theories which are discussed in this paper. The class is parametrized by three integer numbers $N_c$, $N_f$, $k$ and goes under the name of 5d SQCD. The gauge group is $SU(N_c)$ with a CS level $k$ and $N_f$ flavors of fundamental matter. $N_c>1$ for a non Abelian gauge theory, $N_f\ge0$ is non negative, and the CS level $k$ is any integer or half-integer number satisfying $k-N_f/2\in {\mathbb Z}$, but it turns out that the resulting Higgs branches depend only on its absolute value. 
For generic $N_c$, the brane webs put an additional restriction that $|k|\le N_c-N_f/2+2$ \cite{Bergman:2014kza}. For certain small values of $N_c$, other values of $k$ may lead to 5d fixed points \cite{Jefferson:2017ahm, Jefferson:2018irk}. Among such cases, we also study the case $|k| = N_c-N_f/2+3$ with $N_c=3$ in this paper. As for the remaining exceptional cases, the corresponding 5-brane web diagrams are known to include orientifold 5-plane for some cases \cite{Hayashi:2018lyv}, which makes the analysis of the Higgs branch doable but more involved, while the corresponding 5-brane webs are not even known for other cases. We do not study such cases in this paper.
The theory has one gauge coupling which is set to infinity at the UV fixed point, and adjoint valued real masses that transform under the flavor symmetry. For simplicity all these masses are set to 0. The classical flavor symmetry is $SU(N_f) \times U(1)_B \times U(1)_I$ where $U(1)_B$ is the baryonic symmetry which acts on the Higgs branch for $N_f\ge N_c$ and $U(1)_I$ is the instanton symmetry. The flavor symmetry at infinite coupling for $N_c \ge 3$ is summarized in table \ref{Table:global} \cite{Bergman:2013aca, Hayashi:2015fsa, Yonekura:2015ksa, Gaiotto:2015una}.

\begin{table}
\label{global}
\makebox[\textwidth][c]{ 
\begin{tabular}{c|c|c}
Parameter region & Global symmetry at UV & Global symmetries on components\\
\hline
$N_c - \frac{1}{2} N_f > |k| = 0, \frac{1}{2}$  & 
$SU(N_f) \times U(1) \times U(1)$ &
$SU(N_f) \times U(1)$
\\
$N_c - \frac{1}{2} N_f > |k| > \frac{1}{2}$  & 
$SU(N_f) \times U(1) \times U(1)$ &
$SU(N_f) \times U(1)$,  $\quad SU(N_f)$ 
\\
\hline
$N_c - \frac{1}{2} N_f = |k| = 0$ & 
$SU(N_f) \times SU(2) \times SU(2)$ &
$SU(N_f)  \times SU(2) \times SU(2)$
\\
$N_c - \frac{1}{2} N_f = |k| = \frac{1}{2}$ & 
$SU(N_f) \times SU(2) \times U(1)$ &
$SU(N_f)  \times SU(2) \times U(1)$
\\
$N_c - \frac{1}{2} N_f = |k| = 1$ & 
$SU(N_f) \times SU(2) \times U(1)$ &
$SU(N_f)  \times SU(2) \times U(1)$, $\quad SU(N_f)$
\\
$N_c - \frac{1}{2} N_f = |k| > 1$ & 
$SU(N_f) \times SU(2) \times U(1)$ &
$SU(N_f)  \times SU(2) \times U(1)$, $\quad SU(N_f)  \times SU(2)$ 
\\
\hline
$N_c - \frac{1}{2} N_f + 1 = |k| = 0$ & 
$SU(N_f+2)$ &
$SU(N_f+2)$ 
\\
$N_c - \frac{1}{2} N_f + 1 = |k| = \frac{1}{2}$ & 
$SU(N_f+1) \times SU(2)$ &
$SU(N_f+1) \times SU(2)$
\\
$N_c - \frac{1}{2} N_f + 1 = |k| = 1, \frac{3}{2}$ & 
$SU(N_f+1) \times U(1)$ &
$SU(N_f+1) \times U(1)$
\\
$N_c - \frac{1}{2} N_f + 1 = |k| > \frac{3}{2}$ & 
$SU(N_f+1) \times U(1)$ &
$SU(N_f+1) \times U(1)$, $\quad SU(N_f+1)$ 
\\
\hline
$N_c - \frac{1}{2} N_f + 2 = |k| = 0$ & 
(6d) &
(6d) 
\\
$N_c - \frac{1}{2} N_f + 2 = |k| = \frac{1}{2}$ & 
$SO(2N_f+2)$ & 
$SO(2N_f+2)$ 
\\
$N_c - \frac{1}{2} N_f + 2 = |k| = 1$ & 
$SO(2N_f) \times SU(2)$ & 
$SO(2N_f) \times SU(2)$
\\
$N_c - \frac{1}{2} N_f + 2 = |k| = \frac{3}{2},2$ & 
$SO(2N_f) \times U(1)$ & 
$SO(2N_f) \times U(1)$
\\
$N_c - \frac{1}{2} N_f + 2 = |k| > 2$ & 
$SO(2N_f) \times U(1)$ & 
$SO(2N_f) \times U(1)$, $\quad SO(2N_f)$
\end{tabular}
}
\caption{Global Symmetry on the Higgs branch. In cases that the Higgs branch at infinite coupling contains up to 3 components, the global symmetry acts on each component in a possibly different way. All components share mesons, hence share the non Abelian part of the global symmetry. Abelian symmetries act in different ways on each component.}
\label{Table:global}
\end{table}

We are led to look for combinatorial data in the form of a graph with nodes and edges. Such an approach is taken in \cite{Ferlito:2017xdq} and successfully describes ${\cal H}_\infty$ for low enough values of $|k|$ and high enough values of $N_f$. The answer is given in a form of exceptional sequences and is derived by applying two methods, one of which is looking at maximal sub algebras of exceptional algebras and generalizing to higher values of $N_c$ by using the symmetry properties of the associated quivers. The other method for some of the cases uses compactification to three dimensions, and mirror symmetry, but is restricted to cases where the process is understood.  All the results of \cite{Ferlito:2017xdq} are included below, with some different way of presentation, since they constitute a subset of the full set of results of the present paper.

A striking feature of the solutions given in \cite{Ferlito:2017xdq} is that they are all given by a single graph and the Higgs branch at infinite coupling admits a single cone structure. There is, however, a phenomenon which already shows up at finite coupling \cite{Ferlito:2016grh} where for some gauge theories the Higgs branch is a union of two cones. Henceforth, a cone on the Higgs branch will be also called a component of the Higgs branch. It should be stressed that for lower supersymmetry this phenomenon is rather frequent, but for 8 supercharges it is a rather unstudied feature of the Higgs branch, and relatively rare.

The techniques in \cite{Ferlito:2017xdq} form a beautiful set of new results, and have shortcomings due to two reasons. The first is that they are covering special regions of the three parameter family $N_c$,  $N_f$, $k$ for which the Higgs branch ${\cal H}_\infty$ is a single cone, and do not cover cases in which the Higgs branch is a union of two or more cones. The second is that they do not cover cases with high value of $|k|$ and low values of $N_f$. This is due to a new feature of quivers with edge multiplicities that is not taken into account in \cite{Ferlito:2017xdq}. In fact, all edges in the quivers of \cite{Ferlito:2017xdq} have a multiplicity of either 1 or 0. Both of these issues are resolved in the present paper and we give a complete description of ${\cal H}_\infty$ for all values of $|k|\le N_c -N_f/2+2$. As mentioned before, the results of this paper are fully consistent with those of \cite{Ferlito:2017xdq} and provide a complete generalization of them, as well as another (third) consistency check for the validity of the whole approach of representing ${\cal H}_\infty$ as the space of dressed monopole operators.

The simplest case where ${\cal H}_\infty$ is a union of 2 cones is for $N_c=2, N_f=2$ \cite{Seiberg:1996bd,Morrison:1996xf} and has a brane realization which is discussed in subsection \ref{E3}. The corresponding five brane web has two inequivalent ways of dividing into sub webs. This physically translates to having a Higgs branch with two components, each is a cone, and both intersect at the origin. Even more surprising, and certainly new in the physics of 5d gauge theories and their UV fixed points, is the behavior of the generic case in this 3 parameter family. There are 3 different components to the Higgs branch with non trivial intersections that are discussed in detail below. This again features by having 3 inequivalent ways of sub dividing a brane web.

The appearance of sub division of brane webs leads to a new question in five dimensional physics, and in brane physics. Given a web that characterizes a five dimensional fixed point, we ask in how many inequivalent ways can it be sub divided. The answer to this question is the answer to the number of components that the Higgs branch has at the fixed point. There is a corresponding toric diagram and we ask what are the possible ways to get such a toric diagram by using Minkowski sums of smaller toric diagrams. This is an interesting question which is to be addressed in future studies. The notion of Minkowski sums, taken from \cite{1994alg.geom..3004A,1994alg.geom..5008A}, is applied in \cite{Franco:2005zu} for the study of supersymmetry breaking in branes at singularities, with particular emphasis on toric diagrams for theories with $N_f=0$, the so called $Y^{p,q}$ in the paper. It should be stressed that the methods of \cite{1994alg.geom..3004A,1994alg.geom..5008A} are restricted to cases where 7-branes at ends of 5-branes are not needed for generating more Higgs branch moduli. Hence a generalization to include the effects of 7-branes is in order.

It is crucial to point out that parallel external legs in the brane web are not presenting any particular complication in the approach of this paper. This is due to having a 7-brane at the end of each external 5-brane. Any previous claims that parallel legs lead to some 6d states which may or may not couple to the system are avoided in this work by having finite 5-brane webs between 7-branes, none of them extend to infinity in the plane of the brane web, hence lead to 5d fixed points.

Multiple edges in a quiver are not common in gauge theories with this amount of supersymmetry. In addition they do not show up in perturbative open strings, again with this amount of supersymmetry. Nevertheless, such quivers do show up in various non perturbative cases. In \cite{Boalch2012} Philip Boalch discusses the notion of complete graphs in the context of studies of Painlev\'e equations. This is applied to Higgs branches of Argyres Douglas theories in \cite{Xie:2012hs} and are used in \cite{DelZotto:2014kka} for an evaluation of the Higgs branch as an algebraic variety, using the construction of the space of dressed monopole operators \cite{Cremonesi:2013lqa}. It should be pointed out that while complete graphs have all edges with the same multiplicity, in the present paper only one or two edges have multiplicity greater than 1, while the other edges are the usual edges with multiplicity 1.

So how does one get an edge with multiplicity in the quiver? This follows from the association of a quiver node to a brane web, where the rank of the node is equal to the number of copies of such a brane web. Given two different such brane webs, they have an intersection number, which is called \emph{stable intersection} in the tropical geometry literature \cite{2013arXiv1311.2360B}. As a reminder, the physical object which we call a brane web, is the mathematical object which is called a \emph{tropical curve}. To explain this point further, we need to recall the usual way an edge shows up in perturbative open strings.
Given two D branes, represented by nodes in the quiver, one can stretch an F1 between them and this gives rise to a bi fundamental hyper multiplet that is represented by an edge in the quiver. By using standard S and T dualities, we find that a D3 brane stretched between two five brane webs, each represented by a node in the quiver, leads to such an edge connecting the two nodes. There is however, a crucial difference. Two five brane webs can have multiple intersection points and hence we can have one D3 brane stretched per such point, leading to one edge of the quiver per each intersection point. The resulting edge of the quiver, has a multiplicity which is equal to the intersection number between the two five brane webs. With this rule in hand, we are now able to compute all of the missing cases that were not possible with the techniques in \cite{Ferlito:2017xdq}. Furthermore, we are able to check that all the cases computed in \cite{Ferlito:2017xdq} are consistent with having multiplicity 1 to all edges in the quiver, and consist of one component on the Higgs branch. Thus we have a third way 
of deriving the results of \cite{Ferlito:2017xdq}. The computation is demonstrated below with a collection of explicit examples.

Classical Higgs branches are known to be trivial for $N_f<2$, but as one tunes the gauge coupling to infinity, new flat directions show up and the Higgs branch emerges from these new moduli. A simple example of such a phenomenon is 5d SYM with gauge group $G$ with CS level $k=0$ as discussed in \cite{Cremonesi:2015lsa}. There are no matter fields, hence no classical moduli space, but there are new flat directions due to the appearance of massless instantons at infinite coupling. The resulting moduli space is $\co^2/{\mathbb Z}_h$ where $h$ is the dual Coxeter number of the gauge group $G$. Restricting $G$ to be $SU(N_c)$ with level $k=0$ we find that there is a new Higgs branch at infinite coupling, contrary to the classical intuition which associates Higgs branches to matter fields (alternatively, one should start thinking about instanton operators as generating new matter degrees of freedom, which differ from free hypermultiplets). The results of this paper show that there is another $N_f=0$ theory with a non trivial ${\cal H}_\infty$. If we set $|k|=N_c$ we find a new moduli space of the form $\co^2/{\mathbb Z}_2$ for any $N_c$.
Similar observations arise for $N_f=1$. For example there is an interesting pattern of Higgs branches for $SU(3)$ with 1 flavor. For CS levels $1/2$, $3/2$, $5/2$, $7/2$ and $9/2$ we find that ${\cal H}_\infty$ is trivial for $|k|=3/2$, is $\co^2/{\mathbb Z}_3$ for $|k|=1/2,~9/2$, and is $\co^2/{\mathbb Z}_2$ for $|k|=5/2,~7/2$. A very rich pattern of ${\cal H}_\infty$ as the CS level is varied.

The paper is organized as follows. In section \ref{sec:knownExamples} we start our analysis by re deriving some known results from the brane webs, thus establishing tools which allow the computation of the combinatorial (quiver) data for cases that were not known so far. In section \ref{sec:conjecture} we establish the general conjecture that explains how to obtain the quiver data from the 5-brane webs. In section \ref{sec:newNotions} we explore new notions derived from the application of the conjecture to 5d brane webs whose corresponding $\mathcal{H}_\infty$ was not previously known. These are the \emph{union of three cones} and the \emph{intersection of several cones}. In section \ref{sec:classification} we provide a full classification of the Higgs branch at infinite coupling $\mathcal{H}_\infty$ of the three parameter family of 5d $\mathcal{N}=1$ SQCD theories with gauge group $SU(N_c)$, $N_f$ fundamental flavors, and CS level $k$. These results were not known before and have been obtained by applying the conjecture defined in section \ref{sec:conjecture}. Section \ref{sec:conclusions} contains some concluding statements. An appendix \ref{sec:app} has been included with an example that illustrates a detailed computation.

\section{Known Examples}\label{sec:knownExamples}

\subsection{$E_3$ -- Union of Two Cones}
\label{E3}

\begin{figure}
	\centering
		\begin{tikzpicture}[scale=0.75]
			\node[toric] at (0,0) {};
			\node[toric] at (0,1) {};
			\node[toric] at (1,0) {};
			\node[toric] at (1,1) {};
			\node[toric] at (1,2) {};
			\node[toric] at (2,1) {};
			\node[toric] at (2,2) {};
			\draw (0,0)--(0,1)--(1,2)--(2,2)--(2,1)--(1,0)--(0,0);
			\draw (1,1)--(0,0);
			\draw (1,1)--(0,1);
			\draw (1,1)--(1,2);
			\draw (1,1)--(2,2);
			\draw (1,1)--(2,1);
			\draw (1,1)--(1,0);
		\end{tikzpicture}
	\caption{Toric diagram corresponding to the 5d SQCD theory with SU(2) gauge group and $N_f=2$.}
	\label{fig:E3toric}
\end{figure}
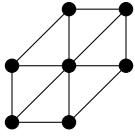

\begin{figure}
	\centering
		\begin{tikzpicture}[scale=0.75]
			\draw
				(0,3)--(0,2)--(0.5,1.5)--(3,1.5)--(3,2.5)--(2.5,3)--(0,3)
				(-1,4)--(0,3)
				(-1,2)--(0,2)
				(0.5,0.5)--(0.5,1.5)
				(4,0.5)--(3,1.5)
				(4,2.5)--(3,2.5)
				(2.5,4)--(2.5,3);
			\node[7brane] at (-1,4) {};
			\node[7brane] at (-1,2) {};
			\node[7brane] at (0.5,0.5) {};
			\node[7brane] at (4,0.5) {};
			\node[7brane] at (4,2.5) {};
			\node[7brane] at (2.5,4) {};
			\draw
				(5,2)--(8,2)
				(7.8,2.2)--(8,2)
				(7.8,1.8)--(8,2);
			\node[] at (6.5,2.5) {$\frac{1}{g^2}, m_i, a \rightarrow 0$};
			\draw
				(9,2)--(11,2)
				(9,3)--(11,1)
				(10,3)--(10,1);
			\node[7brane] at (9,2) {};
			\node[7brane] at (11,2) {};
			\node[7brane] at (9,3) {};
			\node[7brane] at (11,1) {};
			\node[7brane] at (10,3) {};
			\node[7brane] at (10,1) {};
		\end{tikzpicture}
	\caption{Five brane webs (dual to the toric diagram in figure \ref{fig:E3toric}) corresponding to the 5d SQCD theory with SU(2) gauge group and $N_f=2$ before and after taking the gauge coupling $g$ to infinity and all the masses $m_i$ as well as the VEV $a$ of the adjoint scalar field to zero. The horizontal lines represent D5-branes, the vertical lines represent NS5-branes and the diagonal lines represent $(1,-1)$ five branes. Each circle represents a seven brane of the same type as the five brane that ends on it.}
	\label{fig:E3finite}
\end{figure}
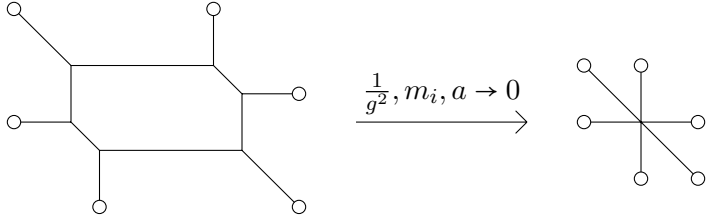

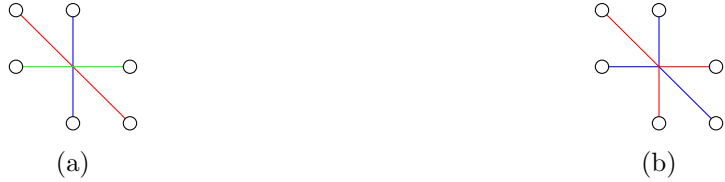
\begin{figure}
	\centering
	\begin{subfigure}[t]{.49\textwidth}
		\centering
		\begin{tikzpicture}[scale=0.75]
			\draw[blue](0,-1)--(0,1);
			\draw[red](-1,1)--(1,-1);
			\draw[green](1,0)--(-1,0);
			\node[7brane] at (0,1) {};
			\node[7brane] at (0,-1) {};
			\node[7brane] at (-1,1) {};
			\node[7brane] at (1,-1) {};
			\node[7brane] at (1,0) {};
			\node[7brane] at (-1,0) {};
		\end{tikzpicture}
        \caption{}
    \end{subfigure}
    \hfill
	\begin{subfigure}[t]{.49\textwidth}
    		\centering
		\begin{tikzpicture}[scale=0.75]
			\draw[blue](0,0)--(0,1);
			\draw[blue](0,0)--(1,-1);
			\draw[blue](0,0)--(-1,0);
			\draw[red](0,0)--(0,-1);
			\draw[red](0,0)--(-1,1);
			\draw[red](0,0)--(1,0);
			\node[7brane] at (0,1) {};
			\node[7brane] at (0,-1) {};
			\node[7brane] at (-1,1) {};
			\node[7brane] at (1,-1) {};
			\node[7brane] at (1,0) {};
			\node[7brane] at (-1,0) {};
		\end{tikzpicture}
        \caption{}
    \end{subfigure}
	\caption{Two distinct phases of brane webs, representing two different components of the Higgs branch of $SU(2)$ with $N_f=2$ at infinite coupling. (a) Corresponds to a component of the Higgs branch with quaternionic dimension $d_{\mathbb{H}}=3-1=2$. (b) Corresponds to a component of the Higgs branch with quaternionic dimension $d_{\mathbb{H}}=2-1=1$. Compare with Figure 30 of \cite{Franco:2005zu}. }
	\label{fig:E3}
\end{figure}

Let us discuss a well known example: $N_c=2$, $N_f=2$. The level is not crucial here, as the CS density is identically 0 for an $SU(2)$ gauge theory.
The Higgs branch at infinite coupling is the union \cite{Seiberg:1996bd,Morrison:1996xf}:
\begin{equation}
\label{E3instanton}
	\mathcal{H}_{\infty}\left(\node{\flav 2}{SU(2)_0}\right) = \orbit{min}{A_2}\cup \orbit{min}{A_1}
\end{equation}
Where $\orbit{min}{A_k}$ denotes the closure of the minimal nilpotent orbit\footnote{The closure of the minimal nilpotent orbit of $\mathfrak{sl}(k+1,\mathbb{C})$ is normally denoted as $\bar{\mathcal O}_{(2,1^{k-1})}$ in the mathematical literature \cite{collingwood1993nilpotent}, or also as $a_k$ in \cite{Kraft1982}. We started using the notation $\orbit{min}{A_k}$ in \cite{Hanany:2018uhm,Cabrera:2018ann}, since it can be extended to exceptional groups which don't have partition data like the partition $(2,1^{k-1})$ in $\bar{\mathcal O}_{(2,1^{k-1})}$, and it can also be extended to non minimal orbits of small dimensions, i.e. the closure of the next to minimal orbit is denoted as $\orbit{n.min}{A_k}$.} of $\mathfrak{sl}(k+1,\mathbb{C})$, and both cones $\orbit{min}{A_2}$ and $\orbit{min}{A_1}$ intersect at the origin.
Physically, this moduli space is the moduli space of 1 $E_3$ instanton. The single instanton can either be an $SU(3)$ instanton or an $SU(2)$ instanton, but not both. Hence this leads to the union structure \ref{E3instanton}.
For each of these cones there is a different 3d $\mathcal N = 4$ quiver, for which the cone is the Coulomb branch. Note that the cone $\orbit{min}{A_2}$ (resp. $\orbit{min}{A_1}$) is isomorphic to the reduced moduli space of one $A_2$ (resp. $A_1$) instanton on $\mathbb{C}^2$. Hence, the 3d quivers are just the corresponding affine Dynkin diagrams \cite{Intriligator:1996ex}:
\begin{align}
	\orbit{min}{A_2} & = \mathcal{C}^{3d}\left( \overset{\overset{\displaystyle\overset 1 \circ}{\diagup~~\diagdown}}{\node{}{1}~-~\node{}{1}}\right) \label{eq:A2}\\
	\orbit{min}{A_1} & = \mathcal{C}^{3d}\left(~ \overset{\overset{\displaystyle\overset 1 \circ}{\parallel}}{\node{}{1}}~\right)\label{eq:A1}
\end{align}
where $\mathcal C^{3d}()$ denotes the 3d Coulomb branch. One can see that both cones are recognizable as different phases on the brane diagram. In order to obtain the brane system, remember that the toric diagram \cite{Aharony:1997bh} can be obtained, and it is represented\footnote{See for example figure 2 in \cite{Taki:2014pba} which contains all toric diagrams employed for the $E_n$ cases in the present section.} in figure \ref{fig:E3toric}. Figure \ref{fig:E3finite} represents the five brane web corresponding to this theory (dual to the toric diagram). On the left of figure \ref{fig:E3finite} the gauge coupling is finite and the masses and the VEV of the adjoint scalar field are different from zero. On the right of the same figure the gauge coupling is taken to infinity and all the masses as well as the VEV of the adjoint scalar field is set to zero (i.e. at the origin of the Coulomb branch). Before taking 
this limit, there is a single web that can move along the 7-branes, and hence the Higgs branch is trivial, remembering to factor out overall position moduli. After taking 
this limit, there are new possibilities of breaking the web into sub webs. In particular, there are two possibilities, represented in figure \ref{fig:E3}, where different colors correspond to different sub webs. In phase (a) there are three different segments that move along the perpendicular directions to the paper, spanned by the seven branes. In phase (b) there are two different sub webs. The transition from (a) to (b) can only take place when all sub webs realign and combine into a single web. This web corresponds to the origin of the cones, indicating that the intersection between the two cones is at a single point -- the origin.

The new perspective that was missing until the present work is that the quivers in equations \ref{eq:A2} and \ref{eq:A1} can be read directly from the brane webs. The goal of this paper is to establish the tools that allow such reading and to put them to use in the analysis of the three parameter family of 5d $\mathcal{N}=1$ theories with gauge group $SU(N_c)$, number of fundamental flavors $N_f$ and Chern Simons level $k$.

For the current example one can deduce the following:
\begin{enumerate}
	\item Each separate brane sub web corresponds to a different gauge node with group $U(1)$ in the quiver.
	\item The links between the nodes in the quiver (corresponding to hypermultiplets of the 3d $\mathcal{N}=4$ theory) are given by the intersection numbers between the branes.
\end{enumerate}
Let us discuss the second point in more detail. For phase (a) the intersection number $I$ between two 5-branes $(p_1,q_1)$ and $(p_2,q_2)$ has already been defined as the absolute value of the determinant\footnote{The intersection number is used as a CS term for a 3d theory on the world volume of the D3 brane in \cite{Bergman:1999na}. In the present work we focus on the world volume theory living on the five branes.}:
\begin{equation}\label{eq:det}
	I\{(p_1,q_1),(p_2,q_2)\}=Abs\left(\begin{vmatrix}
		p_1 & q_1\\ 
 		p_2 & q_2
	\end{vmatrix}\right).
\end{equation}

\begin{figure}
	\centering
		\begin{tikzpicture}[scale=0.75]
			\draw[blue]
				(1,1)--(2,1)
				(2,2)--(2,1)
				(3,0)--(2,1);
			\draw[red]
				(1,2)--(2,1)
				(2,0)--(2,1)
				(3,1)--(2,1);
			\draw
				(4,1)--(5,1)
				(4.8,1.2)--(5,1)
				(4.8,.8)--(5,1);
			\draw[blue]
				(6,0.5)--(8,0.5)
				(8,2.5)--(8,0.5)
				(9,-0.5)--(8,0.5);
			\draw[red]
				(6,2.5)--(7,1.5)
				(7,-0.5)--(7,1.5)
				(9,1.5)--(7,1.5);
			\draw
				(10,1)--(11,1)
				(10.8,1.2)--(11,1)
				(10.8,.8)--(11,1);
			\draw[red]
				(12,1)--(13,1)--(13,2)--(12,1)
				(12,0)--(13,0)
				(14,1)--(14,2);
			\draw[blue]
				(13,1)--(14,1)--(13,0)--(13,1)
				(12,0)--(12,1)
				(13,2)--(14,2);
			\node[toric] at (12,0) {};
			\node[toric] at (12,1) {};
			\node[toric] at (13,0) {};
			\node[toric] at (13,1) {};
			\node[toric] at (13,2) {};
			\node[toric] at (14,1) {};
			\node[toric] at (14,2) {};
		\end{tikzpicture}
	\caption{Stable intersection between two tropical curves. Left: the original tropical curves. Center: a small deformation that moves the curves away from each other. Right: dual toric diagram. The stable intersection is the sum over the areas of polygons in the toric diagram that have edges of both colours. In this case is $1+1=2$.}
	\label{fig:stableIntersection}
\end{figure}
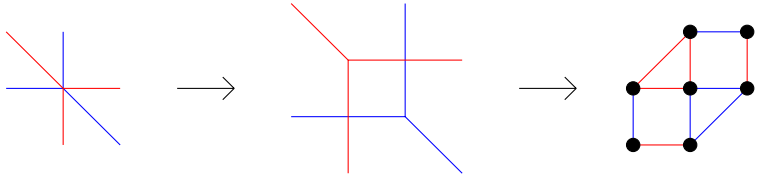

Hence, the intersection between any pair of 5-branes from the set (1,0), (0,1) and $(1,-1)$ is always of value $1$. Therefore, the 3d quiver corresponding to phase (a) is a complete graph \cite{Boalch2012} with three nodes, and edge multiplicity 1, as depicted in equation \ref{eq:A2}. Phase (b) is more complicated because one needs to compute the intersection between the two sub webs. This can be done by introducing an idea from \emph{tropical geometry} \citep{2013arXiv1311.2360B}. In tropical geometry each of the brane sub webs can be seen as a tropical curve. Then, their \emph{stable intersection} can be defined as in \citep{2013arXiv1311.2360B}. Let us review the idea of stable intersection by computing it in the case of the sub webs of phase (b). This is represented in figure \ref{fig:stableIntersection}. The left diagram in figure \ref{fig:stableIntersection} depicts the two \emph{curves}. The diagram in the center represents the same curves after they have been moved a small distance apart from each other. Now, the points at which the curves intersect can be treated as the intersection of two different 5-branes of the form $(p_1,q_1)$ and $(p_2,q_2)$. This can be computed by using the determinant in equation \ref{eq:det}. The \emph{stable intersection} is defined as the sum over all such intersection numbers. Note that a different deformation of the initial brane system, where the sub webs are moved apart in a different direction, always results in the same value for the stable intersection. Alternatively, the intersection numbers can be computed from the dual toric diagram. The dual toric diagram of the displaced sub webs is displayed at the right of figure \ref{fig:stableIntersection}. This toric diagram has four different polygons: two squares and two triangles. The stable intersection is given by the total area of all the polygons that have edges of both colors. In this case, the triangles do not contribute, since their edges are of a single color. The two squares contribute, and the sum of their areas is 2. 

Hence, the stable intersection between the two sub webs in phase (b) has value of 2. This value corresponds to the number of edges (or hypermultiplets) between the corresponding gauge groups of the quiver in equation \ref{eq:A1}. Equivalently one can think of 1 D3 brane which is stretched between the sub webs per each intersection point. 

On this particular example, we have found a way to read the 3d quivers that describe (via equations \ref{E3instanton}, \ref{eq:A2} and \ref{eq:A1}) the $\mathcal{H}_\infty$ of the 5d theory. Let us explore some more examples where the $\mathcal{H}_\infty$ is already known in the next sections. After that we provide the final answer on reading the 3d quiver for the whole 3 parameter family of 5d SQCD theories, an answer which is actually more general and applies for any five brane web.

\subsection{$E_4$ -- Necklace Quiver}

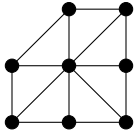
\begin{figure}
	\centering
		\begin{tikzpicture}[scale=0.75]
			\node[toric] at (0,0) {};
			\node[toric] at (0,1) {};
			\node[toric] at (1,0) {};
			\node[toric] at (1,1) {};
			\node[toric] at (1,2) {};
			\node[toric] at (2,0) {};
			\node[toric] at (2,1) {};
			\node[toric] at (2,2) {};
			\draw (0,0)--(0,1)--(1,2)--(2,2)--(2,1)--(2,0)--(1,0)--(0,0);
			\draw (1,1)--(0,0);
			\draw (1,1)--(0,1);
			\draw (1,1)--(1,2);
			\draw (1,1)--(2,2);
			\draw (1,1)--(2,1);
			\draw (1,1)--(2,0);
			\draw (1,1)--(1,0);
		\end{tikzpicture}
	\caption{Toric diagram corresponding to the 5d $\mathcal N=1$ SQCD theory with SU(2) gauge group and $N_f=3$.}
	\label{fig:E4toric}
\end{figure}

%
\begin{figure}
	\centering
		\begin{tikzpicture}[scale=0.75]
			\draw
				(1,4)--(1,2)--(2,1)--(3,1)--(4,2)--(4,3)--(3,4)--(1,4)
				(0,5)--(1,4)
				(0,2)--(1,2)
				(2,0)--(2,1)
				(3,0)--(3,1)
				(5,2)--(4,2)
				(5,3)--(4,3)
				(3,5)--(3,4);
			\node[7brane] at (0,5) {};
			\node[7brane] at (0,2) {};
			\node[7brane] at (2,0) {};
			\node[7brane] at (3,0) {};
			\node[7brane] at (5,2) {};
			\node[7brane] at (5,3) {};
			\node[7brane] at (3,5) {};
			\draw
				(6,2)--(9,2)
				(8.8,2.2)--(9,2)
				(8.8,1.8)--(9,2);
			\node[] at (7.5,2.5) {$\frac{1}{g^2}, m_i, a \rightarrow 0$};
			
		\draw(11.1,2-1)--(11.1,3);
		\draw[](12,2-0.1)--(10,2-0.1);
		\draw[](11,2)--(11,2-1);
		\draw[](11,2)--(10,3);
		\draw[](11,2)--(12,2);
		\draw[](11.05,2-1)--(11.05,2-2);
		\draw[](12,2-0.05)--(13,2-0.05);
			\node[7brane] at (11.1,3) {};
			\node[7brane] at (10,2-0.1) {};
			\node[7brane] at (10,3) {};
			\node[7brane] at (11.05,2-1) {};
			\node[7brane] at (11.05,2-2) {};
			\node[7brane] at (12,2-0.05) {};
			\node[7brane] at (13,2-0.05) {};
		\end{tikzpicture}
	\caption{Five brane webs corresponding to the 5d SQCD theory with SU(2) gauge group and $N_f=3$ before and after taking the gauge coupling $g$ to infinity and all the masses $m_i$ as well as the VEV $a$ of the adjoint scalar field to zero. Note that coincident 5-branes are depicted slightly apart in the right-hand diagram. This is done to make the diagram easier to read, but the branes at the center of the diagram that look parallel to each other should be considered as fully coincident.}
	\label{fig:E4finite}
\end{figure}
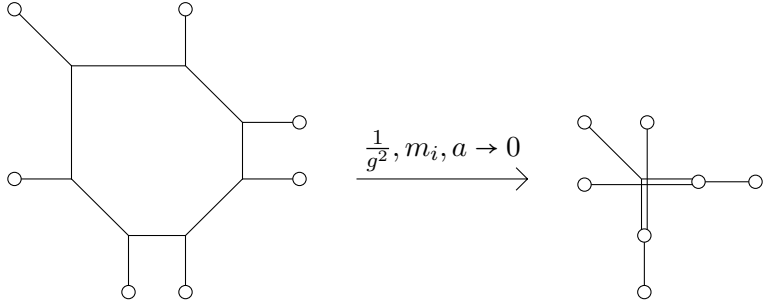

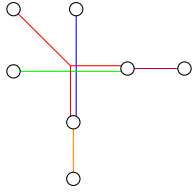
\begin{figure}
	\centering
	\begin{tikzpicture}[scale=0.75]
		\draw[blue](0.1,-1)--(0.1,1);
		\draw[green](1,-0.1)--(-1,-0.1);
		\draw[red](0,0)--(0,-1);
		\draw[red](0,0)--(-1,1);
		\draw[red](0,0)--(1,0);
		\draw[orange](0.05,-1)--(0.05,-2);
		\draw[purple](1,-0.05)--(2,-0.05);
			\node[7brane] at (0.1,1) {};
			\node[7brane] at (-1,-0.1) {};
			\node[7brane] at (-1,1) {};
			\node[7brane] at (0.05,-1) {};
			\node[7brane] at (0.05,-2) {};
			\node[7brane] at (1,-0.05) {};
			\node[7brane] at (2,-0.05) {};
	\end{tikzpicture}
	\caption{Brane web of $SU(2)$ with $N_f=3$ at infinite coupling. There are 5 different sub webs depicted in different colors.}
	\label{fig:E4}
\end{figure}

Let us now turn to $N_c=2$, $N_f=3$. The corresponding toric diagram is represented in figure \ref{fig:E4toric}. The brane system at finite and infinite coupling is depicted in figure \ref{fig:E4finite}. This case is different from the previous example, in the sense that there is a unique way of maximally dividing the system at the infinite coupling limit into sub webs. Correspondingly, the Higgs branch has one component. The subdivision is depicted in figure \ref{fig:E4}. It is known \cite{Seiberg:1996bd} that the Higgs branch of the theory at infinite coupling is the closure of the minimal nilpotent orbit of $\mathfrak{sl}(5,\mathbb{C})$:
\begin{equation}\label{E4instanton}
	\mathcal{H}_{\infty}\left(\node{\flav 3}{SU(2)}\right) = \orbit{min}{A_4}
\end{equation}
This space is isomorphic to the reduced moduli space of one $A_4$ instanton on $\mathbb{C}^2$ and can also be written as the Coulomb branch of a 3d $\mathcal N=4$ quiver gauge theory where the quiver is the affine Dynkin diagram of $A_4$ \cite{Intriligator:1996ex}:
\begin{equation}\label{eq:A4}
	\orbit{min}{A_4} = \mathcal{C}^{3d}\left(~\overset{\displaystyle\node{1}{}}{\overset{\diagup ~~~~~~~~\diagdown}{\node{}1-\node{}1-\node{}1-\node{}1}} ~\right)
\end{equation}

\begin{table}
	\centering
	\begin{tabular}{|C|C|C|}
		\hline 
		Displaced Branes & Toric Diagram & Stable Intersection \\ \hline
		\begin{tikzpicture}[scale=0.75]
			\draw[red](0,2)--(1,1)--(1,0) (1,1)--(2,1);
			\draw[blue](1.3,0)--(1.3,2); 
			\addvmargin{2mm}
		\end{tikzpicture} &
		\begin{tikzpicture}[scale=0.75]
			\draw[red](0,0)--(1,0)--(1,1)--(0,0) (2,0)--(2,1);
			\draw[blue](1,0)--(2,0) (1,1)--(2,1); 
			\node[toric]at(0,0){};
			\node[toric]at(1,0){};
			\node[toric]at(2,0){};
			\node[toric]at(1,1){};
			\node[toric]at(2,1){};
			\addvmargin{2mm}
		\end{tikzpicture} &
		1
			\\ \hline
		\begin{tikzpicture}[scale=0.75]
			\draw[red](0,2)--(1,1)--(1,0) (1,1)--(2,1);
			\draw[green](0,1-0.3)--(2,1-.3); 
			\addvmargin{2mm}
		\end{tikzpicture} &
		\begin{tikzpicture}[scale=0.75]
			\draw[red](0,0)--(1,0)--(1,1)--(0,0) (0,-1)--(1,-1);
			\draw[green](0,0)--(0,-1) (1,0)--(1,-1); 
			\node[toric]at(0,0){};
			\node[toric]at(0,-1){};
			\node[toric]at(1,1){};
			\node[toric]at(1,0){};
			\node[toric]at(1,-1){};
			\addvmargin{2mm}
		\end{tikzpicture}&
		1	
		 	\\ \hline
		\begin{tikzpicture}[scale=0.75]
			\draw[blue](1,2)--(1,0);
			\draw[green](0,1)--(2,1);
			\addvmargin{2mm}
		\end{tikzpicture} &
		\begin{tikzpicture}[scale=0.75]
			\draw[green](1,0)--(1,1) (2,0)--(2,1);
			\draw[blue](1,0)--(2,0) (1,1)--(2,1); 
			\node[toric]at(1,0){};
			\node[toric]at(2,0){};
			\node[toric]at(1,1){};
			\node[toric]at(2,1){};
			\addvmargin{2mm}
		\end{tikzpicture} &
		1
			\\ \hline
	\end{tabular}
	\caption{Stable Intersections of the sub webs of $E_4$.}
	\label{tab:E4stableIntersections}
\end{table}

In the sub division of branes of figure \ref{fig:E4} there are 5 different sub webs that can move between seven branes, generating the Higgs branch at infinite coupling (represented with different colors: red, blue, green, orange and purple). Once again, each of them corresponds to a different gauge node in equation \ref{eq:A4}, each with multiplicity 1 as there is one copy of each sub web. Equation \ref{eq:A4} indicates that each of the sub webs needs to be connected to two and only two other sub webs. Let us see how it works. The orange segment can only connect with the red web and the blue segment, and the purple segment can only connect with the red web and the green segment. This leaves only one possibility: the green segment and the blue segment need to be connected by a single link, and the red web needs to be disconnected from the blue segment and the green segment. The stable intersection between any pair (red, blue), (red, green) and (blue, green) is of value 1, see the corresponding dual toric diagrams in table \ref{tab:E4stableIntersections}. The new feature introduced in this example is the possibility of different sub webs ending on the same 7-brane. We see that if two sub webs end on the same 7-brane from opposite sides, we consider the corresponding gauge nodes connected by a link (i.e. the connection between orange and blue, or between orange and red). However, if two sub webs end on the same 7-brane on the same side, this contributes with $-1$ to the number of links between the corresponding gauge nodes. This effect removes the link between red and blue (similarly, red and green) that would arise due to their stable intersection, leaving their corresponding gauge group nodes disconnected. One can summarize these observations in the following:
\begin{summary}[Quiver edge multiplicity -- Stable Intersection]
	The number of edges between two gauge nodes corresponding to two different brane sub webs is equal to the stable intersection between the sub webs plus the contribution from the 7-branes. The contribution from a 7-brane is positive if the sub webs end on it from opposite sides, and negative otherwise.
\end{summary}

%

With this rule, the 3d quiver as read from the diagram in figure \ref{fig:E4} is:

\begin{equation}\label{eq:quiverE4}
\overset{\displaystyle{\color{red}\node{1}{}}}{\overset{\diagup ~~~~~~~~\diagdown}{{\color{orange}\node{}1}-{\color{blue}\node{}1}-{\color{green}\node{}1}-{\color{purple}\node{}1}}} 
\end{equation}
Hence, one recovers the quiver that describes the $\mathcal{H}_\infty$ via equations \ref{E4instanton} and \ref{eq:A4}. It is interesting to see that even though the brane system does not resemble a necklace, the intersection numbers between brane webs do form a necklace quiver. This quiver has the feature that all edges have multiplicity 1 and the brane system has only one way to maximally divide it into sub webs, hence is part of the quivers which were computed in \cite{Ferlito:2017xdq} in full agreement.

\subsection{$E_5$ -- Node Multiplicity}

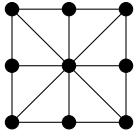
\begin{figure}
	\centering
		\begin{tikzpicture}[scale=0.75]
			\node[toric] at (0,0) {};
			\node[toric] at (0,1) {};
			\node[toric] at (0,2) {};
			\node[toric] at (1,0) {};
			\node[toric] at (1,1) {};
			\node[toric] at (1,2) {};
			\node[toric] at (2,0) {};
			\node[toric] at (2,1) {};
			\node[toric] at (2,2) {};
			\draw (0,0)--(0,1)--(0,2)--(1,2)--(2,2)--(2,1)--(2,0)--(1,0)--(0,0);
			\draw (1,1)--(0,0);
			\draw (1,1)--(0,1);
			\draw (1,1)--(0,2);
			\draw (1,1)--(1,2);
			\draw (1,1)--(2,2);
			\draw (1,1)--(2,1);
			\draw (1,1)--(2,0);
			\draw (1,1)--(1,0);
		\end{tikzpicture}
	\caption{Toric diagram corresponding to the 5d SQCD theory with SU(2) gauge group and $N_f=4$.}
	\label{fig:E5toric}
\end{figure}

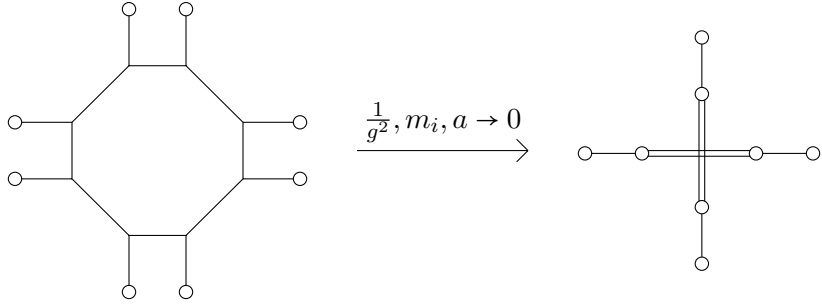
\begin{figure}
	\centering
		\begin{tikzpicture}[scale=0.75]
			\draw
				(1,3)--(1,2)--(2,1)--(3,1)--(4,2)--(4,3)--(3,4)--(2,4)--(1,3)
				(2,5)--(2,4)
				(0,3)--(1,3)
				(0,2)--(1,2)
				(2,0)--(2,1)
				(3,0)--(3,1)
				(5,2)--(4,2)
				(5,3)--(4,3)
				(3,5)--(3,4);
			\node[7brane] at (0,3) {};
			\node[7brane] at (0,2) {};
			\node[7brane] at (2,0) {};
			\node[7brane] at (3,0) {};
			\node[7brane] at (5,2) {};
			\node[7brane] at (5,3) {};
			\node[7brane] at (3,5) {};
			\node[7brane] at (2,5) {};
			\draw
				(6,2.5)--(9,2.5)
				(8.8,2.7)--(9,2.5)
				(8.8,2.3)--(9,2.5);
			\node[] at (7.5,3) {$\frac{1}{g^2}, m_i, a \rightarrow 0$};
		\draw[](12.1,2.5-1)--(12.1,3.5);
		\draw[](13,2.5-0.1)--(12-1,2.5-0.1);
		\draw[](12,3.5)--(12,2.5-1);
		\draw[](12-1,2.5)--(13,2.5);
		\draw[](12.05,2.5-1)--(12.05,0.5);
		\draw[](13,2.5-0.05)--(14,2.5-0.05);
		\draw[](12.05,3.5)--(12.05,4.5);
		\draw[](12-1,2.5-0.05)--(12-2,2.5-0.05);
			\node[7brane] at (12.05,2.5-1) {};
			\node[7brane] at (12.05,2.5-2) {};
			\node[7brane] at (13,2.5-0.05) {};
			\node[7brane] at (14,2.5-0.05) {};
			\node[7brane] at (12.05,3.5) {};
			\node[7brane] at (12.05,4.5) {};
			\node[7brane] at (12-1,2.5-0.05) {};
			\node[7brane] at (12-2,2.5-0.05) {};
		\end{tikzpicture}
	\caption{Five brane webs corresponding to the 5d SQCD theory with SU(2) gauge group and $N_f=4$ before and after taking the gauge coupling $g$ to infinity and all the masses $m_i$ as well as the VEV $a$ of the adjoint scalar field to zero.}
	\label{fig:E5finite}
\end{figure}

Let us study the $SU(N_c)$ 5d SQCD theory with parameter $N_c=2$ and number of flavors $N_f=4$. The toric diagram is depicted in figure \ref{fig:E5toric} and the brane system at finite and infinite coupling is depicted in figure \ref{fig:E5finite}. The Higgs branch at infinite coupling is the reduced moduli space of one $D_5$ instanton on $\mathbb{C}^2$. This space is a hyperK\"ahler cone, isomorphic to the closure of the minimal nilpotent orbit of $\mathfrak{so}(10,\mathbb{C})$:
\begin{equation}\label{E5instanton}
	\mathcal{H}_{\infty}\left(\node{\flav 4}{SU(2)}\right) = \orbit{min}{D_5}
\end{equation}
This space can be found as the Coulomb branch of a 3d $\mathcal{N}=4$ quiver which is the affine Dynkin diagram of $D_5$ \cite{Intriligator:1996ex}:
\begin{equation}\label{eq:E5}
	\orbit{min}{D_5} = \mathcal{C}^{3d}\left(~\node{}1-\node{\upnode 1}2-\node{\upnode 1}2-\node{}1 ~\right)
\end{equation}

\begin{figure}
	\centering
	\begin{tikzpicture}[scale=0.75]
		\draw[blue](0.1,-1)--(0.1,1);
		\draw[green](1,-0.1)--(-1,-0.1);
		\draw[blue](0,1)--(0,-1);
		\draw[green](-1,0)--(1,0);
		\draw[orange](0.05,-1)--(0.05,-2);
		\draw[purple](1,-0.05)--(2,-0.05);
		\draw[red](0.05,1)--(0.05,2);
		\draw[brown](-1,-0.05)--(-2,-0.05);
			\node[7brane] at (0.05,-1) {};
			\node[7brane] at (0.05,-2) {};
			\node[7brane] at (1,-0.05) {};
			\node[7brane] at (2,-0.05) {};
			\node[7brane] at (0.05,1) {};
			\node[7brane] at (0.05,2) {};
			\node[7brane] at (-1,-0.05) {};
			\node[7brane] at (-2,-0.05) {};
	\end{tikzpicture}
	\caption{Maximal subdivision of the brane web of $SU(2)$ with $N_f=4$ at infinite coupling.}
	\label{fig:E5}
\end{figure}
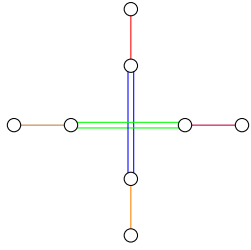

The new feature of this example that does not occur in the previous two examples is the appearance of nodes of rank higher than 1 in the 3d quiver of \ref{eq:E5}. This is easy to relate to the brane web: a number of $n$ identical sub webs corresponds to a single node of rank $n$. Figure \ref{fig:E5} depicts the subdivision of the brane system into the maximal number of sub webs. There are two segments in blue that are identical and correspond to one of the nodes of rank 2, and two segments in green that are identical and correspond to the other rank 2 node. In order to establish the links in the 3d quiver, consider a single copy of each sub web. For example, a single blue segment has stable intersection of 1 with a single green segment, therefore there is a single link with multiplicity one between them (there are no extra contributions to the number of links since they do not share any common 7-branes). Similarly, a single blue segment ends on the same 7-brane as the orange segment, and they do it on opposite sides, hence there is a single link between them (just as between the blue and orange segments in figure \ref{fig:E4}). Therefore, the quiver as read from the brane system in figure \ref{fig:E5} is:
\begin{equation}
	\cnode{orange}{}{1}-\cnode{blue}{\cupnode{red}{1}}{2}-\cnode{green}{\cupnode{brown}{1}}{2}-\cnode{purple}{}{1}
\end{equation}
This quiver indeed describes the correct Higgs branch, according to equations \ref{E5instanton} and \ref{eq:E5}.

\subsection{$E_6$ - 7-Brane Contributions without Stable Intersection}

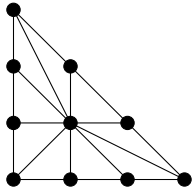
\begin{figure}
	\centering
		\begin{tikzpicture}[scale=0.75]
			\node[toric] at (0,0) {};
			\node[toric] at (0,1) {};
			\node[toric] at (0,2) {};
			\node[toric] at (0,3) {};
			\node[toric] at (1,0) {};
			\node[toric] at (1,1) {};
			\node[toric] at (1,2) {};
			\node[toric] at (2,0) {};
			\node[toric] at (2,1) {};
			\node[toric] at (3,0) {};
			\draw (0,0)--(0,1)--(0,2)--(0,3)--(1,2)--(2,1)--(3,0)--(2,0)--(1,0)--(0,0);
			\draw (1,1)--(0,0);
			\draw (1,1)--(0,1);
			\draw (1,1)--(0,2);
			\draw (1,1)--(0,3);
			\draw (1,1)--(1,2);
			\draw (1,1)--(2,1);
			\draw (1,1)--(2,0);
			\draw (1,1)--(1,0);
			\draw (1,1)--(3,0);
		\end{tikzpicture}
	\caption{Toric diagram corresponding to the 5d SQCD theory with SU(2) gauge group and $N_f=5$.}
	\label{fig:E6toric}
\end{figure}

\begin{figure}
	\centering
		\begin{tikzpicture}[scale=0.75]
			\draw
				(1,3)--(1,2)--(2,1)--(3,1)--(4,2)--(5,4)--(5,5)--(4,5)--(2,4)--(1,3)
				(0,4)--(2,4)
				(0,3)--(1,3)
				(0,2)--(1,2)
				(2,0)--(2,1)
				(3,0)--(3,1)
				(4,0)--(4,2)
				(6.5,5.5)--(5,4)
				(6,6)--(5,5)
				(5.5,6.5)--(4,5);
			\node[7brane] at (0,4) {};
			\node[7brane] at (0,3) {};
			\node[7brane] at (0,2) {};
			\node[7brane] at (2,0) {};
			\node[7brane] at (3,0) {};
			\node[7brane] at (4,0) {};
			\node[7brane] at (6.5,5.5) {};
			\node[7brane] at (6,6) {};
			\node[7brane] at (5.5,6.5) {};
			\draw
				(7,3)--(10,3)
				(9.8,3.2)--(10,3)
				(9.8,2.8)--(10,3);
			\node[] at (8.5,3.5) {$\frac{1}{g^2}, m_i, a \rightarrow 0$};
		\draw[](14-3,3)--(14-2,3);
		\draw[](14-2,3.05)--(14-1,3.05);
		\draw[](14-2,3-0.05)--(14-1,3-0.05);
		\draw[](14-1,3)--(14,3);
		\draw[](14-1,3.1)--(14,3.1);
		\draw[](14-1,3-0.1)--(14-0.1,3-0.1);
		\node[7brane] at (14-3,3) {};
		\node[7brane] at (14-2,3) {};
		\node[7brane] at (14-1,3) {};
		\draw[](14,3-3)--(14,3-2);
		\draw[](14.05,3-2)--(14.05,3-1);
		\draw[](14-0.05,3-2)--(14-0.05,3-1);
		\draw[](14,3-1)--(14,3);
		\draw[](14.1,3-1)--(14.1,3);
		\draw[](14-0.1,3-1)--(14-0.1,3-0.1);
		\node[7brane] at (14,3-3){};
		\node[7brane] at (14,3-2){};
		\node[7brane] at (14,3-1){};
		\draw[](17,6)--(16,5);
		\draw[](16.05,5)--(15.05,4); 
		\draw[](15.95,5)--(14.95,4);
		\draw[](14,3)--(15,4);
		\draw[](14.1,3)--(15.1,4);
		\draw[](14,3.1)--(15,4.1);
		\node[7brane] at (17,6){};
		\node[7brane] at (16,5){};
		\node[7brane] at (15,4){};
		\end{tikzpicture}
	\caption{Five brane webs corresponding to the 5d SQCD theory with SU(2) gauge group and $N_f=5$ before and after taking the gauge coupling $g$ to infinity all the masses $m_i$ as well as the VEV $a$ of the adjoint scalar field to zero.}
	\label{fig:E6finite}
\end{figure}
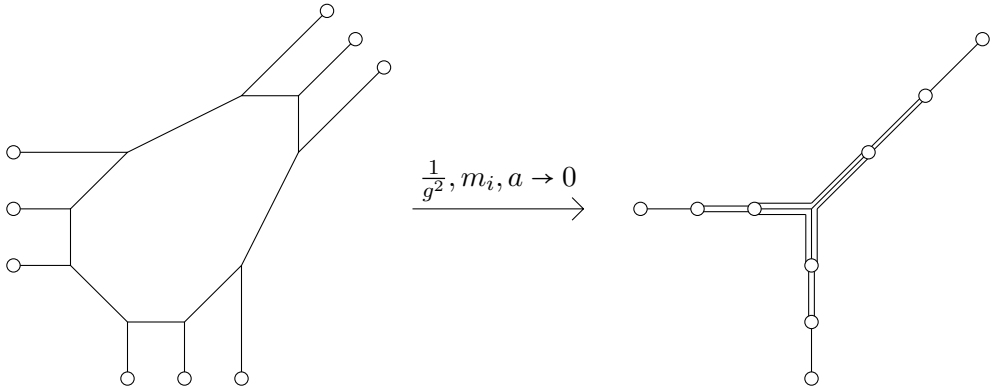

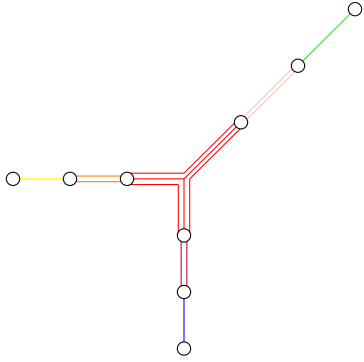
\begin{figure}
	\centering
	\begin{tikzpicture}[scale=0.75]
		\draw[yellow](-3,0)--(-2,0);
		\draw[orange](-2,0.05)--(-1,0.05);
		\draw[orange](-2,-0.05)--(-1,-0.05);
		\draw[red](-1,0)--(0,0);
		\draw[red](-1,0.1)--(0,0.1);
		\draw[red](-1,-0.1)--(-0.1,-0.1);
		\node[7brane] at (-3,0) {};
		\node[7brane] at (-2,0) {};
		\node[7brane] at (-1,0) {};
		\draw[blue](0,-3)--(0,-2);
		\draw[purple](0.05,-2)--(0.05,-1);
		\draw[purple](-0.05,-2)--(-0.05,-1);
		\draw[red](0,-1)--(0,0);
		\draw[red](0.1,-1)--(0.1,0);
		\draw[red](-0.1,-1)--(-0.1,-0.1);
		\node[7brane] at (0,-3){};
		\node[7brane] at (0,-2){};
		\node[7brane] at (0,-1){};
		\draw[green](3,3)--(2,2);
		\draw[pink](2.05,2)--(1.05,1);
		\draw[pink](1.95,2)--(.95,1);
		\draw[red](0,0)--(1,1);
		\draw[red](0.1,0)--(1.1,1);
		\draw[red](0,0.1)--(1,1.1);
		\node[7brane] at (3,3){};
		\node[7brane] at (2,2){};
		\node[7brane] at (1,1){};
	\end{tikzpicture}
	\caption{Maximal subdivision of the brane web of $SU(2)$ with $N_f=5$ at infinite coupling.}
	\label{fig:E6}
\end{figure}

The next example considers the case of $N_c=2$, $N_f=5$. In this case there is no need to compute \emph{stable intersections} between sub webs, since all the contributions to the number of links between two gauge nodes in the 3d quiver are given by the ending of the different sub webs on shared 7-branes. The toric diagram is depicted in figure \ref{fig:E6toric} and the brane system is represented in figure \ref{fig:E6finite}. The Higgs branch at infinite coupling is isomorphic to the reduced moduli space of one $E_6$ instanton on $\mathbb{C}^2$ (alternatively, it can also be described as the closure of the minimal nilpotent orbit of $\mathfrak{e}_6$):
\begin{equation}
	\mathcal{H}_{\infty}\left(\node{\flav 5}{SU(2)}\right) = \orbit{min}{E_6}
\end{equation}
Hence the $\mathcal{H}_\infty$ has a single component, as can be seen from the fact that there is a unique maximal subdivision of the brane system, figure \ref{fig:E6}. This space can be found as the Coulomb branch of a 3d $\mathcal{N}=4$ quiver which is the affine Dynkin diagram of $E_6$ \cite{Intriligator:1996ex}:
\begin{equation}\label{eq:E6}
	\orbit{min}{E_6} = \mathcal{C}^{3d}\left(~\node{}1-\node{}2-\node{\topnode{\topnode{}1}2}3-\node{}2-\node{}1 ~\right)
\end{equation}
According to the rules developed in the previous examples, the quiver can be read directly from the brane configuration in figure \ref{fig:E6}:
\begin{equation}\label{eq:quiverE6}
	\cnode{yellow}{}1-\cnode{orange}{}2-\cnode{red}{\ctopnode{pink}{\ctopnode{green}{}1}2}3-\cnode{purple}{}2-\cnode{blue}{}1 
\end{equation}
This is precisely the expected result. Note that once more, $n$ multiple identical copies of the same sub web correspond to a single gauge node with a rank $n$. In this case there is no need to compute stable intersections, since none of the different sub webs intersect. The number of links between the different gauge nodes of equation \ref{eq:quiverE6} are only determined by different sub webs ending on the same 7-brane from opposite directions.

\subsection{Super Yang-Mills -- Edge Multiplicity}

\begin{figure}
	\centering
		\begin{tikzpicture}[scale=0.5]
			\draw
				(3,5)--(5,4)--(7,4)--(6,5)--(3,5)
				(6,2)--(7,1)--(10,1)--(8,2)--(6,2)
				(0,6)--(3,5)
				(6,6)--(6,5)
				(7,0)--(7,1)
				(13,0)--(10,1)
				(5,4)--(5.1,3.9)
				(7,4)--(7.2,3.9)
				(6,2)--(5.8,2.1)
				(8,2)--(7.9,2.1);
			\node[7brane,label=above:{$(-N_c,1)$}] at (0,6) {};
			\node[7brane,label=above:{$(0,1)$}] at (6,6) {};
			\node[7brane,label=below:{$(0,1)$}] at (7,0) {};
			\node[7brane,label=below:{$(N_c,-1)$}] at (13,0) {};
			\node[dot] at (6.5,2.5) {};
			\node[dot] at (6.5,3) {};
			\node[dot] at (6.5,3.5) {};
			\draw
				(13-0.5,3)--(15+0.5,3)
				(14.8+0.5,3.2)--(15+0.5,3)
				(14.8+0.5,2.8)--(15+0.5,3);
			\node[] at (14,3.7) {$\frac{1}{g^2},a_i \rightarrow 0$};
			\draw (18,4)--(24,2);
			\draw (21,4)--(21,2);
			\node[7brane,label=above:{$(-N_c,1)$}] at (18,4) {};
			\node[7brane,label=above:{$(0,1)$}] at (21,4) {};
			\node[7brane,label=below:{$(0,1)$}] at (21,2) {};
			\node[7brane,label=below:{$(N_c,-1)$}] at (24,2) {};
		\end{tikzpicture}
	\caption{5d brane web corresponding to Super Yang-Mills theory with gauge group $SU(N_c)$, $N_f = 0$ and CS level $k=0$. The diagram on the right corresponds to the limit where the gauge coupling is taken to infinity and the VEVs $a_i$ of the adjoint scalar field to zero.}
	\label{fig:YMfinite}
\end{figure}
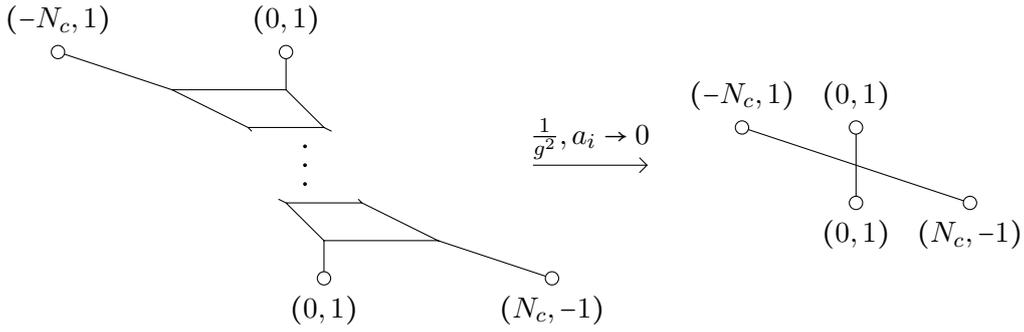

Another well known result \cite{Cremonesi:2015lsa} is the Higgs branch at infinite coupling of Super Yang-Mills $SU(N_c)$ with no flavors, $N_f=0$, and CS level $k=0$. The brane system is depicted in figure \ref{fig:YMfinite}. The Higgs branch at infinite coupling is \cite{Cremonesi:2015lsa}\footnote{For $N_c=2$ this was identified in \cite{Seiberg:1996bd} and confirmed by the brane picture in \cite{Aharony:1997bh}.}:
\begin{equation}
	\mathcal{H}_{\infty}\left(\node{\flav 0}{SU(N_c)_0}\right) = \mathbb{C}^2/\mathbb{Z}_{N_c}
\end{equation}

This can be written as the Coulomb branch of a 3d quiver with two gauge nodes of rank 1 and a number of $N_c$ hypermultiplets between them. Let us write it as the complete graph:
\begin{equation}
	\mathbb{C}^2/\mathbb{Z}_{N_c} = \mathcal{C}^{3d}\left(~\node{}1\xlongequal{N_c}\node{}1 ~\right)
\end{equation}

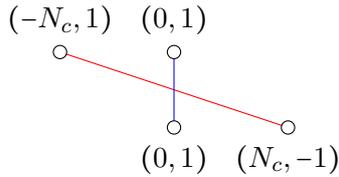
\begin{figure}
	\centering
	\begin{tikzpicture}[scale=0.5]
			\draw[red] (18,4)--(24,2);
			\draw[blue] (21,4)--(21,2);
			\node[7brane,label=above:{$(-N_c,1)$}] at (18,4) {};
			\node[7brane,label=above:{$(0,1)$}] at (21,4) {};
			\node[7brane,label=below:{$(0,1)$}] at (21,2) {};
			\node[7brane,label=below:{$(N_c,-1)$}] at (24,2) {};
	\end{tikzpicture}
	\caption{Maximal subdivision of the brane web of $SU(N_c)$ with $N_f=0$ and $k=0$ at infinite coupling.}
	\label{fig:YM}
\end{figure}

The same quiver is read from the brane picture in figure \ref{fig:YM}, since there are only two segments, corresponding to the two different nodes, and the stable intersection between them is just their intersection number $I$:
\begin{equation}
	I\{(-N_c,1),(0,1)\} = Abs\left(\begin{vmatrix}
		-N_c & 1\\ 
 		0 & 1
	\end{vmatrix}\right) = N_c
\end{equation}

Hence, the 3d quiver read directly from the branes is:
\begin{equation}
 \cnode{red}{}{1}\xlongequal{N_c}\cnode{blue}{}{1}
\end{equation}

Equipped with the examples above, we are now ready to generalize to the main result of this paper for the 3 parameter family of SQCD theories.

\section{Conjecture}\label{sec:conjecture}

In this section we present the main conjecture that contains all the information on reading the combinatorial data for the Higgs branch of the 5d theory from the brane web. Note that this technique is not restricted to infinite coupling, or to 5d theories where the gauge group is a single factor. Furthermore, it can also be used to obtain subspaces of the Higgs branch, if the subdivision of the brane web is not maximal. In section \ref{sec:classification} we use this conjecture to obtain the Higgs branches at infinite coupling for the 3 parameter family of 5d $\mathcal N = 1$ SQCD theories with gauge group $SU(N_c)$, number of colors $N_f$ and CS level $k$.

Before stating the conjecture, let us define three quantities that can be computed for any pair of brane sub webs in a given five brane web. The first thing that can be computed is the \emph{stable intersection} (which in this section is denoted as $SI$), as explained in section \ref{sec:knownExamples} and in \cite{2013arXiv1311.2360B}, by slightly displacing the sub webs with respect to each other, and adding the area of polygones with two colors in the toric dual diagram (see the example in figure \ref{fig:stableIntersection}). The next quantity that can be computed is the contribution to the number of hypermultiplets from 5-branes ending on the same 7-brane. Let the 7-branes shared by two different brane sub webs be denoted as $A_i$ ($i=1,2,3,...$). For each $A_i$, one can compute the following two quantities:
	\begin{itemize}
		\item $X_i =$ number of combinations of two 5-branes from the different brane sub webs which are attached to $A_i$ on \emph{opposite} sides.
		\item $Y_i = $ number of combinations of two 5-branes from the different brane sub webs which are attached to $A_i$ on the \emph{same} sides.
	\end{itemize}

Please see appendix \ref{sec:app} for an explicit computation of the quantities $SI$, $X_i$ and $Y_i$ in a given example.

\begin{conjecture}\label{con:main}
	Given a five brane web divided into sub webs that can move along the directions spanned by the 7-branes placed at the end of each $(p,q)$ brane, the moduli space generated by this motion is given as the moduli space of dressed monopole operators of a quiver. The quiver can be obtained in the following way. Each set of $m$ identical sub webs corresponds to a different gauge node with group $U(m)$ in the quiver. Given a pair of gauge nodes in the quiver, the number of edges $E$ between them is determined by selecting two sub webs, one corresponding to each node, and computing for them their stable intersection $SI$, as well as the quantities $X_i$ and $Y_i$ (defined above) for all the 7-branes $A_i$ shared by the different sub webs. The number of edges $E$ is given by
	\begin{equation}\label{eq:conj_E}
		E = SI + \sum_i X_i - \sum_i Y_i~.
	\end{equation}
\end{conjecture}

In order to connect this conjecture to the previous examples, note that the contribution $X_i$ gives rise to edges between the nodes in the quiver \ref{eq:quiverE6}. On the other hand, the contribution $Y_i$ makes sure that the red node in quiver \ref{eq:quiverE4} remains disconnected from the blue node and the green node.

\subsection{Comments}

Here, we make several comments on this conjecture.

\begin{figure}
\centering
\begin{minipage}{6cm}
\includegraphics[width=4.5cm]{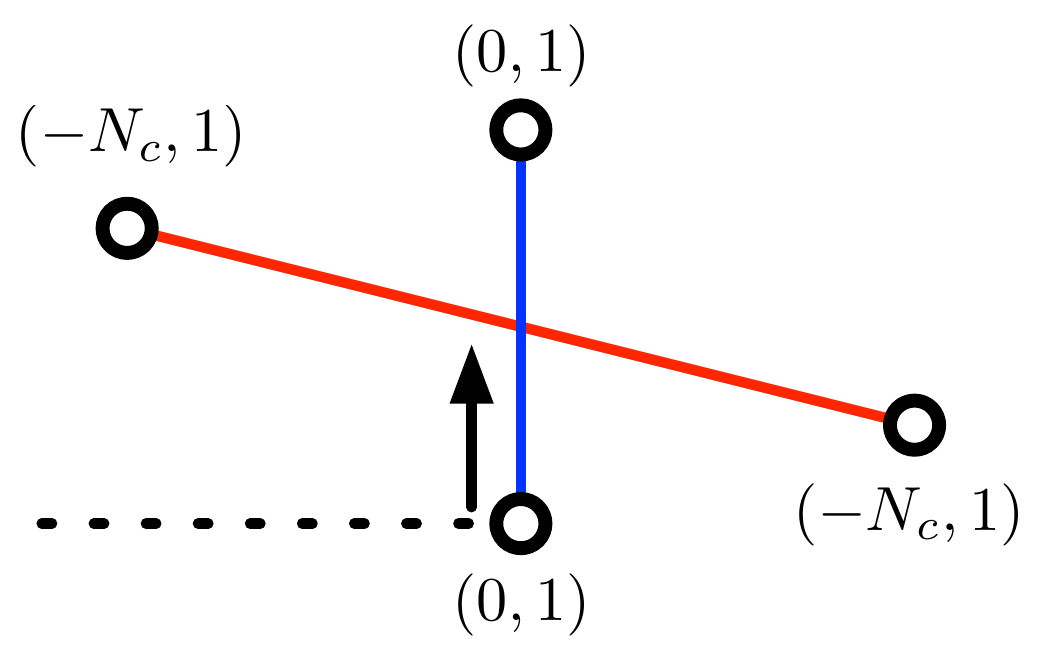}
\caption{Before Hanany-Witten transition. $SI=N_c, \sum_i X_i = 0$.}
\label{Fig:BeforeHW}
\end{minipage}
\begin{minipage}{6cm}
\includegraphics[width=4.5cm]{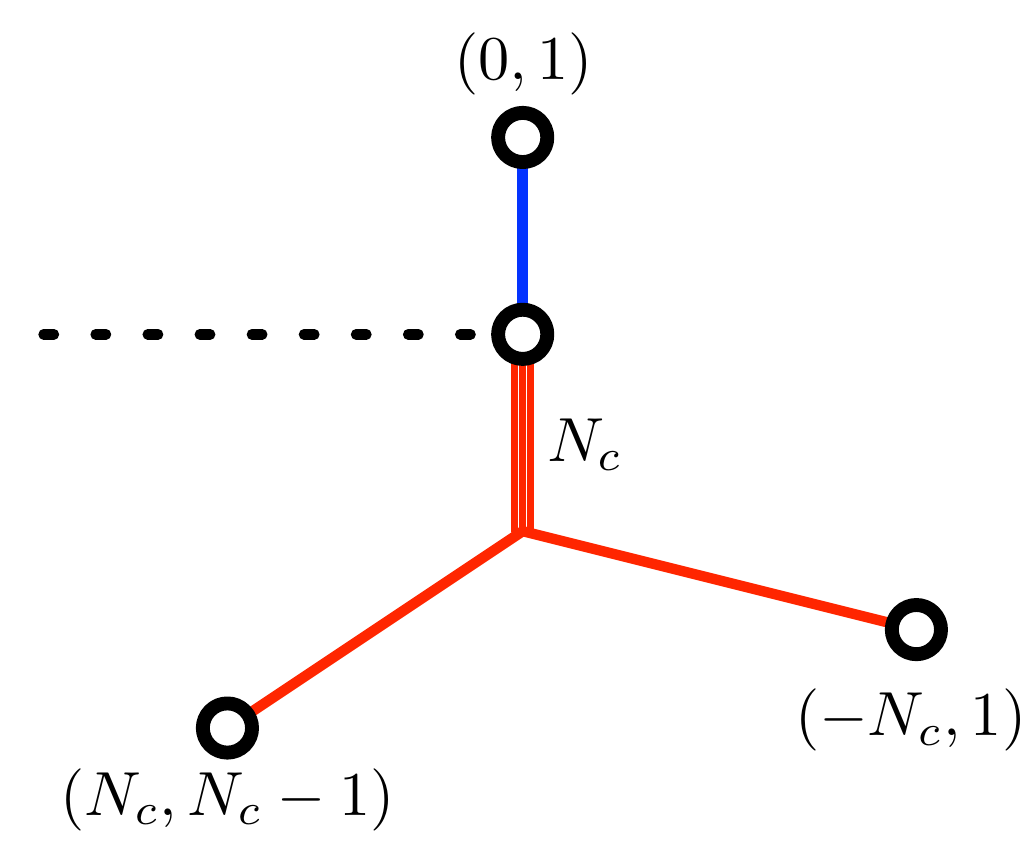}
\caption{After Hanany-Witten transition. $SI=0, \sum_i X_i = N_c$.}
\label{Fig:AfterHW}
\end{minipage}
\end{figure}

First, we discuss that the combination in equation \ref{eq:conj_E} can be interpreted as a natural generalization of the stable intersection defined in the tropical geometry literature. In this paper, we have generalized the tropical curve by introducing 7-branes so that the 5-branes are terminated at the 7-branes instead of extended to infinity. In this case, we can deform the sub web in such a way that the stable intersection discontinuously changes by moving a 7-brane across a 5-brane. As an example, we consider figure \ref{fig:YM}. As discussed before, the stable intersection for the (0,1) 5-brane and the $(-N_c,1)$ 5-brane is given by $N_c$. However, when we move one of the (0,1) 7-brane as depicted in figure \ref{Fig:BeforeHW} until it goes across the original $(-N_c,1)$ 5-brane, we find that the naively computed stable intersection changes from $SI=N_c$ to $SI=0$ in a sense that the two 5-brane sub webs are not intersecting any more if we ignore the contribution from the (0,1) 7-branes. Since this is not the desired property for ``stable'' intersection, we need a generalized quantity which is kept invariant even after the Hanany-Witten transition. The web diagram after Hanany-Witten transition is given in figure \ref{Fig:AfterHW}. Part of the of the original $(-N_c,1)$ 5-brane is changed to $(N_c,N_c-1)$ 5-brane by going across the monodromy cut created by the $(0,1)$ 7-brane. The important point is that new (0,1) 5-branes are created by Hanany-Witten transition.
In general, when $(p,q)$ 7-brane goes across the $(p',q')$ 5-brane, the number of $(p,q)$ 5-branes created by the Hanany-Witten effect is given by the intersection number of $(p,q)$ and $(p',q')$ \cite{Bergman:1998ej}, which, in our case, is $N_c$. 
Due to this effect, the contribution from the 7-branes is given by $\sum_i X_i =N_c$ in figure \ref{Fig:AfterHW}, which makes the combination in equation \ref{eq:conj_E} invariant under the Hanany-Witten transition. 
Necessity of the term $Y_i$ in equation \ref{eq:conj_E} could be analogously understood. If $n (\le N_c)$ (0,1) 5-branes are attached to the $(0,1)$ 7-brane before the Hanany-Witten transition, these will disappear and new $N_c - n$ $(0,1)$ 5-branes will be created after the Hanany-Witten transition. Again, the combination in equation \ref{eq:conj_E} is invariant also in this case.
Therefore, we claim that this combination is the natural generalization of the stable intersection number in a sense that it is invariant under the deformation of the web diagram after including the Hanany-Witten transition. We expect this property to be true for any 5-brane web diagram.

Next, we discuss the physical interpretation of this stability. 
It is often claimed that the Higgs branch is ``stable'' in a sense that it does not receive any quantum corrections and is independent of any gauge coupling. However, this statement should be interpreted in a careful way because new Higgs branch directions can open up at infinite coupling. 
Since the generalized stable intersection in equation \ref{eq:conj_E} does not change by continuous deformations of the brane webs once we fix the sub webs, it indicates that the quiver is stable under deformation of the external parameters like masses and inverse gauge couplings.
We expect that the stability of the Higgs branch can be correctly interpreted in terms of this stability of the quiver ensured by the property of the stable intersection.
We keep the detailed study of this point as our future work.

\section{New Notions}\label{sec:newNotions}

We proceed with examples which demonstrate new features that arise after utilizing Conjecture \ref{con:main} in the analysis of 5d Higgs branches.

\subsection{Union of Three Cones}

Let us study the case of 5d $\mathcal{N} = 1$ SQCD with gauge group $SU(5)$, with $N_f=6$ fundamental flavors and CS level $k=1$. The corresponding toric diagram is depicted in figure \ref{fig:5,6,1,toric}. The brane web of the theory and the limit where the coupling is taken to infinity are depicted in figure \ref{fig:5,6,1}. The Higgs branch at infinite coupling $\mathcal{H}_\infty$ was not known before the present paper. Now we are able to compute it as the moduli space of dressed monopole operators, by maximally dividing the brane system into sub webs and then applying Conjecture \ref{con:main}. We see in the example of $SU(2)$ with two flavors that there are two different ways of maximally subdividing the brane system (figure \ref{fig:E3}), and this implies that $\mathcal{H}_\infty$ is the union of two cones (it has two components). In the present case, there are three different ways of maximally subdividing the brane web, see figure \ref{fig:5,6,1,phases}. This means that the Higgs branch at infinite coupling is a union of three cones:
\begin{equation}\label{eq:union}
	\mathcal{H}_{\infty}\left(\node{\flav 6}{SU(5)_1}\right) = C_1 \cup C_2 \cup C_3
\end{equation}
The three cones $C_1$, $C_2$ and $C_3$ are computed utilizing Conjecture \ref{con:main} on each different maximal subdivision, to obtain a particular quiver. These quivers are depicted on the bottom of figure \ref{fig:5,6,1,phases}. Hence:
\begin{align}
	C_1 & = \mathcal{C}^{3d}\left(~\cnode{black}{}1 ~-~\cnode{black}{}{2}~-~\overset{\overset{\displaystyle{\color{black}\overset 1 \circ}}{\parallel}}{\cnode{black}{}{3}}~-~\cnode{black}{}{2}~-~\cnode{black}{}1~\right) \\
	C_2 & = \mathcal{C}^{3d}\left(~\cnode{black}{}1 ~-~\overset{\overset{\displaystyle{\color{black}\overset 1 \circ}\xlongequal{3}{\color{black}\overset 1 \circ}}{\diagup~~~~~~~~\diagdown}}{\cnode{black}{}{2}~-~\cnode{black}{}{2}~-~\cnode{black}{}{2}}~-~\cnode{black}{}1~\right)\\
	C_3 & = \mathcal{C}^{3d}\left(~\cnode{black}{}1 \overset{\overset{\displaystyle{\color{black}\overset 1 \circ}~~~\xlongequal{4}~~~{\color{black}\overset 1 \circ}}{\diagup~~~~~~~~~~~~~~~~~~\diagdown}}{~-~\cnode{black}{}{1}~-~\cnode{black}{}{1}~-~\cnode{black}{}{1}~-~}\cnode{black}{}1~\right)
\end{align}

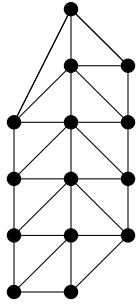
\begin{figure}
	\centering
	\begin{tikzpicture}[scale=0.75]
		\node[toric] at(0, 0){};
		\node[toric] at(0, 1){};
		\node[toric] at(0, 2){};
		\node[toric] at(0, 3){};
		\node[toric] at(0, 4){};
		\node[toric] at(0, 5){};
		\node[toric] at(-1, 0){};
		\node[toric] at(1, 4){};
		\node[toric] at(-1, 1){};
		\node[toric] at(1, 3){};
		\node[toric] at(-1, 2){};
		\node[toric] at(1, 2){};
		\node[toric] at(-1, 3){};
		\node[toric] at(1, 1){};
		\draw 	(0,0)--
				(-1,0)--
				(-1,1)--
				(-1,2)--
				(-1,3)--
				(0,5)--
				(1,4)--
				(1,3)--
				(1,2)--
				(1,1)--(0,0);
		\draw	(0,1)--(0,0);
		\draw	(0,1)--(-1,0);
		\draw	(0,1)--(-1,1);
		\draw	(0,1)--(1,1);
		\draw	(0,2)--(0,1);
		\draw	(0,2)--(-1,1);
		\draw	(0,2)--(-1,2);
		\draw	(0,2)--(1,1);
		\draw	(0,2)--(1,2);
		\draw	(0,3)--(0,2);
		\draw	(0,3)--(-1,2);
		\draw	(0,3)--(-1,3);
		\draw	(0,3)--(1,2);
		\draw	(0,3)--(1,3);
		\draw	(0,4)--(0,3);
		\draw	(0,4)--(-1,3);
		\draw	(0,4)--(1,3);
		\draw	(0,4)--(1,4);
		\draw	(0,5)--(0,4);
		\draw	(0,5)--(-1,3);
		\draw	(0,5)--(1,4);
	\end{tikzpicture}
	\caption{Toric diagram corresponding to the 5d SQCD theory with SU(5) gauge group, number of flavors $N_f=6$ and CS level $k=1$.}
	\label{fig:5,6,1,toric}
\end{figure}

\begin{figure}
	\centering
	\begin{tikzpicture}[scale=0.75]
		\draw (1,5)--(2,4)--(2,3.5)--(2.5,3)--(2.5,2.5)
		--(3,2)--(3,1.5)--(3.5,1)--(4,1)--(4,2)--(4.5,2.5)--(4.5,3)
		--(5,3.5)--(5,4)--(5.5,4.5)--(5.5,5)--(1,5);
		\draw (2,4)--(5,4);
		\draw (2.5,3)--(4.5,3);
		\draw (3,2)--(4,2);
		\draw (0,5.5)--(1,5);
		\draw (1,3.5)--(2,3.5);
		\draw (1.5,2.5)--(2.5,2.5);
		\draw (2,1.5)--(3,1.5);
		\draw (3.5,0)--(3.5,1);
		\draw (5,0)--(4,1);
		\draw (5.5,2.5)--(4.5,2.5);
		\draw (6,3.5)--(5,3.5);
		\draw (6.5,4.5)--(5.5,4.5);
		\draw (6.5,6)--(5.5,5);
		\node[7brane] at (0,5.5) {};
		\node[7brane] at (1,3.5) {};
		\node[7brane] at (1.5,2.5) {};
		\node[7brane] at (2,1.5) {};
		\node[7brane] at (3.5,0) {};
		\node[7brane] at (5,0) {};
		\node[7brane] at (5.5,2.5) {};
		\node[7brane] at (6,3.5) {};
		\node[7brane] at (6.5,4.5){};
		\node[7brane] at (6.5,6) {};
			\draw
				(8,3)--(10,3)
				(9.8,3.2)--(10,3)
				(9.8,2.8)--(10,3);
			\node[] at (9,3.5) {$\frac{1}{g^2}\rightarrow 0$};
		\draw[](11,3)--(14-2,3);
		\draw[](12,3.05)--(13,3.05);
		\draw[](12,3-0.05)--(13,3-0.05);
		\draw[](13,3)--(15,3);
		\draw[](13,3.1)--(15,3.1);
		\draw[](13,3-0.1)--(15,3-0.1);
		\draw (15,3.05)--(16,3.05);
		\draw (15,3-.05)--(16,3-.05);
		\draw (16,3)--(17,3);
		\draw (12,4)--(14,3);
		\draw (15,4)--(14,3);
		\draw (14,2)--(14,3);
		\draw (15,2)--(14,3);
		\node[7brane] at (11,3) {};
		\node[7brane] at (12,3) {};
		\node[7brane] at (13,3) {};
		\node[7brane] at (15,3) {};
		\node[7brane] at (16,3) {};
		\node[7brane] at (17,3) {};
		\node[7brane] at (12,4) {};
		\node[7brane] at (15,4) {};
		\node[7brane] at (14,2) {};
		\node[7brane] at (15,2) {};
	\end{tikzpicture}
	\caption{Brane system for 5d SQCD with gauge group $SU(5)$, $N_f=6$ flavors and CS level $k=1$ before and after taking the gauge coupling to infinity.}
	\label{fig:5,6,1}
\end{figure}
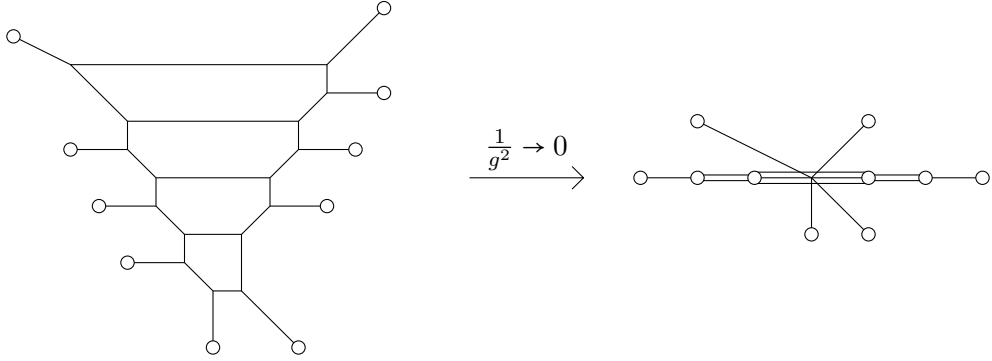

\begin{figure}
	\centering
	\begin{subfigure}[t]{.30\textwidth}\centering
	\begin{tikzpicture}[scale=0.75]
		\draw[green](11,3)--(14-2,3);
		\draw[](12,3.05)--(13,3.05);
		\draw[](12,3-0.05)--(13,3-0.05);
		\draw[green] (13,3)--(14,3);
		\draw[green] (14,3)--(15,3);
		\draw[green](13,3.1)--(15,3.1);
		\draw[green](13,3-0.1)--(15,3-0.1);
		\draw[] (15,3.05)--(16,3.05);
		\draw[] (15,3-.05)--(16,3-.05);
		\draw[green] (16,3)--(17,3);
		\draw[red] (12,4)--(14,3);
		\draw[red] (15,4)--(14,3);
		\draw[red] (14,2)--(14,3);
		\draw[red] (15,2)--(14,3);
		\node[7brane] at (11,3) {};
		\node[7brane] at (12,3) {};
		\node[7brane] at (13,3) {};
		\node[7brane] at (15,3) {};
		\node[7brane] at (16,3) {};
		\node[7brane] at (17,3) {};
		\node[7brane] at (12,4) {};
		\node[7brane] at (15,4) {};
		\node[7brane] at (14,2) {};
		\node[7brane] at (15,2) {};
	\end{tikzpicture}
		\caption*{$\cnode{green}{}1 ~-~\cnode{black}{}{2}~-~\overset{\overset{\displaystyle{\color{red}\overset 1 \circ}}{\parallel}}{\cnode{green}{}{3}}~-~\cnode{black}{}{2}~-~\cnode{green}{}1$}
	\end{subfigure}
	\hfill
	\begin{subfigure}[t]{.30\textwidth}\centering
	\begin{tikzpicture}[scale=0.75]
		\draw[green](11,3)--(14-2,3);
		\draw[black](12,3.05)--(13,3.05);
		\draw[black](12,3-0.05)--(13,3-0.05);
		\draw[blue] (13,3)--(14,3);
		\draw[red] (14,3)--(15,3);
		\draw[green](13,3.1)--(15,3.1);
		\draw[green](13,3-0.1)--(15,3-0.1);
		\draw[black] (15,3.05)--(16,3.05);
		\draw[black] (15,3-.05)--(16,3-.05);
		\draw[green] (16,3)--(17,3);
		\draw[red] (12,4)--(14,3);
		\draw[blue] (15,4)--(14,3);
		\draw[blue] (14,2)--(14,3);
		\draw[red] (15,2)--(14,3);
		\node[7brane] at (11,3) {};
		\node[7brane] at (12,3) {};
		\node[7brane] at (13,3) {};
		\node[7brane] at (15,3) {};
		\node[7brane] at (16,3) {};
		\node[7brane] at (17,3) {};
		\node[7brane] at (12,4) {};
		\node[7brane] at (15,4) {};
		\node[7brane] at (14,2) {};
		\node[7brane] at (15,2) {};
	\end{tikzpicture}
		\caption*{$\cnode{green}{}1 ~-~\overset{\overset{\displaystyle{\color{blue}\overset 1 \circ}\xlongequal{3}{\color{red}\overset 1 \circ}}{\diagup~~~~~~~~\diagdown}}{\cnode{black}{}{2}~-~\cnode{green}{}{2}~-~\cnode{black}{}{2}}~-~\cnode{green}{}1$}
	\end{subfigure}
	\hfill
	\begin{subfigure}[t]{.30\textwidth}\centering
	\begin{tikzpicture}[scale=0.75]
		\draw[green](11,3)--(14-2,3);
		\draw[black](12,3.05)--(13,3.05);
		\draw[blue](12,3-0.05)--(13,3-0.05);
		\draw[blue] (13,3)--(14,3);
		\draw[red] (14,3)--(15,3);
		\draw[green](13,3.1)--(15,3.1);
		\draw[blue](13,3-0.1)--(14,3-0.1);
		\draw[red](14,3-0.1)--(15,3-0.1);
		\draw[black] (15,3.05)--(16,3.05);
		\draw[red] (15,3-.05)--(16,3-.05);
		\draw[green] (16,3)--(17,3);
		\draw[red] (12,4)--(14,3);
		\draw[blue] (15,4)--(14,3);
		\draw[red] (14,2)--(14,3);
		\draw[blue] (15,2)--(14,3);
		\node[7brane] at (11,3) {};
		\node[7brane] at (12,3) {};
		\node[7brane] at (13,3) {};
		\node[7brane] at (15,3) {};
		\node[7brane] at (16,3) {};
		\node[7brane] at (17,3) {};
		\node[7brane] at (12,4) {};
		\node[7brane] at (15,4) {};
		\node[7brane] at (14,2) {};
		\node[7brane] at (15,2) {};
	\end{tikzpicture}
		\caption*{$\cnode{green}{}1 \overset{\overset{\displaystyle{\color{blue}\overset 1 \circ}~~~\xlongequal{4}~~~{\color{red}\overset 1 \circ}}{\diagup~~~~~~~~~~~~~~~~~~\diagdown}}{~-~\cnode{black}{}{1}~-~\cnode{green}{}{1}~-~\cnode{black}{}{1}~-~}\cnode{green}{}1$}
	\end{subfigure}
	\caption{Different components of the Higgs branch at infinite coupling of 5d SQCD with gauge group $SU(5)$, $N_f=6$ flavors and CS level $k=1$. None of the three web sub divisions can be a sub division of the other two. 
Note that the red and the blue sub webs in the rightmost web cannot be further subdivided due to s-rule.
The quiver which is obtained by applying Conjecture \ref{con:main} is depicted underneath each phase. Note that in table \ref{tab:region1} these three different phases receive labels III, I and II from left to right.}
	\label{fig:5,6,1,phases}
\end{figure}
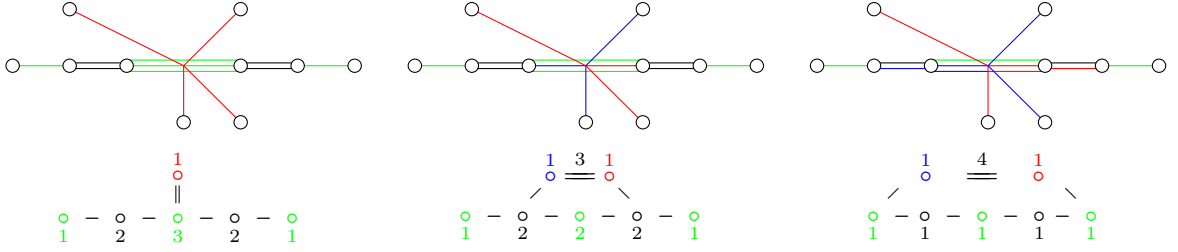

In this way, Conjecture \ref{con:main} can be utilized to derive new properties of 5d theories at infinite coupling that were not understood before.

\subsection{The Intersection of Several Cones}\label{sec:intersectionOfCones}

In fact, Conjecture \ref{con:main} can also be used to specify the intersection between any pair of cones in equation \ref{eq:union}. Given two sub divisions, for example the first and the second from the left in figure \ref{fig:5,6,1,phases}, find the maximal subdivision $S$ such that both brane systems are subdivisions of $S$. $S$ is depicted in figure \ref{fig:5,6,1,intersec1}. The intersection of both cones is then the moduli space of dressed monopole operators given by the quiver associated to $S$ via Conjecture \ref{con:main}.

Hence, we have:
\begin{equation}
	C_1\cap C_2 = \mathcal{C}^{3d}\left(~\cnode{black}{}1 -\overset{\overset{\displaystyle{\color{black}\overset 1 \circ}}{\diagup~~~\diagdown}}{\cnode{black}{}{2}-\cnode{black}{}{2}-\cnode{black}{}{2}}-\cnode{black}{}1~\right)
\end{equation}

Note that this space is the closure of the \emph{next to minimal nilpotent orbit} of $\mathfrak{sl}(6,\mathbb{C})$:
\begin{equation}
	\mathcal{C}^{3d}\left(~\cnode{black}{}1 -\overset{\overset{\displaystyle{\color{black}\overset 1 \circ}}{\diagup~~~\diagdown}}{\cnode{black}{}{2}-\cnode{black}{}{2}-\cnode{black}{}{2}}-\cnode{black}{}1~\right) = \mathcal{C}^{3d}\left(~\node{}1-\node{\flav 1}2-\node{}2-\node{\flav 1}2-\node{}1~\right) =\orbit{n.min}{A_5}
\end{equation}
Therefore, we obtain the result:
\begin{equation}
	C_1\cap C_2 =\orbit{n.min}{A_5}
\end{equation}

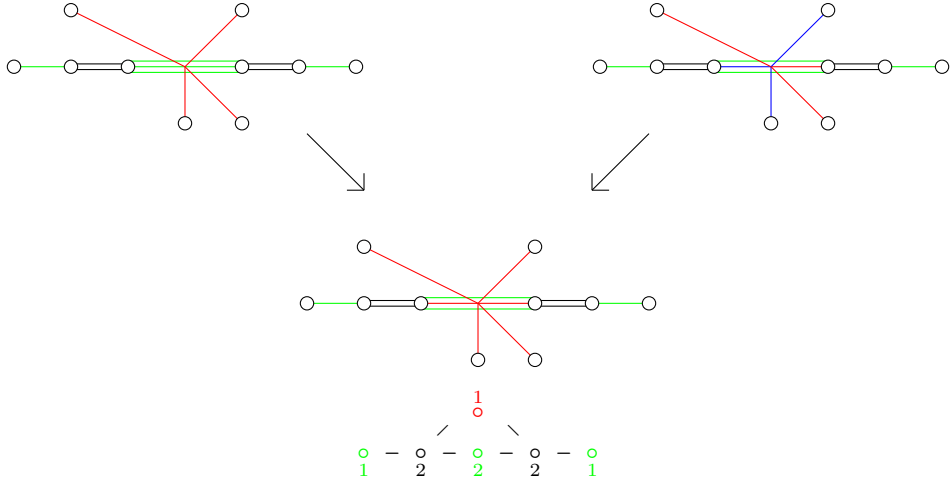
\begin{figure}
	\centering
	\begin{subfigure}[t]{.49\textwidth}\centering
	\begin{tikzpicture}[scale=0.75]
		\draw[green](11,3)--(14-2,3);
		\draw[](12,3.05)--(13,3.05);
		\draw[](12,3-0.05)--(13,3-0.05);
		\draw[green] (13,3)--(14,3);
		\draw[green] (14,3)--(15,3);
		\draw[green](13,3.1)--(15,3.1);
		\draw[green](13,3-0.1)--(15,3-0.1);
		\draw[] (15,3.05)--(16,3.05);
		\draw[] (15,3-.05)--(16,3-.05);
		\draw[green] (16,3)--(17,3);
		\draw[red] (12,4)--(14,3);
		\draw[red] (15,4)--(14,3);
		\draw[red] (14,2)--(14,3);
		\draw[red] (15,2)--(14,3);
		\node[7brane] at (11,3) {};
		\node[7brane] at (12,3) {};
		\node[7brane] at (13,3) {};
		\node[7brane] at (15,3) {};
		\node[7brane] at (16,3) {};
		\node[7brane] at (17,3) {};
		\node[7brane] at (12,4) {};
		\node[7brane] at (15,4) {};
		\node[7brane] at (14,2) {};
		\node[7brane] at (15,2) {};
	\end{tikzpicture}
	\end{subfigure}
	\hfill
	\begin{subfigure}[t]{.49\textwidth}\centering
	\begin{tikzpicture}[scale=0.75]
		\draw[green](11,3)--(14-2,3);
		\draw[black](12,3.05)--(13,3.05);
		\draw[black](12,3-0.05)--(13,3-0.05);
		\draw[blue] (13,3)--(14,3);
		\draw[red] (14,3)--(15,3);
		\draw[green](13,3.1)--(15,3.1);
		\draw[green](13,3-0.1)--(15,3-0.1);
		\draw[black] (15,3.05)--(16,3.05);
		\draw[black] (15,3-.05)--(16,3-.05);
		\draw[green] (16,3)--(17,3);
		\draw[red] (12,4)--(14,3);
		\draw[blue] (15,4)--(14,3);
		\draw[blue] (14,2)--(14,3);
		\draw[red] (15,2)--(14,3);
		\node[7brane] at (11,3) {};
		\node[7brane] at (12,3) {};
		\node[7brane] at (13,3) {};
		\node[7brane] at (15,3) {};
		\node[7brane] at (16,3) {};
		\node[7brane] at (17,3) {};
		\node[7brane] at (12,4) {};
		\node[7brane] at (15,4) {};
		\node[7brane] at (14,2) {};
		\node[7brane] at (15,2) {};
	\end{tikzpicture}
	\end{subfigure}
	\hfill
	\begin{subfigure}[t]{.9\textwidth}
	\centering
	\begin{tikzpicture}[scale=0.75]
		\draw (11,6)--(12,5);
		\draw (11.7,5)--(12,5);
		\draw (12,5.3)--(12,5);
		\draw (17,6)--(16,5);
		\draw (16.3,5)--(16,5);
		\draw (16,5.3)--(16,5);
		\draw[green](11,3)--(14-2,3);
		\draw[black](12,3.05)--(13,3.05);
		\draw[black](12,3-0.05)--(13,3-0.05);
		\draw[red] (13,3)--(14,3);
		\draw[red] (14,3)--(15,3);
		\draw[green](13,3.1)--(15,3.1);
		\draw[green](13,3-0.1)--(15,3-0.1);
		\draw[black] (15,3.05)--(16,3.05);
		\draw[black] (15,3-.05)--(16,3-.05);
		\draw[green] (16,3)--(17,3);
		\draw[red] (12,4)--(14,3);
		\draw[red] (15,4)--(14,3);
		\draw[red] (14,2)--(14,3);
		\draw[red] (15,2)--(14,3);
		\node[7brane] at (11,3) {};
		\node[7brane] at (12,3) {};
		\node[7brane] at (13,3) {};
		\node[7brane] at (15,3) {};
		\node[7brane] at (16,3) {};
		\node[7brane] at (17,3) {};
		\node[7brane] at (12,4) {};
		\node[7brane] at (15,4) {};
		\node[7brane] at (14,2) {};
		\node[7brane] at (15,2) {};
	\end{tikzpicture}
		\caption*{$\cnode{green}{}1 ~-~\overset{\overset{\displaystyle{\color{red}\overset 1 \circ}}{\diagup~~~~~~\diagdown}}{\cnode{black}{}{2}~-~\cnode{green}{}{2}~-~\cnode{black}{}{2}}~-~\cnode{green}{}1$}
	\end{subfigure}
	\caption{Intersection of two cones.}
	\label{fig:5,6,1,intersec1}
\end{figure}

We believe that this result nicely illustrates the power behind Conjecture \ref{con:main}. Similarly, the triple intersection between all three components $C_1\cap C_2\cap C_3$ can also be identified. The maximal subdivision $S'$ such that all brane systems in figure \ref{fig:5,6,1,phases} are subdivisions of $S'$ is depicted in figure \ref{fig:5,6,1,3intersec}. The corresponding quiver can be obtained via Conjecture \ref{con:main}, such that the triple intersection is defined as the space of dressed monopole operators:

\begin{equation}\label{eq:minA5}
	C_1\cap C_2\cap C_3 =\mathcal{C}^{3d}\left(~\overset{\displaystyle\node{1}{}}{\overset{\diagup ~~~~~~~~~~~~\diagdown}{\node{}1-\node{}1-\node{}1-\node{}1-\node{}1}} ~\right)
\end{equation}

The space of dressed monopole operators in equation \ref{eq:minA5} can be identified with the reduced moduli space of one $A_5$ instanton on $\mathbb{C}^2$ \cite{Intriligator:1996ex}, or equivantley with the closure of the minimal nilpotent orbit of $\mathfrak{sl}(6,\mathbb{C})$:
\begin{equation}
	C_1\cap C_2\cap C_3=\orbit{min}{A_5}
\end{equation}

\begin{figure}
	\centering
	\begin{subfigure}[t]{.9\textwidth}
	\centering
	\begin{tikzpicture}[scale=0.75]
		\draw[green](11,3)--(14-2,3);
		\draw[black](12,3.05)--(13,3.05);
		\draw[red](12,3-0.05)--(13,3-0.05);
		\draw[red] (13,3)--(14,3);
		\draw[red] (14,3)--(15,3);
		\draw[green](13,3.1)--(15,3.1);
		\draw[red](13,3-0.1)--(15,3-0.1);
		\draw[black] (15,3.05)--(16,3.05);
		\draw[red] (15,3-.05)--(16,3-.05);
		\draw[green] (16,3)--(17,3);
		\draw[red] (12,4)--(14,3);
		\draw[red] (15,4)--(14,3);
		\draw[red] (14,2)--(14,3);
		\draw[red] (15,2)--(14,3);
		\node[7brane] at (11,3) {};
		\node[7brane] at (12,3) {};
		\node[7brane] at (13,3) {};
		\node[7brane] at (15,3) {};
		\node[7brane] at (16,3) {};
		\node[7brane] at (17,3) {};
		\node[7brane] at (12,4) {};
		\node[7brane] at (15,4) {};
		\node[7brane] at (14,2) {};
		\node[7brane] at (15,2) {};
	\end{tikzpicture}
		\caption*{$\overset{{\color{red}\displaystyle\node{1}{}}}{\overset{\diagup ~~~~~~~~~~~~\diagdown}{\cnode{green}{}1-\node{}1-\cnode{green}{}1-\node{}1-\cnode{green}{}1}}$}
	\end{subfigure}
	\caption{Intersection of the three components depicted in figure \ref{fig:5,6,1,phases}.}
	\label{fig:5,6,1,3intersec}
\end{figure}
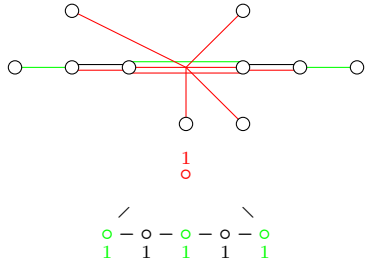

\section{Computation of $\mathcal{H}_{\infty}$ for SQCD}\label{sec:classification}

In this section we present a classification of the Higgs branch at infinite coupling of the class of 5d $\mathcal{N}=1$ SQCD theories with gauge group $SU(N_c)$, number of flavors $N_f$ and CS level $k$, represented by the following 5d quiver:
\begin{equation}
	\node{\flav{N_f}}{SU(N_c)_k}
\end{equation}
For generic $N_c$, the brane web construction of such theories imposes the restriction:
\begin{equation}\label{eq:condition}
	|k|\leq N_c - N_f/2 + 2 .
\end{equation}
We have applied the techniques developed in this paper to all theories that satisfy condition \ref{eq:condition}, and have found that the 3 parameter family of theories is divided into 4 different regions, given by:
\begin{align}
		1)~~|k|&<N_c - \frac{N_f}{2} \\
		2)~~|k|&=N_c - \frac{N_f}{2}  \\
		3)~~|k|&=N_c - \frac{N_f}{2} +1 \\
		4)~~|k|&=N_c - \frac{N_f}{2} +2
\end{align}
These four regions were previously identified as regions which differ by the pattern of symmetry enhancement. It turns out that these regions are also consistent with the pattern of different Higgs branches which is computed in the present paper.
In each of the regions the Higgs branch at infinite coupling contains several components, with non trivial intersections. Let us express the components and their intersections by providing their construction as spaces of dressed monopole operators, computed from quivers obtained employing Conjecture \ref{con:main} (alternatively one can say that the different components are constructed as Coulomb branches of 3d $\mathcal{N} = 4$ quivers, as it was done in the examples of sections \ref{sec:knownExamples} and \ref{sec:newNotions}).

\subsection{First Region:  $|k|<N_c - \frac{N_f}{2}$}

\paragraph{General case: $|k|>\frac{1}{2}$.} In this region there are degenerate cases for $|k|=0$ and $|k|=\frac 1 2$. Let us first consider the general case where:
\begin{equation}
	|k|>\frac{1}{2}.
\end{equation}

\begin{figure}
\centering
\includegraphics[width=15cm]{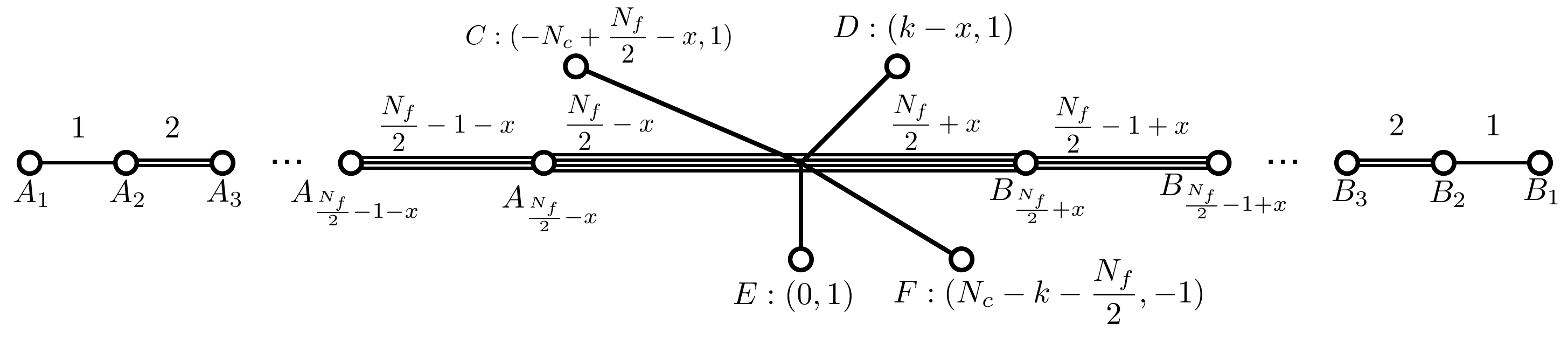}
\caption{5-brane web for $0 \le k <N_c - \frac{N_f}{2}$. $x$ as defined in \ref{evenodd}, makes a distinction between even and odd $N_f$.}
\label{Fig:k<Nc-Nf2}
\end{figure}
\begin{figure}
\centering
\includegraphics[width=5cm]{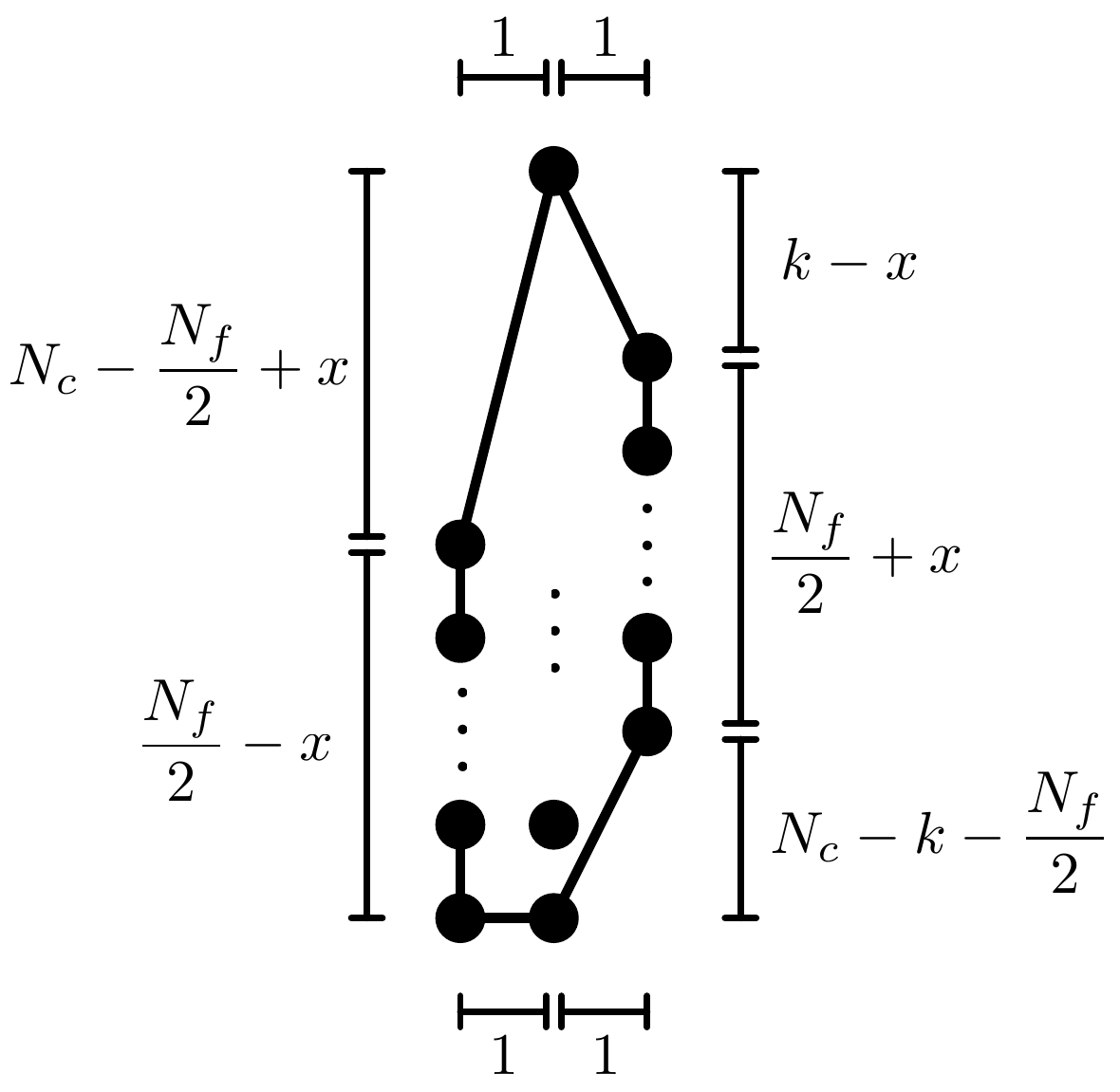}
\caption{Toric diagram for $0 \le k <N_c - \frac{N_f}{2}$.}
\label{Fig:toric_k<Nc-Nf2}
\end{figure}
\begin{figure}
\centering
\includegraphics[width=15cm]{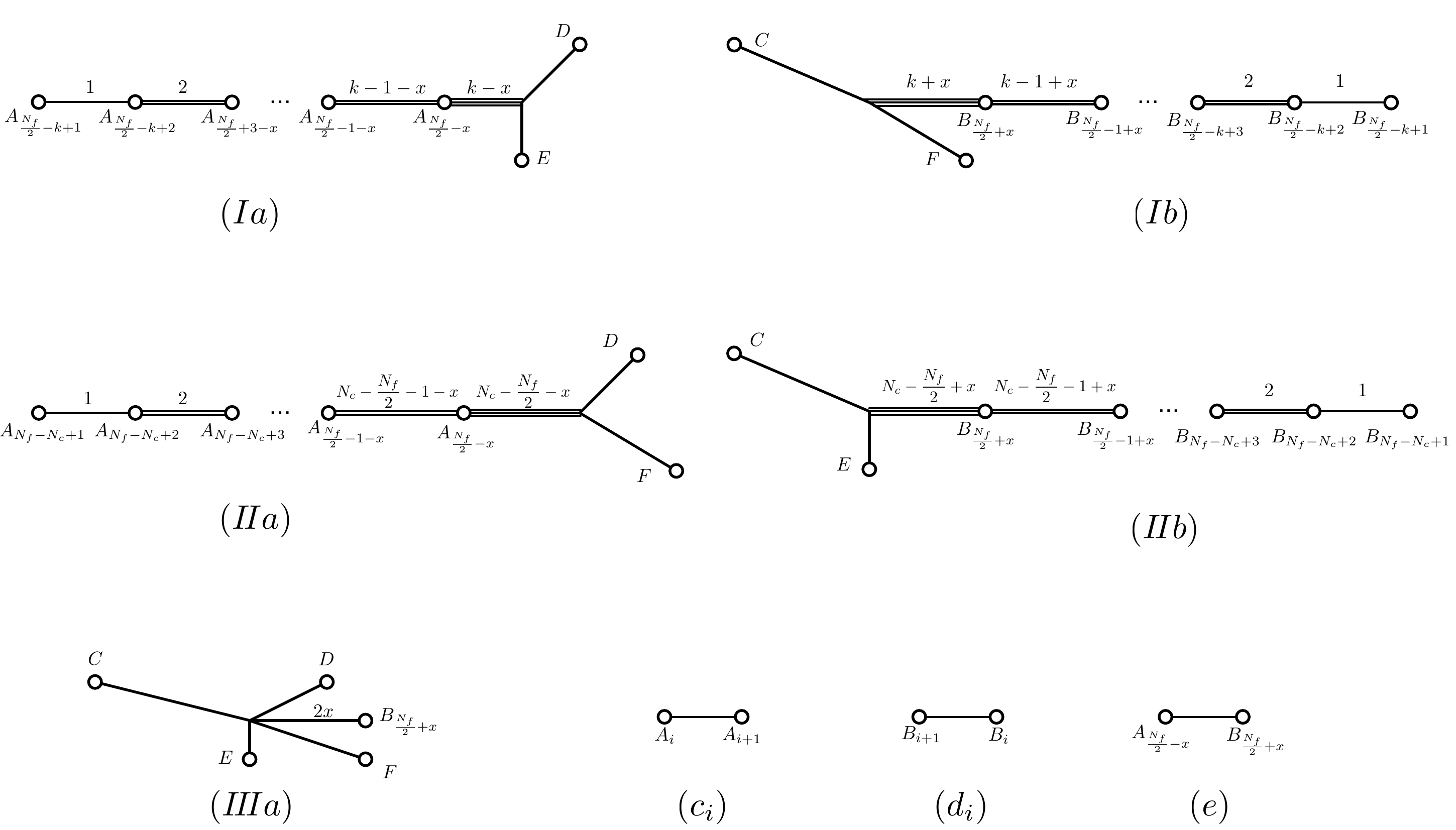}
\caption{The sub webs in the Higgs phase for $0 \le k <N_c - \frac{N_f}{2}$.}
\label{Fig:k<Nc-Nf2_subgraph}
\end{figure}

The 5-brane web for this case is depicted in figure \ref{Fig:k<Nc-Nf2}, with the corresponding dual toric diagram in figure \ref{Fig:toric_k<Nc-Nf2}. The length of each edge is written in figure \ref{Fig:toric_k<Nc-Nf2} supposing that the distance between each dot is one. (Therefore, the number of dots included in the corresponding edge is one more than the written number.)
The triangulation of this toric diagram is omitted since it does not affect the analysis of the Higgs branch.
In the Higgs branch, the brane system in figure \ref{Fig:k<Nc-Nf2} decomposes into sub webs depicted in figure \ref{Fig:k<Nc-Nf2_subgraph}. Here it is convenient to introduce a variable $x$ defined as
\begin{equation}
\label{evenodd}
x = 0 \quad (\text{ for } N_f \text{: even} ), 
\qquad
x = \frac{1}{2} \quad (\text{ for } N_f \text{: odd} ).
\end{equation}


There are three different patterns of sub dividing the original 5-brane web, corresponding to three different components that form the Higgs branch,
which we denote I, II and III, respectively.

\begin{align}
\text{Phase I: } &(Ia) \times 1 + (Ib) \times 1 
+ \sum_{i=1}^{\frac{N_f}{2}-k} (c_i) \times i 
+ \sum_{i=\frac{N_f}{2}-k+1}^{\frac{N_f}{2}-1-x} (c_i) \times \left( \frac{N_f}{2} - k \right)
\cr
&+ \sum_{i=1}^{\frac{N_f}{2}-k} (d_i) \times i 
+ \sum_{i=\frac{N_f}{2}-k+1}^{\frac{N_f}{2}-1+x} (d_i) \times \left( \frac{N_f}{2} - k \right)
+ (e) \times \left( \frac{N_f}{2} - k \right)
\cr
\text{Phase II: } &(I\! Ia) \times 1 + (I\! Ib) \times 1 
+ \sum_{i=1}^{N_f - N_c} (c_i) \times i 
+ \sum_{i=N_f - N_c+1}^{\frac{N_f}{2}-1-x} (c_i) \times ( N_f - N_c )
\cr
&+ \sum_{i=1}^{N_f - N_c} (d_i) \times i 
+ \sum_{i=N_f - N_c+1}^{\frac{N_f}{2}-1+x} (d_i) \times ( N_f - N_c )
+ (e) \times ( N_f - N_c )
\cr
\text{Phase III: } &(I\! I\! I a) \times 1 
+ \sum_{i=1}^{\frac{N_f}{2}-1-x} (c_i) \times i 
+ \sum_{i=1}^{\frac{N_f}{2}-1+x} (d_i) \times i 
+ (e) \times \left( \frac{N_f}{2}-x \right)
\end{align}

Each different pattern gives rise to a different quiver, obtained via Conjecture \ref{con:main}. These are depicted in table \ref{tab:region1}. The rightmost column of Table \ref{tab:region1} contains the quaternionic dimension of each component. The global symmetry of the 5d Higgs branch at the UV fixed point is:

\begin{equation}
	SU(N_f)\times U(1) \times U(1).
\end{equation}

Only some factors of the global symmetry act non-trivially on each phase\footnote{It can be computed from the subset of nodes that are balanced.}. This is also specified in a column of table \ref{tab:region1}.

\begin{table}
	\makebox[\textwidth][c]{ 
	\begin{tabular}{|c|c|c|c|}
		\hline
		Phase & Quiver & Global Symmetry & Dimension \\ \hline
		I &  $\underbrace{\node{}1-\dots -\overset{\overset{\displaystyle\overset 1 \circ\xlongequal{N_c - \frac{N_f}{2}+|k|}\overset 1 \circ}{\diagup~~~~~~~~~~~~~~~~~~\diagdown}}{\node{}{\frac{N_f}{2}-|k|}-~~~~\dots~~~~-\node{}{\frac{N_f}{2}-|k|}}-\dots-\node{}1}_{N_f - 1}$ & $SU(N_f)\times U(1)$ & $\frac{N_f^2}{4}-k^2+1$\\ \hline
		II &  $\underbrace{\node{}1-\dots -\overset{\overset{\displaystyle\overset 1 \circ\xlongequal{2N_c - N_f}\overset 1 \circ}{\diagup~~~~~~~~~~~~~~~\diagdown}}{\node{}{N_f-N_c}-~~\dots~~-\node{}{N_f-N_c}}-\dots-\node{}1}_{N_f - 1}$ & $SU(N_f)\times U(1)$ & $N_c N_f - N_c^2 + 1$\\ \hline
		III ($N_f$ even) &  $\underbrace{\node{}1-\node{}2-\dots-\node{\overset{\displaystyle \overset{1}{\circ}}{\parallel}}{\frac{N_f }2}-\dots-\node{}2-\node{}1}_{N_f - 1}$ & $SU(N_f)$ & $\frac{N_f^2}{4}$\\ \hline
		III ($N_f$ odd) & $\underbrace{\node{}1-\node{}2-\dots -\overset{\overset{\displaystyle\overset 1 \circ}{\diagup~~\diagdown}}{\node{}{\frac{N_f-1}2}-\node{}{\frac{N_f-1}2}}-\dots-\node{}2-\node{}1}_{N_f - 1}$ & $SU(N_f)$ & $\frac{N_f^2-1}{4}$\\ \hline
	\end{tabular}
	}
	\caption{Components of $\mathcal{H}_\infty$ for  $\frac{1}{2}<|k|<N_c - \frac{N_f}{2}$. 
	Component I is present for $\frac {N_f}{2} \geq |k|$, Component II is present for $N_f \geq N_c$, and Component III is present for $N_f \ge 2$. 
	The three dots in the quivers indicate a chain of balanced gauge nodes (i.e. the number of colors that each node sees is twice its rank). For example, in phase I, the ranks of the gauge nodes in the main chain start with 1, then 2, then keep increasing until $\frac{N_f}{2}-|k|$ is reached, then the ranks stay constant at $\frac{N_f}{2}-|k|$ in the middle of the chain, and then start decreasing all the way down to 1. The number underneath the brace in each quiver indicates the total number of balanced nodes in the main chain of the quiver.}
	\label{tab:region1}
\end{table}

%

Note that component I is present for values:
\begin{equation}
	\frac {N_f}2 \geq |k| .
\end{equation}
This range of values is new and was not observed in any previous analysis of five dimensional gauge theories. It is to be considered as purely an effect of infinite coupling and to be discovered by analysing the pattern of components on the Higgs branch. Symmetry enhancement computations, for example, do not reveal this special regime of the 3 parameter family. This component contains mesons with spin 1 under $SU(2)_R$ and baryon-like objects with spin $N_c/2$ under $SU(2)_R$. The spectrum of operators in the chiral ring, and the chiral ring relations depends on all three values of $N_c, N_f$ and $k$.

Component II is present for the regime:
\begin{equation}
	N_f \geq N_c ,
\end{equation}
which is the typical regime where baryons exist. Such a phase is known from classical physics, sometimes called the baryonic branch, containing both mesons and baryons. The dimension of the moduli space does not change from the classical one. The details, however, are different and this component of the Higgs branch should not be confused as already being present at finite coupling. The mesons and baryons are certainly present in the classical theory, but the chiral ring relations they satisfy receive corrections from instaton operators at infinite coupling. 

Component III is present for:
\begin{equation}
	N_f \geq 2,
\end{equation}
which is the familiar classical condition for the presence of a Higgs branch. The moduli space of this component is a closure of a nilpotent orbit of height 2 and consists of mesons only. It is still different from the classical moduli space, as the rank of the meson matrix $M$ is maximal for a nilpotency of degree 2, $M^2=0$, namely $r(M)\le N_f/2$. This is a rank condition which is independent of $N_c$. The classical rank condition $r(M)\le N_c$ is relaxed by instanton operator corrections.

\paragraph{Intersections of the different components}
When specifying different components of the moduli space, it is important to specify how these components intersect. This is crucial, for example, for the purpose of counting operators on the chiral ring, making sure that each operator is counted exactly once.
We notice that since $|k|<N_c - N_f/2$, the conditions $N_f - N_c < N_f/2-|k|$ and $N_f-N_c<N_f/2$ are always satisfied.
For the present case, the intersection between component I, II and III are given by table \ref{tab:region1.intersect}. These intersections are closures of nilpotent orbits of height 2, containing mesons only. The meson matrix on the intersection $M$ satisfies $M^2=0$ and the rank of $M$ is at most the maximal rank of a node in the quiver, either $N_f-N_c$, or $N_f/2-|k|$.

\begin{table}
	\centering
	\begin{tabular}{|c|c|c|c|}
	\hline
	Intersection & Quiver & Global Symmetry & Dimension \\ \hline
	I $\cap$ II,  II $\cap$ III  & $\underbrace{\node{}1-\dots -\overset{\overset{\displaystyle\overset 1 \circ}{\diagup~~~~~~~~\diagdown}}{\node{}{N_f-N_c}-\dots-\node{}{N_f-N_c}}-\dots-\node{}1}_{N_f - 1}$ & $SU(N_f)$ & $N_c(N_f-N_c)$\\ \hline
	I $\cap$ III & $\underbrace{\node{}1-\dots -\overset{\overset{\displaystyle\overset 1 \circ}{\diagup~~~~~~~~\diagdown}}{\node{}{\frac{N_f}{2}-|k|}-\dots-\node{}{\frac{N_f}{2}-|k|}}-\dots-\node{}1}_{N_f - 1}$ & $SU(N_f)$ & $\frac{N_f^2}{4}-k^2$ \\ \hline
	I $\cap$ II $\cap$ III  & $\underbrace{\node{}1-\dots -\overset{\overset{\displaystyle\overset 1 \circ}{\diagup~~~~~~~~\diagdown}}{\node{}{N_f-N_c}-\dots-\node{}{N_f-N_c}}-\dots-\node{}1}_{N_f - 1}$ & $SU(N_f)$ & $N_c(N_f-N_c)$\\ \hline
	\end{tabular}
	\caption{Intersection of the different components of $\mathcal{H}_\infty$ for $\frac 1 2 <|k|<N_c - \frac{N_f}{2}$.}		\label{tab:region1.intersect}
\end{table}


\paragraph{Exceptional cases: $|k| \leq \frac 1 2$.}

The brane system and the toric diagram for the exceptional cases does not change. They are still depicted by figures \ref{Fig:k<Nc-Nf2} and \ref{Fig:toric_k<Nc-Nf2}. Let us consider the first case:
\begin{align}
	|k| = \frac 1 2 .
\end{align}

The global symmetry does not change:
\begin{equation}
	SU(N_f)\times U(1) \times U(1) .
\end{equation}

For this case, the sub web (IIIa) in Figure \ref{Fig:k<Nc-Nf2_subgraph} is decomposed into (Ia) and (Ib). 
Therefore, phase III is fully included in phase I, so there are only two distinct components in the Higgs branch, given in table \ref{tab:region1kHalf}.

\begin{table}
	\centering
	\begin{tabular}{|c|c|c|}
		\hline
		Phase & Quiver & Global Symmetry \\ \hline
		I & $\underbrace{\node{}1-\dots -\overset{\overset{\displaystyle\overset 1 \circ\xlongequal{N_c - \frac{N_f}{2}+\frac 1 2}\overset 1 \circ}{\diagdown~~~~~~~~~~~\diagup}}{\node{}{\frac{N_f-1}{2}}-\node{}{\frac{N_f-1}{2}}}-\dots-\node{}1}_{N_f - 1}$  & $SU(N_f)\times U(1)$ \\ \hline
		II & $\underbrace{\node{}1-\dots -\overset{\overset{\displaystyle\overset 1 \circ\xlongequal{2N_c - N_f}\overset 1 \circ}{\diagup~~~~~~~~~~~~~~~\diagdown}}{\node{}{N_f-N_c}-~~\dots~~-\node{}{N_f-N_c}}-\dots-\node{}1}_{N_f - 1}$  & $SU(N_f)\times U(1)$ \\ \hline
		I $\cap$ II & $\underbrace{\node{}1-\dots -\overset{\overset{\displaystyle\overset 1 \circ}{\diagup~~~~~~~~\diagdown}}{\node{}{N_f-N_c}-\dots-\node{}{N_f-N_c}}-\dots-\node{}1}_{N_f - 1}$ & $SU(N_f)$ \\ \hline
	\end{tabular}
	\caption{Components of $\mathcal{H}_\infty$ for  $\frac 1 2 = |k|<N_c - \frac{N_f}{2}$.
	Component I is present for $N_f \geq 1$ and Component II is present for $N_f \geq N_c$.
	}
	\label{tab:region1kHalf}
\end{table}


 The second case is for:
\begin{align}
	|k| = 0 .
\end{align}

The global symmetry does not change:
\begin{equation}
	SU(N_f)\times U(1) \times U(1) .
\end{equation}

For this case phase III is also fully included in phase I, so there are only two distinct components in the Higgs branch, given in table \ref{tab:region1k0}.

\begin{table}
	\centering
	\begin{tabular}{|c|c|c|}
		\hline
		Phase & Quiver & Global Symmetry \\ \hline
		I &  $\underbrace{\node{}1-\node{}2-\dots -\node{\overset{\displaystyle\overset{1}{\circ}\xlongequal{N_c - \frac{N_f}{2}}\overset{1}{\circ}}{\diagdown~~~~~~\diagup}}{\frac{N_f}{2}}-\dots-\node{}2-\node{}1}_{N_f - 1}$ & $SU(N_f)\times U(1)$ \\ \hline
		II &  $\underbrace{\node{}1-\dots -\overset{\overset{\displaystyle\overset 1 \circ\xlongequal{2N_c - N_f}\overset 1 \circ}{\diagup~~~~~~~~~~~~~~~\diagdown}}{\node{}{N_f-N_c}-~~\dots~~-\node{}{N_f-N_c}}-\dots-\node{}1}_{N_f - 1}$ & $SU(N_f)\times U(1)$ \\ \hline
		I $\cap$ II & $\underbrace{\node{}1-\dots -\overset{\overset{\displaystyle\overset 1 \circ}{\diagup~~~~~~~~\diagdown}}{\node{}{N_f-N_c}-\dots-\node{}{N_f-N_c}}-\dots-\node{}1}_{N_f - 1}$ & $SU(N_f)$ \\ \hline
	\end{tabular}
	\caption{Components of $\mathcal{H}_\infty$ for  $0=|k|<N_c - \frac{N_f}{2}$. 
	Component II is present for $N_f \geq N_c$
	}
	\label{tab:region1k0}
\end{table}

%

\subsection{Second Region:  $|k|=N_c - \frac{N_f}{2}$}

\paragraph{General case: $|k|>1$.} Let us consider the general case, which happens for:
\begin{equation}
	|k|>1 .
\end{equation}

The global symmetry of the 5d Higgs branch at the UV fixed point is enhanced to:
\begin{equation}
	SU(N_f)\times SU(2) \times U(1) .
\end{equation}

The brane system is depicted in figure \ref{Fig:k=Nc-Nf2} and the toric diagram is represented in figure \ref{Fig:toric_k=Nc-Nf2}. There are two different maximal sub divisions of the brane system, which means that the Higgs branch $\mathcal{H}_\infty$ is the union of two components. The sub webs that make the two different sub divisions are depicted in figure \ref{Fig:k=Nc-Nf2_subgraph}. They arrange in the two possible subdivisions called phase I and phase III (due to the similarity to the previous case). They are:

%
\begin{figure}
\centering
\includegraphics[width=15cm]{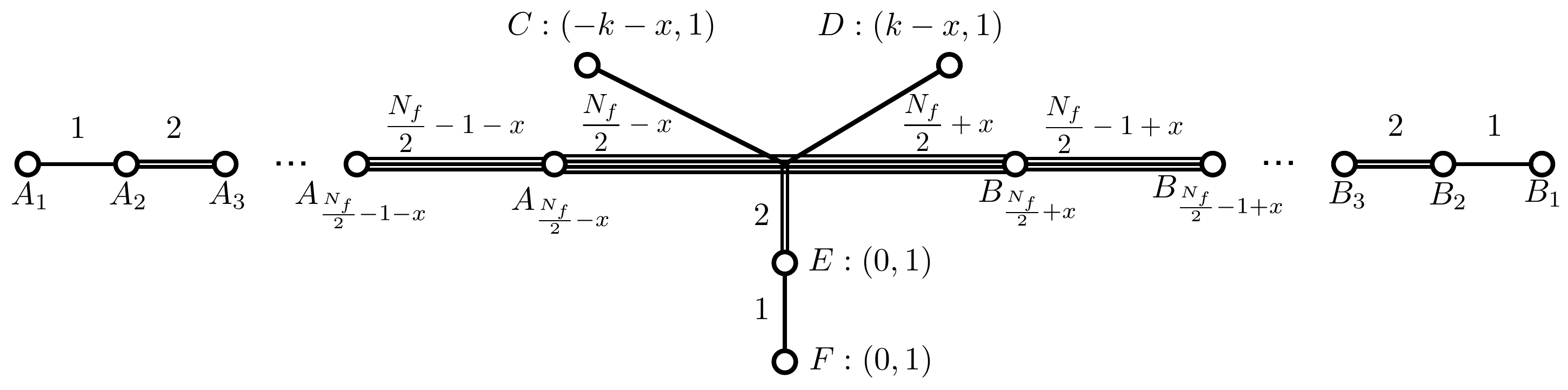}
\caption{5-brane web for $0 \le k=N_c - \frac{N_f}{2}$.}
\label{Fig:k=Nc-Nf2}
\end{figure}
\begin{figure}
\centering
\includegraphics[width=4cm]{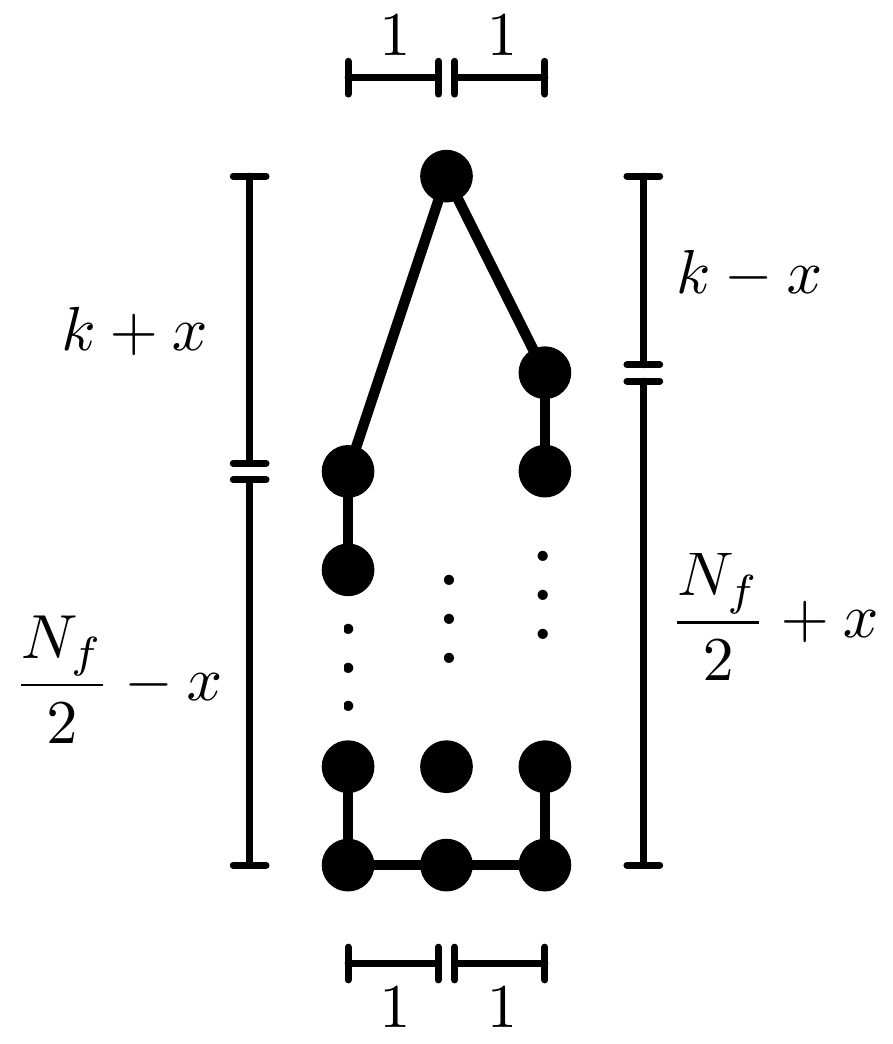}
\caption{Toric diagram for $0 \le k =N_c - \frac{N_f}{2}$.}
\label{Fig:toric_k=Nc-Nf2}
\end{figure}
\begin{figure}
\centering
\includegraphics[width=15cm]{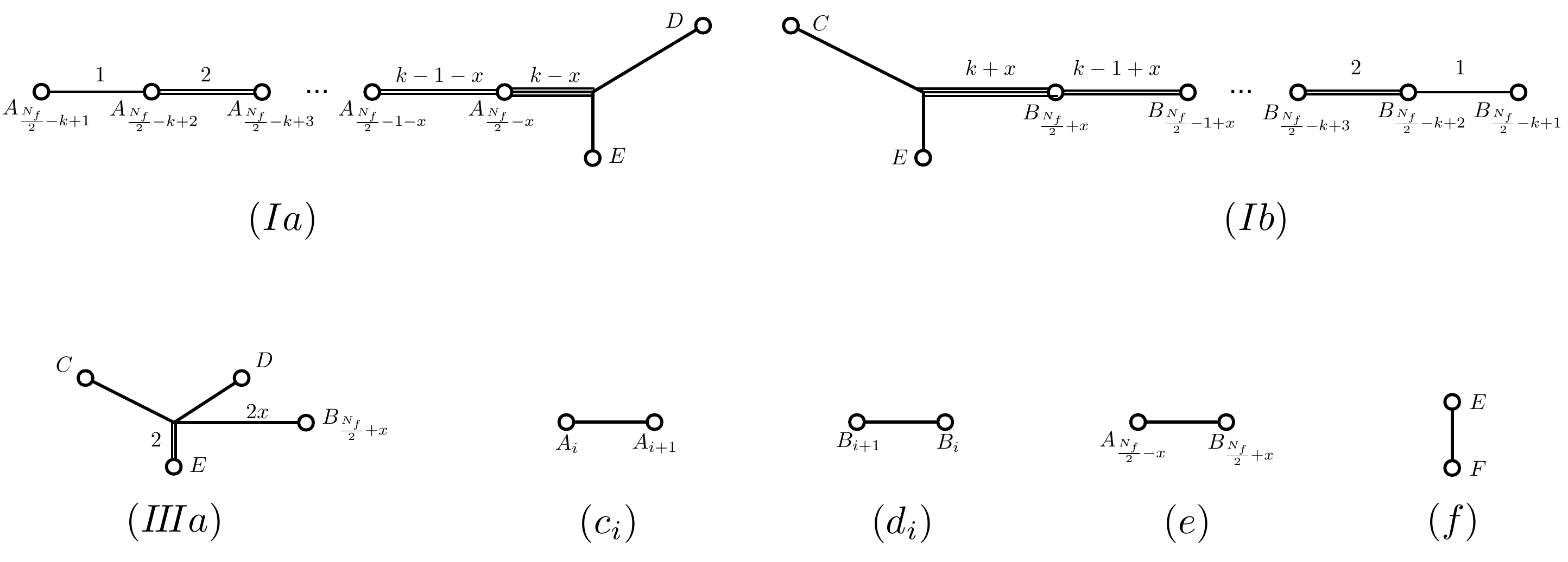}
\caption{The sub webs in the Higgs phase for $0 \le k =N_c - \frac{N_f}{2}$.}
\label{Fig:k=Nc-Nf2_subgraph}
\end{figure}

\begin{align}
\text{Phase I: } &(Ia) \times 1 + (Ib) \times 1 
+ \sum_{i=1}^{\frac{N_f}{2}-k} (c_i) \times i 
+ \sum_{i=\frac{N_f}{2}-k+1}^{\frac{N_f}{2}-1-x} (c_i) \times \left( \frac{N_f}{2} - k \right)
\cr
&+ \sum_{i=1}^{\frac{N_f}{2}-k} (d_i) \times i 
+ \sum_{i=\frac{N_f}{2}-k+1}^{\frac{N_f}{2}-1+x} (d_i) \times \left( \frac{N_f}{2} - k \right)
+ (e) \times \left( \frac{N_f}{2} - k \right)
+ (f) \times 1
\cr
\text{Phase III: } &(I\! I\! I a) \times 1 
+ \sum_{i=1}^{\frac{N_f}{2}-1-x} (c_i) \times i 
+ \sum_{i=1}^{\frac{N_f}{2}-1+x} (d_i) \times i 
+ (e) \times \left( \frac{N_f}{2}-x \right)
+ (f) \times 1
\end{align}

The two components of the Higgs branch are identified with the space of dressed momopole operators of the quivers obtained from the different sub divisions via Conjecture \ref{con:main}, collected in table \ref{tab:region2} (alternatively, one says that the components of $\mathcal{H}_\infty$ are computed as the 3d $\mathcal{N}=4$ Coulomb branches of the quivers in table \ref{tab:region2}). Note the difference with the quivers in table \ref{tab:region1}, where an extra node with rank $1$ and 2 flavors appears at the top, contributing to the global symmetry with a new $SU(2)$ factor. The intersection of the two components is given also as the space of dressed monopole operators of a quiver, following the same procedure as in section \ref{sec:intersectionOfCones}, and it is also included in table \ref{tab:region2}.

\begin{table}
	\centering
	\begin{tabular}{|c|c|c|}
		\hline
		Phase & Quiver & Global Symmetry \\ \hline
		I & $\underbrace{\node{}1-\dots -\overset{\overset{\overset{\overset{\displaystyle \overset 1 \circ}{\diagup~~~\diagdown}}{\displaystyle\overset 1 \circ\xlongequal{2|k|-1}\overset 1 \circ}}{\diagup~~~~~~~~~~~~~~~~~~~\diagdown}}{\node{}{\frac{N_f}2-|k|}-~~~~~\dots~~~~~-\node{}{\frac{N_f}2-|k|}}-\dots-\node{}1}_{N_f - 1}$ & $SU(N_f)\times SU(2)\times U(1)$  \\ \hline
		III ($N_f$ even) & $\underbrace{\node{}1-\node{}2-\dots -\node{\overset{\displaystyle \overset{\overset{\displaystyle {\circ}\rlap{\,\,$\scriptstyle 1$}}{\parallel}}{\circ}\rlap{\,\,$\scriptstyle 1$}}{\parallel}}{\frac{N_f }2}-\dots-\node{}2-\node{}1}_{N_f - 1}$ & $SU(N_f)\times SU(2)$ \\ \hline
		III ($N_f$ odd) & $\underbrace{\node{}1-\node{}2-\dots -\overset{\overset{\overset{\displaystyle \overset{1}{\circ}}{\parallel}}{\overset{\displaystyle {\color{white} 1} \circ {\scriptstyle 1}}{\diagup~~\diagdown}}}{\node{}{\frac{N_f-1}2}-\node{}{\frac{N_f-1}2}}-\dots-\node{}2-\node{}1}_{N_f - 1}$ & $SU(N_f)\times SU(2)$ \\ \hline
		I $\cap$ III  & $\underbrace{\node{}1-\node{}2-\dots -\overset{
		 \overset{\overset{ \displaystyle {\color{white} 1} \circ {\scriptstyle 1}}{\parallel}}{\overset{\displaystyle {\color{white} 1} \circ {\scriptstyle 1}}{\diagup~~~~~~~~\diagdown}}
		 }{\node{}{\frac{N_f}{2}-|k|}-\dots-\node{}{\frac{N_f}{2}-|k|}}-\dots-\node{}2-\node{}1}_{N_f - 1}$ & $SU(N_f)\times SU(2)$ \\ \hline
	\end{tabular}
	\caption{Components of $\mathcal{H}_\infty$ for $1<|k|=N_c - \frac{N_f}{2}$.
	Component I is present for $N_f \geq N_c$, which means $\frac{N_f}{2} \ge |k|$, and Component III is present for $N_f \ge 1$. 
	}
	\label{tab:region2}
\end{table}

\paragraph{Exceptional cases: $|k|\leq 1$.}

The exceptional cases share the same brane system and toric diagram depicted in figures \ref{Fig:k=Nc-Nf2} and \ref{Fig:toric_k=Nc-Nf2}. They differ from the previous cases in the fact that the maximal subdivisions of the brane system are slightly different. The first exceptional case appears for:
\begin{equation}
	|k| = 1 .
\end{equation}

In this case the global symmetry enhancement is the same:
\begin{equation}
	SU(N_f)\times SU(2) \times U(1) .
\end{equation}

However, the phase III does not present the extra node that contributes to the $SU(2)$ factor of the global symmetry in the generic case (see table \ref{tab:region2}). 
This can be seen in the brane web in the fact that supersymmetry prevents from breaking loose the segment that would correspond to this node in the general case.  
That is, we need to combine (IIIa) and (f) to form the minimal sub web in order to avoid breaking the s-rule.
The corresponding quivers are given in table \ref{tab:region2k1}.

\begin{table}
	\centering
	\begin{tabular}{|c|c|c|}
		\hline
		Phase & Quiver & Global Symmetry \\ \hline
		I & $\underbrace{\node{}1-\node{}2-\dots -\overset{\overset{\overset{\overset{\displaystyle \overset 1\circ}{\diagup ~\diagdown}}{\displaystyle {\scriptstyle 1} \circ-  \circ {\scriptstyle1}}}{\diagup~~~~~~~~~~~\diagdown}}{\node{}{\frac{N_f-2}{2}}-\node{}{\frac{N_f-2}{2}}-\node{}{\frac{N_f-2}{2}}}-\dots-\node{}2-\node{}1}_{N_f - 1}$ & $SU(N_f)\times SU(2)\times U(1)$\\ \hline
		III ($N_f$ even) &  $\underbrace{\node{}1-\node{}2-\dots -\node{\overset{\displaystyle \overset{1}{\circ}}{\parallel}}{\frac{N_f}{2}}-\dots-\node{}2-\node{}1}_{N_f - 1}$ & $SU(N_f)$ \\ \hline
		I $\cap$ III & $\underbrace{\node{}1-\dots -\overset{\overset{\displaystyle\overset 1 \circ}{\diagup~~~~~~~~\diagdown}}{\node{}{\frac{N_f-2}{2}}-\node{}{\frac{N_f-2}{2}}-\node{}{\frac{N_f-2}{2}}}-\dots-\node{}1}_{N_f - 1}$ & $SU(N_f)$\\ \hline
	\end{tabular}
	\caption{Components of $\mathcal{H}_\infty$ for  $1=|k|=N_c - \frac{N_f}{2}$.}
	\label{tab:region2k1}
\end{table}

%
%
%

The next case is:
\begin{equation}
	|k|=\frac 1 2 .
\end{equation}

For this case the global symmetry enhancement does not change:
\begin{equation}
	SU(N_f)\times SU(2) \times U(1) .
\end{equation}

The phase III is a subset of phase I, reflecting the fact that (IIIa) is further divided into (Ia) and (Ib).
It givies the single phase, consistent with \cite{Ferlito:2017xdq}. The quiver is depicted in table \ref{tab:region2kHalf}.

\begin{table}
	\centering
	\begin{tabular}{|c|c|c|}
		\hline
		Phase & Quiver & Global Symmetry \\ \hline
		I & $\underbrace{\node{}1-\node{}2-\dots -\overset{\overset{\overset{\overset{\displaystyle \overset 1\circ}{\diagup ~~\diagdown}}{\displaystyle {\scriptstyle 1} \circ ~~~~~~ \circ {\scriptstyle1}}}{\vert~~~~~~~~\vert}}{\node{}{\frac{N_f-1}{2}}-\node{}{\frac{N_f-1}{2}}}-\dots-\node{}2-\node{}1}_{N_f - 1}$ & $SU(N_f)\times SU(2)\times U(1)$ \\ \hline
	\end{tabular}
	\caption{The omponent of $\mathcal{H}_\infty$ for  $\frac 1 2=|k|=N_c - \frac{N_f}{2}$.}
	\label{tab:region2kHalf}
\end{table}

\begin{figure}
\centering
\begin{minipage}{8cm}
\centering
\includegraphics[width=7cm]{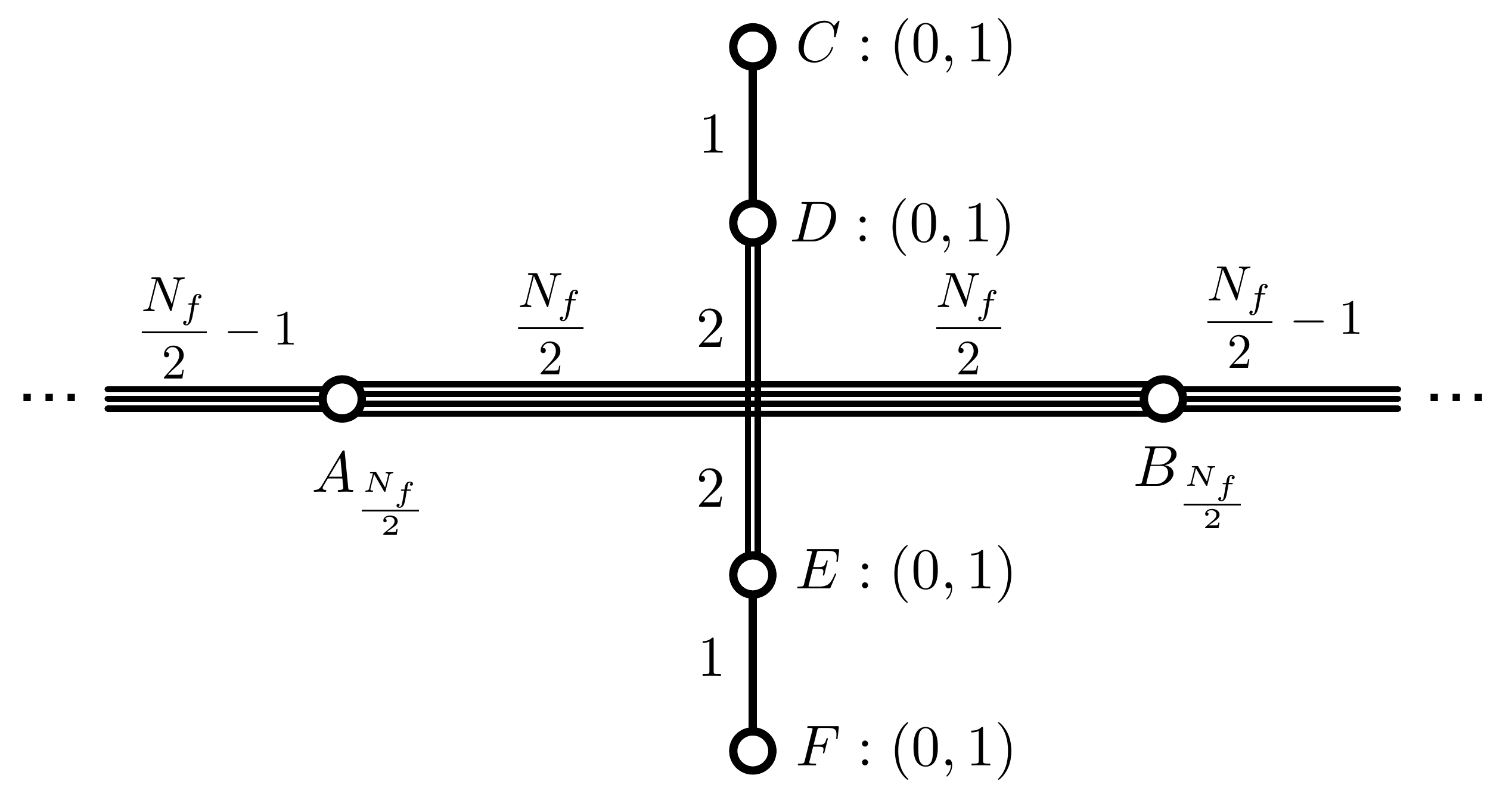}
\caption{5-brane web for $k=N_c - \frac{N_f}{2}=0$ at finite coupling.}
\label{Fig:k=Nc-Nf2=0}
\end{minipage}
\hspace{1cm}
\begin{minipage}{3.5cm}
\centering
\includegraphics[width=2.3cm]{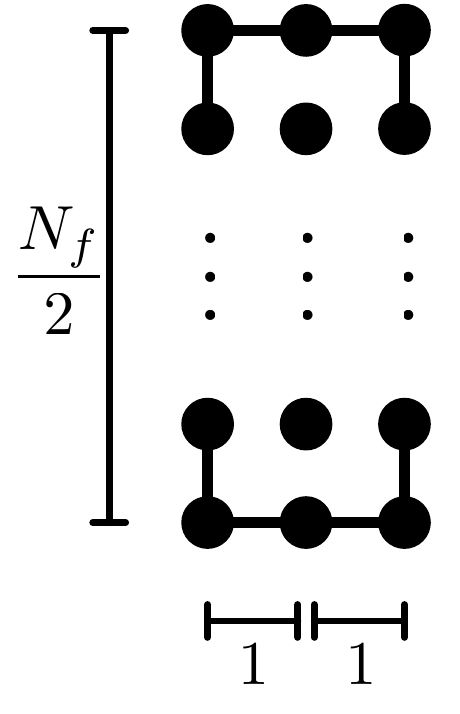}
\caption{Toric for $k=N_c - \frac{N_f}{2}=0$.}
\label{Fig:toric_k=Nc-Nf2=0}
\end{minipage}
\end{figure}

The next case is:
\begin{equation}
	|k| = 0 .
\end{equation}
The global symmetry enhancement is different:
\begin{equation}
	SU(N_f)\times SU(2)\times SU(2) .
\end{equation}
The web diagram at infinite coupling in this case is given in the form in figure \ref{Fig:k=Nc-Nf2=0}
while corresponding toric diagram is in figure \ref{Fig:toric_k=Nc-Nf2=0}.
Due to the fact that the 7-branes labelled by $C$ and $D$ are both (0,1) 7-branes, 
the maximal subdivision of the brane web is different from the generic case.

There is only a single phase \cite{Ferlito:2017xdq}, we denoted it as I$'$, and the corresponding quiver is given in table \ref{tab:region2k0}.

\begin{table}
	\centering
	\begin{tabular}{|c|c|c|}
		\hline
		Phase & Quiver & Global Symmetry \\ \hline
		I$'$ & $\underbrace{\node{}{1}-\node{}{2}-\dots -\node{\overset{{\llap{$\overset{{}}{\overset{{1}}{\circ}} -$}\displaystyle \overset{{2}}{\circ}{\rlap{$-\overset{{}}{\overset{{1}}{\circ}}$}}}}{\scriptstyle\vert}}{\frac{N_f}{2}} -\dots -\node{}{2}-\node{}{1}}_{N_f - 1}$ & $SU(N_f)\times SU(2)\times SU(2)$ \\ \hline
	\end{tabular}
	\caption{The component of $\mathcal{H}_\infty$ for  $0=|k|=N_c - \frac{N_f}{2}$.}
	\label{tab:region2k0}
\end{table}
%

\subsection{Third Region:  $|k|=N_c - \frac{N_f}{2}+1$}

\paragraph{General case: $|k|>\frac{3}{2}$.}

Let us consider the general case with:
\begin{equation}
	|k|>\frac 3 2 
\end{equation}

The global symmetry is enhanced to:
\begin{equation}
	SU(N_f+1)\times U(1)
\end{equation}

\begin{figure}
\centering
\begin{minipage}{7cm}
\centering
\includegraphics[width=7cm]{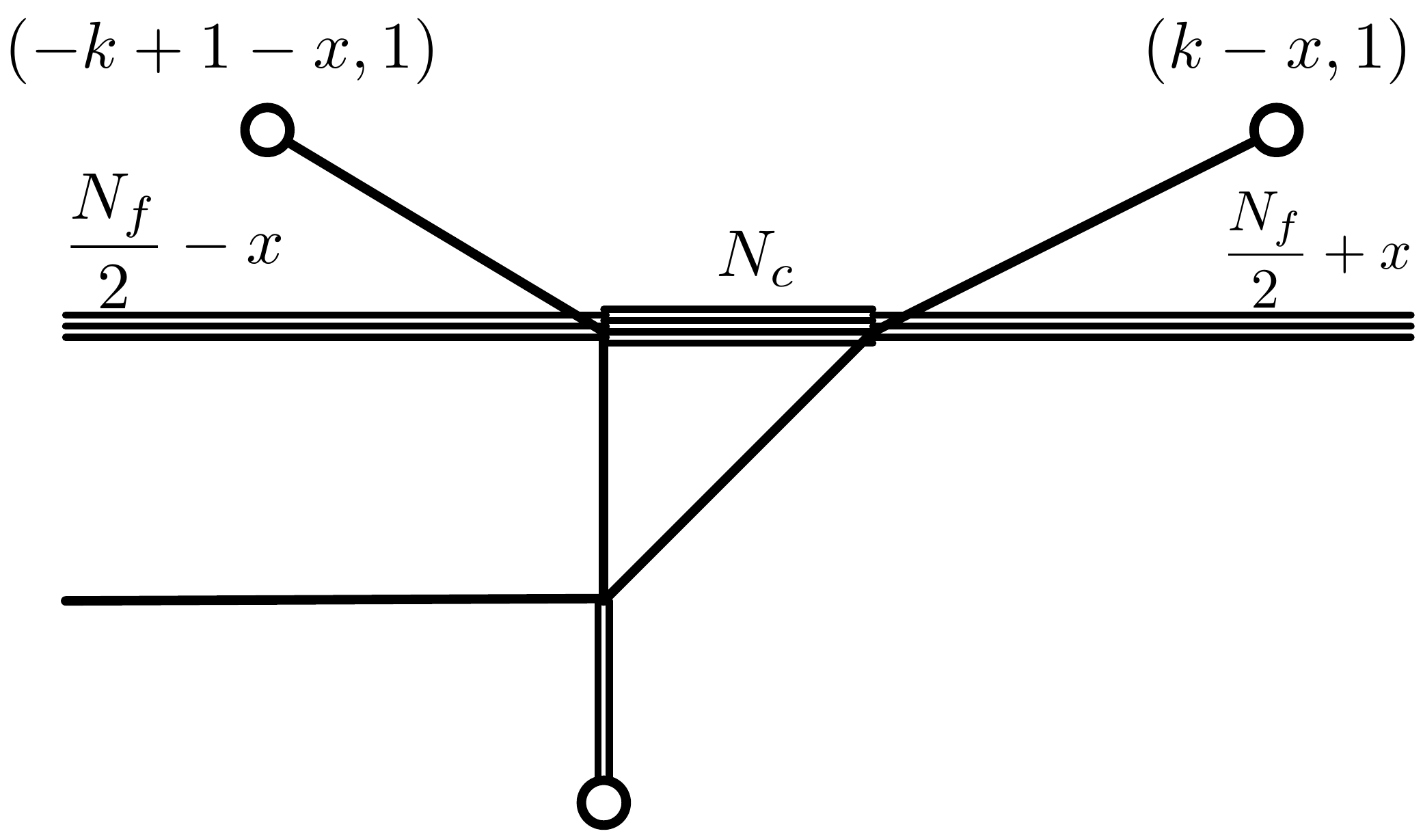}
\caption{5-brane web for $ 0 \le k =N_c - \frac{N_f}{2}+1$ at finite coupling.}
\label{Fig:k=Nc-Nf2+1_weak}
\end{minipage}
\hspace{1cm}
\begin{minipage}{4cm}
\centering
\includegraphics[width=4cm]{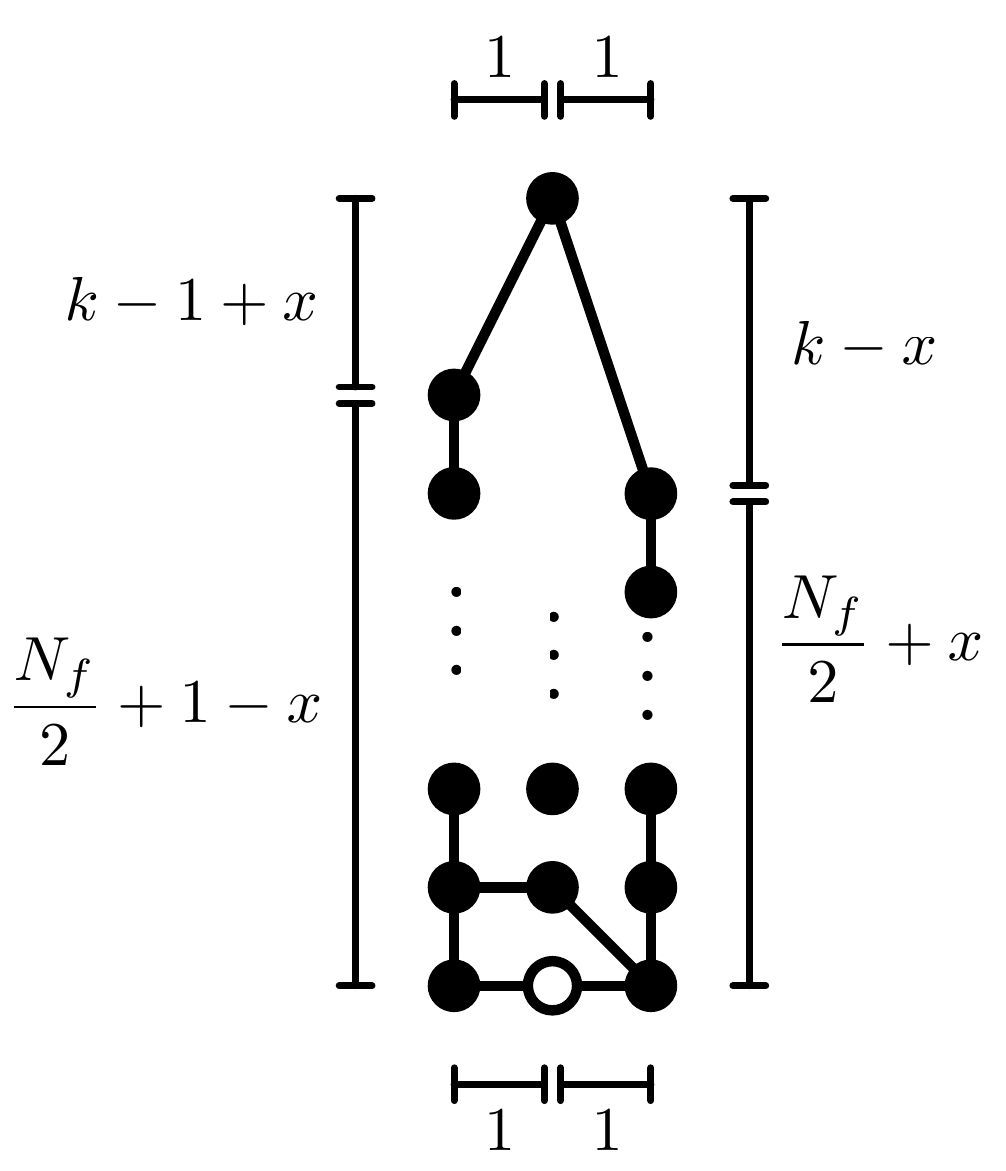}
\caption{Toric for $0 \le k =N_c - \frac{N_f}{2}+1$.}
\label{Fig:toric_k=Nc-Nf2+1}
\end{minipage}
\end{figure}

\begin{figure}
\centering
\includegraphics[width=15cm]{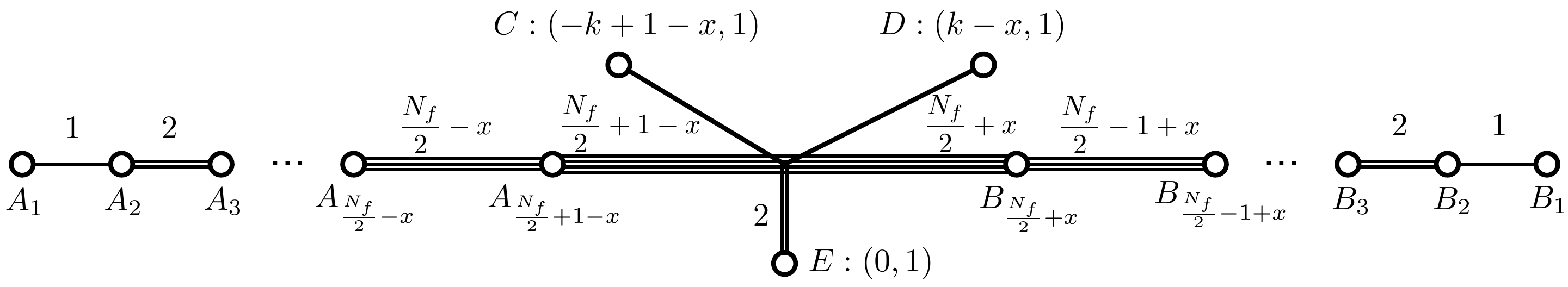}
\caption{5-brane web for $0 \le k =N_c - \frac{N_f}{2}+1$ at infinite coupling.}
\label{Fig:k=Nc-Nf2+1}
\end{figure}

\begin{figure}
\centering
\includegraphics[width=15cm]{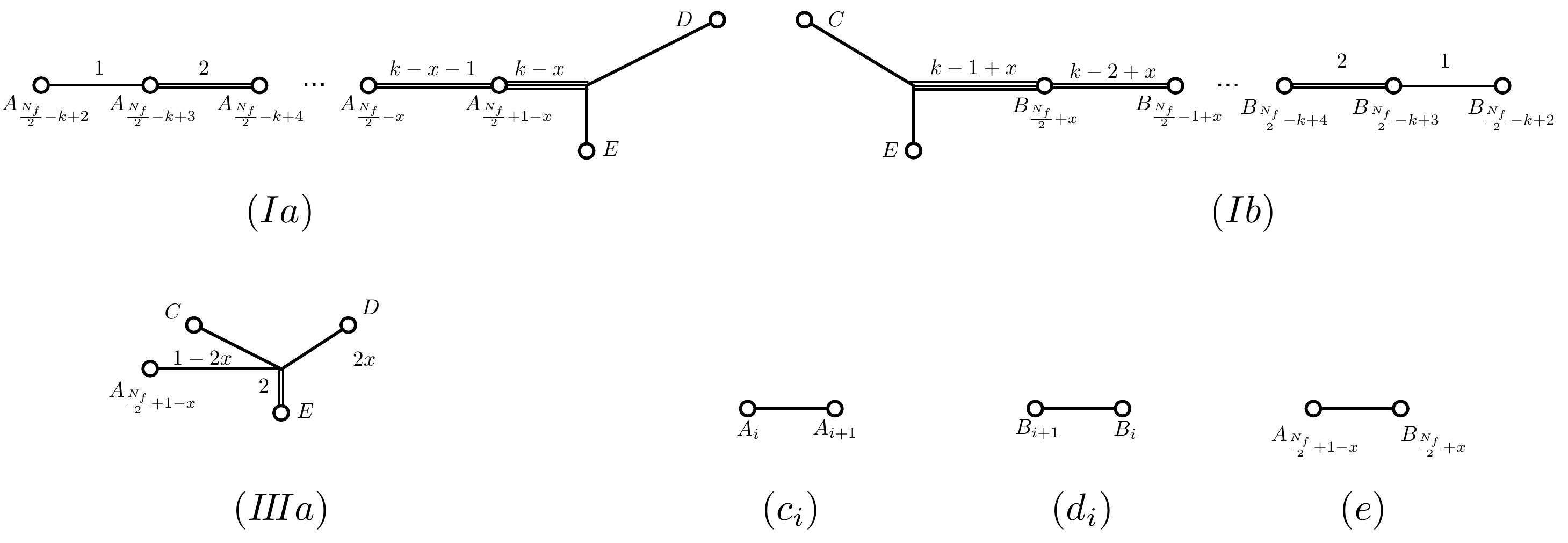}
\caption{The sub webs in the Higgs phase for $ 0 \le k =N_c - \frac{N_f}{2}+1$.}
\label{Fig:k=Nc-Nf2+1_subgraph}
\end{figure}

The brane system at finite coupling is depicted in figure \ref{Fig:k=Nc-Nf2+1_weak} as discussed in \cite{Bergman:2014kza, Bao:2013pwa}, while the corresponding toric diagram%
\footnote{
Although this 5-brane web configuration corresponds to non-toric geometry, we can also introduce the generalization of the toric diagram, which is denoted as ``dot diagram'' in \cite{Benini:2009gi}. In this paper, we simply denote ``toric diagram'' for such diagram. The white dots in the toric diagram correspond to the shrunken faces, which
indicate coincident 5-branes attached to an identical 7-brane, following the convention in \cite{Benini:2009gi}.
}
is in figure \ref{Fig:toric_k=Nc-Nf2+1}. 
When the coupling is taken to be infinite, the web diagram is represented in figure \ref{Fig:k=Nc-Nf2+1}. 
There are only two possible maximal subdivisions of the brane system (figure \ref{Fig:k=Nc-Nf2+1}). The different sub webs that make up these two different phases have been collected in figure \ref{Fig:k=Nc-Nf2+1_subgraph}. The two phases have been labelled I and III due to the similarities with the previous analysis. They are:

\begin{align}
\text{Phase I: } &(Ia) \times 1 + (Ib) \times 1 
+ \sum_{i=1}^{\frac{N_f}{2}-k+1} (c_i) \times i 
+ \sum_{i=\frac{N_f}{2}-k+2}^{\frac{N_f}{2}-x} (c_i) \times \left( \frac{N_f}{2} - k + 1 \right)
\cr
&+ \sum_{i=1}^{\frac{N_f}{2}-k+1} (d_i) \times i 
+ \sum_{i=\frac{N_f}{2}-k+2}^{\frac{N_f}{2}-1+x} (d_i) \times \left( \frac{N_f}{2} - k +1\right)
+ (e) \times \left( \frac{N_f}{2} - k +1 \right)
\cr
\text{Phase III: } &(I\! I\! I a) \times 1 
+ \sum_{i=1}^{\frac{N_f}{2}-x} (c_i) \times i 
+ \sum_{i=1}^{\frac{N_f}{2}-1+x} (d_i) \times i 
+ (e) \times \left( \frac{N_f}{2}+x \right)
\end{align}

 The quivers are depicted in table \ref{tab:region3}. Note that the number of nodes in the balanced chain of the quiver is extended with respect to the previous cases from $N_f-1$ to $N_f$.
%
%

\begin{table}
	\centering
	\begin{tabular}{|c|c|c|}
		\hline
		Phase & Quiver & Global Symmetry \\ \hline
		I & $\underbrace{\node{}1-\dots -\overset{\overset{\displaystyle\overset 1 \circ\xlongequal{2|k|-2}\overset 1 \circ}{\diagup~~~~~~~~~~~~~~~~~~\diagdown}}{\node{}{\frac{N_f}{2}-|k|+1}-~~~~\dots~~~~-\node{}{\frac{N_f}{2}-|k|+1}}-\dots-\node{}1}_{N_f}$ & $SU(N_f+1)\times U(1)$ \\ \hline
		III ($N_f$ even) & $\underbrace{\node{}1-\node{}2-\dots -\overset{\overset{\displaystyle\overset 1 \circ}{\diagup~~\diagdown}}{\node{}{\frac{N_f}2}-\node{}{\frac{N_f}2}}-\dots-\node{}2-\node{}1}_{N_f}$ & 	$SU(N_f + 1 )$ \\ \hline
		III ($N_f$ odd) & $\underbrace{\node{}1-\node{}2-\dots-\node{\overset{\displaystyle \overset{1}{\circ}}{\parallel}}{\frac{N_f+1 }2}-\dots-\node{}2-\node{}1}_{N_f}$  & $SU(N_f + 1 )$ \\ \hline
		I $\cap$ III & $\underbrace{\node{}1-\dots -\overset{\overset{\displaystyle\overset 1 \circ}{\diagup~~~~~~~~\diagdown}}{\node{}{\frac{N_f}{2}-|k|+1}-\dots-\node{}{\frac{N_f}{2}-|k|+1}}-\dots-\node{}1}_{N_f}$ & $SU(N_f+1)$ \\ \hline
	\end{tabular}
	\caption{Components of $\mathcal{H}_\infty$ for  $\frac 3 2<|k|=N_c - \frac{N_f}{2}+1$. Component I is present for $N_f \ge N_c$, which means $\frac{N_f}{2} \ge |k|-1$, and Component III is present for $N_f \ge 1$}
	\label{tab:region3}
\end{table}

%
%
%
%

\paragraph{Exceptional cases: $|k|\leq\frac 3 2$.}

The exceptional cases still have the same brane systems and toric diagram as those in figures \ref{Fig:k=Nc-Nf2+1_weak}, \ref{Fig:k=Nc-Nf2+1} and \ref{Fig:toric_k=Nc-Nf2+1}. Let us consider:

\begin{equation}
	|k|= \frac 3 2 .
\end{equation}

The global symmetry is enhanced to:
\begin{equation}
	SU(N_f+1)\times U(1) .
\end{equation}
In this case, the sub web $(I\!I\!Ia)$ is not allowed due to s-rule.
Thus, there is a unique maximal subdivision of the brane system, denoted as phase I, consistent with \cite{Ferlito:2017xdq}. The quiver obtained via Conjecture \ref{con:main} are given in table \ref{tab:region3k3Halves}.

\begin{table}
	\centering
	\begin{tabular}{|c|c|c|}
		\hline
		Phase & Quiver & Global Symmetry \\ \hline
		I & $\underbrace{\node{}1-\node{}2-\dots -\overset{\overset{\displaystyle\overset 1 \circ~ -~ \overset 1 \circ}{\diagup~~~~~~~~~\diagdown}}{\node{}{\frac{N_f-1}{2}}-\node{}{\frac{N_f-1}{2}}-\node{}{\frac{N_f-1}{2}}}-\dots-\node{}2-\node{}1}_{N_f}$ & $SU(N_f+1)\times U(1)$\\ \hline
	\end{tabular}
	\caption{The component of $\mathcal{H}_\infty$ for  $\frac 3 2 = |k|=N_c - \frac{N_f}{2}+1$.}
	\label{tab:region3k3Halves}
\end{table}

%

The next case is:
\begin{equation}
	|k|= 1 .
\end{equation}

The global symmetry is the same as in the general case:
\begin{equation}
	SU(N_f+1)\times U(1) .
\end{equation}
The phase III from the general case is contained in phase I in this case,
reflecting the property that $(I\!I\!Ia)$ can be further divided into $(Ia)$ and $(Ib)$.
Hence, there is a single phase, consistent with \cite{Ferlito:2017xdq}, given in table \ref{tab:region3k1}.

\begin{table}
	\centering
	\begin{tabular}{|c|c|c|}
		\hline
		Phase & Quiver & Global Symmetry \\ \hline
		I & $\underbrace{\node{}1-\dots -\node{\upnode 1}{\frac{N_f}{2}}-\node{\upnode 1}{\frac{N_f}{2}}-\dots-\node{}1}_{N_f}$ & $SU(N_f+1)\times U(1)$  \\ \hline
	\end{tabular}
	\caption{The component of $\mathcal{H}_\infty$ for  $1=|k|=N_c - \frac{N_f}{2}+1$.}
	\label{tab:region3k1}
\end{table}
%

\begin{figure}
\centering
\begin{minipage}{7cm}
\centering
\includegraphics[width=7cm]{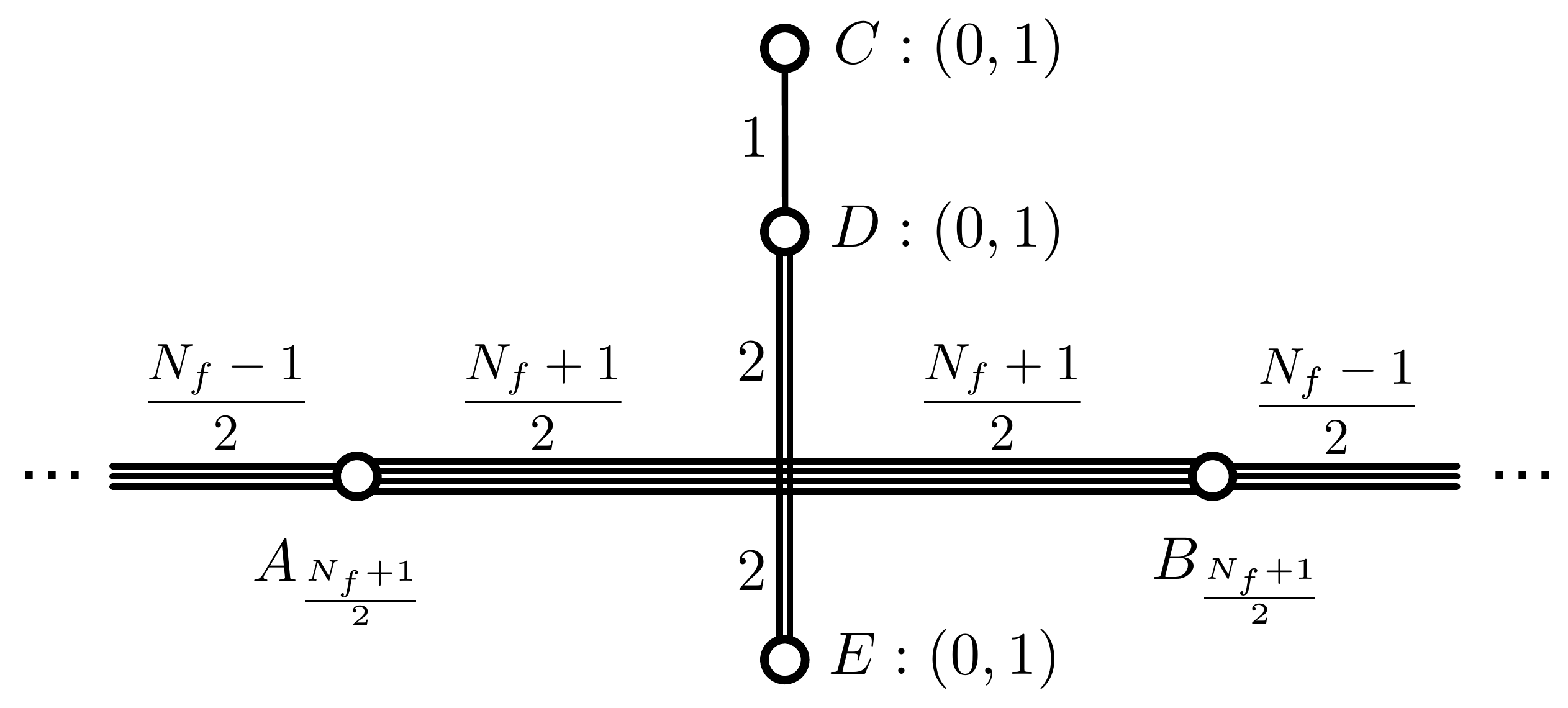}
\caption{5-brane web for $\frac{1}{2}=k=N_c - \frac{N_f}{2}+1$ at finite coupling.}
\label{Fig:k=Nc-Nf2+1=1over2}
\end{minipage}
\hspace{1cm}
\begin{minipage}{3.5cm}
\centering
\includegraphics[width=2.5cm]{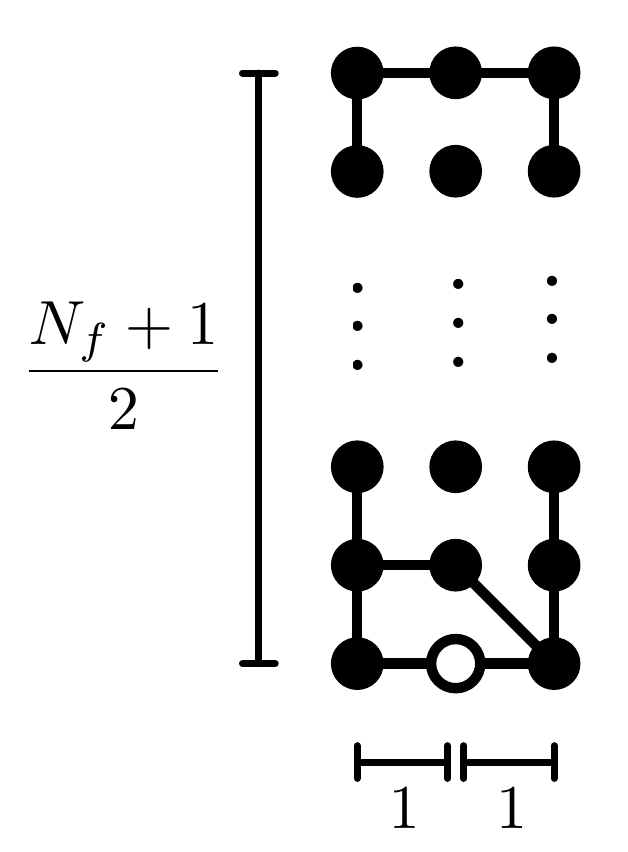}
\caption{Toric for $\frac{1}{2}=k=N_c - \frac{N_f}{2}+1$.}
\label{Fig:toric_k=Nc-Nf2+1=1over2}
\end{minipage}
\end{figure}

The next case is:
\begin{equation}
	|k|= \frac 1 2 .
\end{equation}
The global symmetry is enhanced to:
\begin{equation}
	SU(N_f+1)\times SU(2) .
\end{equation}
In this case, the web diagram at infinite coupling is given as in figure \ref{Fig:k=Nc-Nf2+1=1over2}
while the corresponding toric diagram is in figure \ref{Fig:toric_k=Nc-Nf2+1=1over2}.
%
%
There is a single phase I$'$, consistent with \cite{Ferlito:2017xdq}, given in table \ref{tab:region3kHalf}.
\begin{table}
	\centering
	\begin{tabular}{|c|c|c|}
		\hline
		Phase & Quiver & Global Symmetry \\ \hline
		I$'$ & $\underbrace{\node{}1-\node{}2-\dots -\node{\overset{\displaystyle \overset{\overset{\displaystyle {\circ}\rlap{\,\,$\scriptstyle 1$}}{\vert}}{\circ}\rlap{\,\,$\scriptstyle 2$}}{\vert}}{\frac{N_f+1 }2}-\dots-\node{}2-\node{}1}_{N_f }$ &   $SU(N_f+1)\times SU(2)$\\ \hline
	\end{tabular}
	\caption{The component of $\mathcal{H}_\infty$ for  $\frac 1 2 = |k|=N_c - \frac{N_f}{2}+1$.}
	\label{tab:region3kHalf}
\end{table}

\begin{figure}
\centering
\begin{minipage}{7cm}
\centering
\includegraphics[width=7cm]{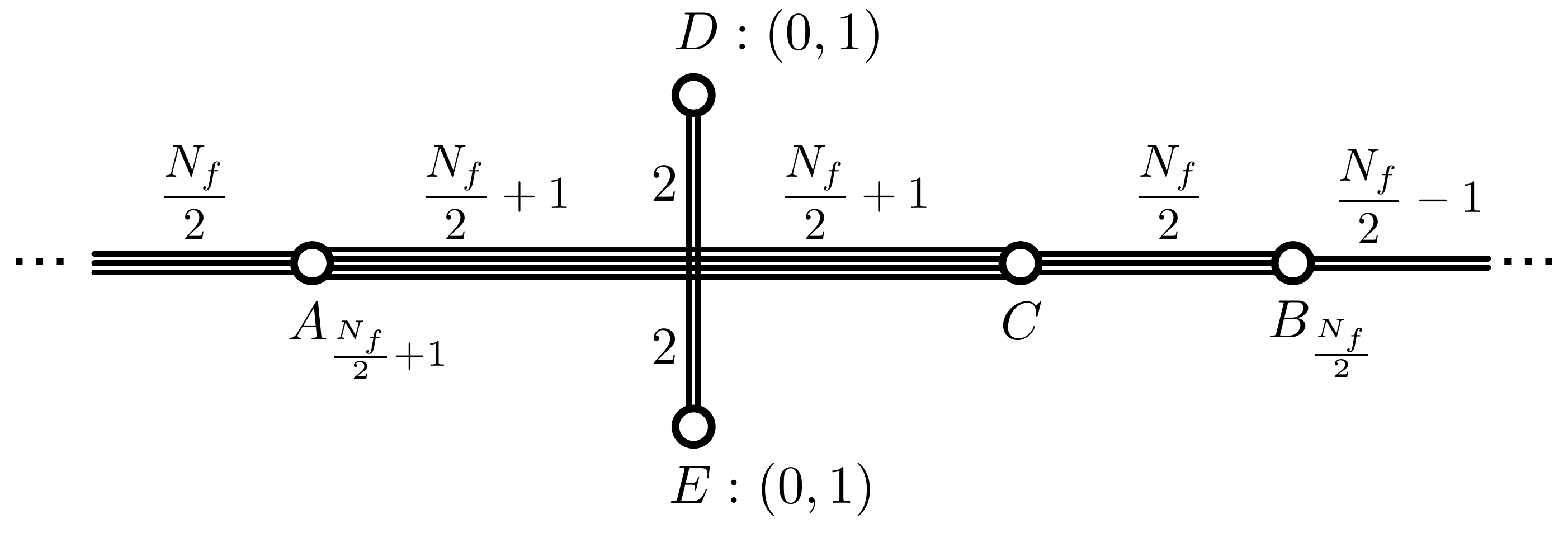}
\caption{5-brane web for $0=k=N_c - \frac{N_f}{2}+1$ at finite coupling.}
\label{Fig:k=Nc-Nf2+1=0}
\end{minipage}
\hspace{1cm}
\begin{minipage}{3.5cm}
\centering
\includegraphics[width=2.5cm]{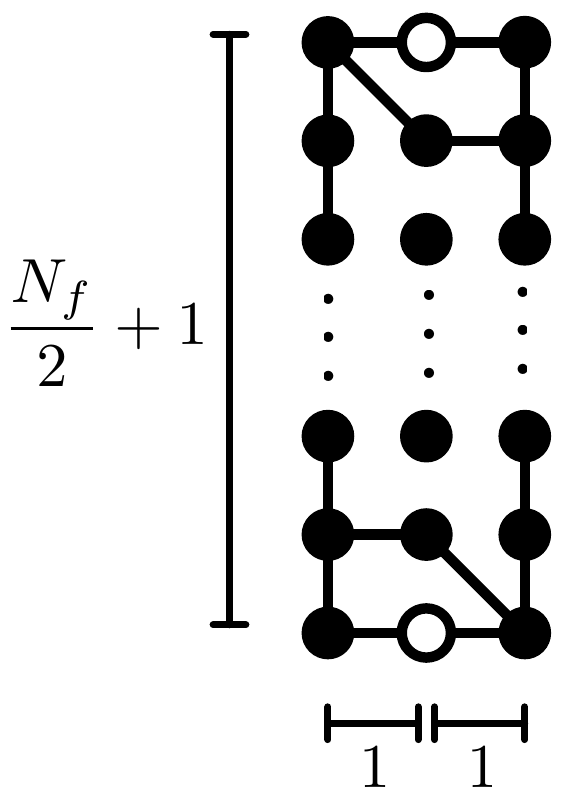}
\caption{Toric for $0=k=N_c - \frac{N_f}{2}+1$.}
\label{Fig:toric_k=Nc-Nf2+1=0}
\end{minipage}
\end{figure}

The next case is:
\begin{equation}
	|k| = 0
\end{equation}
The global symmetry is enhanced to:
\begin{equation}
	SU(N_f+2) .
\end{equation}
The web diagram and the toric diagrams are given in figure \ref{Fig:k=Nc-Nf2+1=0} and
in figure \ref{Fig:toric_k=Nc-Nf2+1=0}, respectively.
%
%
There is a single component I$'$, consistent with \cite{Ferlito:2017xdq}, given in table \ref{tab:region3k0}.

\begin{table}
	\centering
	\begin{tabular}{|c|c|c|}
		\hline
		Phase & Quiver & Global Symmetry \\ \hline
		I$'$ & $\underbrace{\node{}{1}-\node{}{2}-\dots -\node{\upnode{2}}{\frac{N_f+2}{2}} -\dots -\node{}{2}-\node{}{1}}_{N_f + 1}$ & $SU(N_f+2)$ \\ \hline
	\end{tabular}
	\caption{The component of $\mathcal{H}_\infty$ for  $0=|k|=N_c - \frac{N_f}{2}+1$.}
	\label{tab:region3k0}
\end{table}


\subsection{Fourth Region:  $|k|=N_c - \frac{N_f}{2}+2$}

\begin{figure}
\centering
\includegraphics[width=6.5cm]{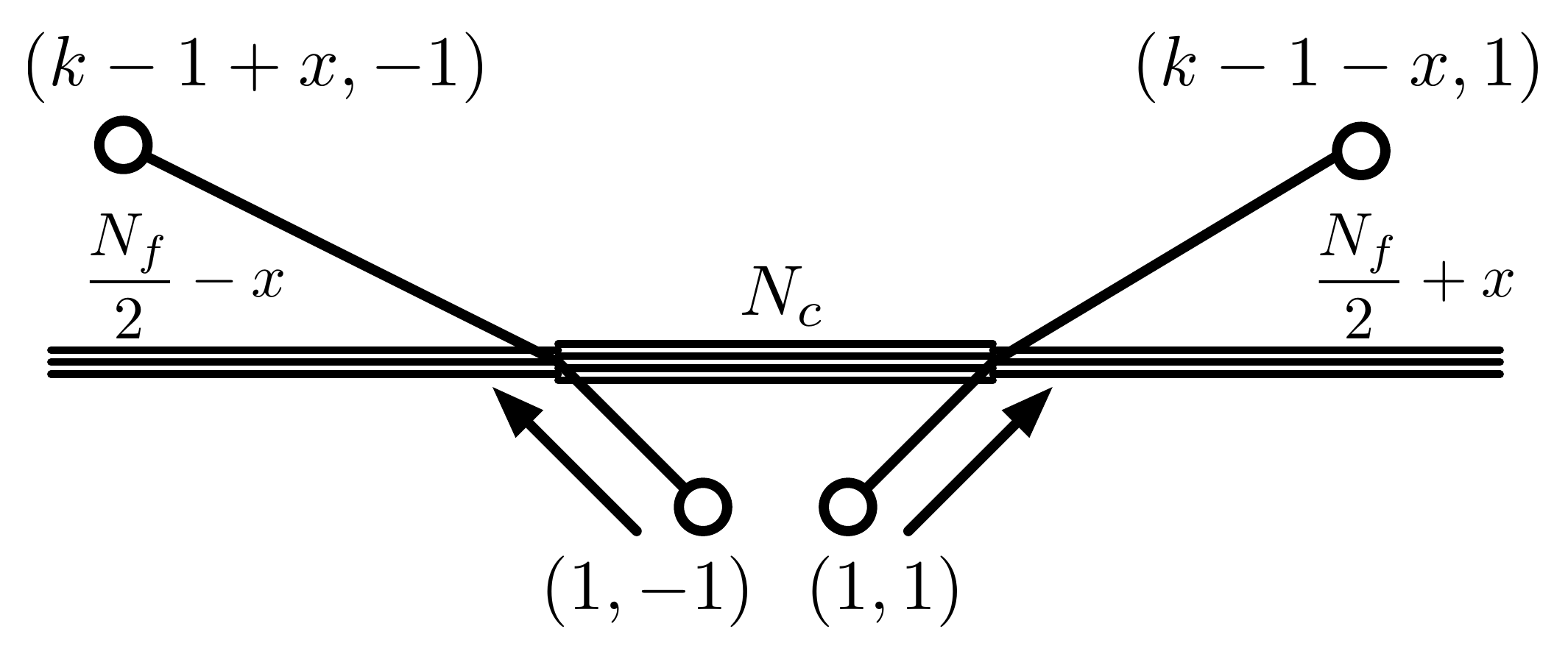}
\caption{Brane system for $0 \le k=N_c - \frac{N_f}{2}+2$.}
\label{Fig:k=Nc-Nf2+2_1}
\end{figure}

\begin{figure}
\centering
\begin{minipage}{6.5cm}
\includegraphics[width=6.5cm]{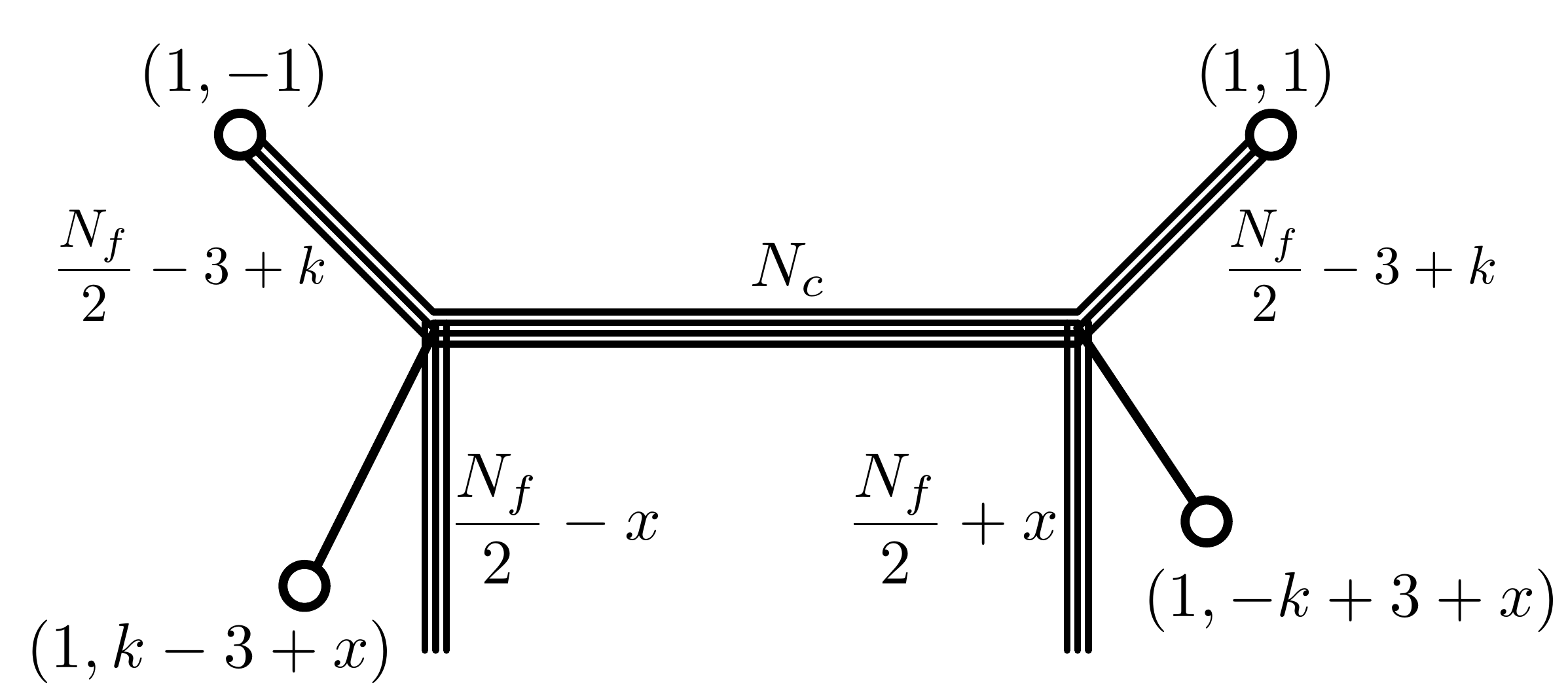}
\caption{$k=N_c - \frac{N_f}{2}+2 > 2$. After Hanany-Witten transition.}
\label{Fig:k=Nc-Nf2+2_2_large}
\end{minipage}
\hspace{5mm}
\begin{minipage}{6.5cm}
\includegraphics[width=6.5cm]{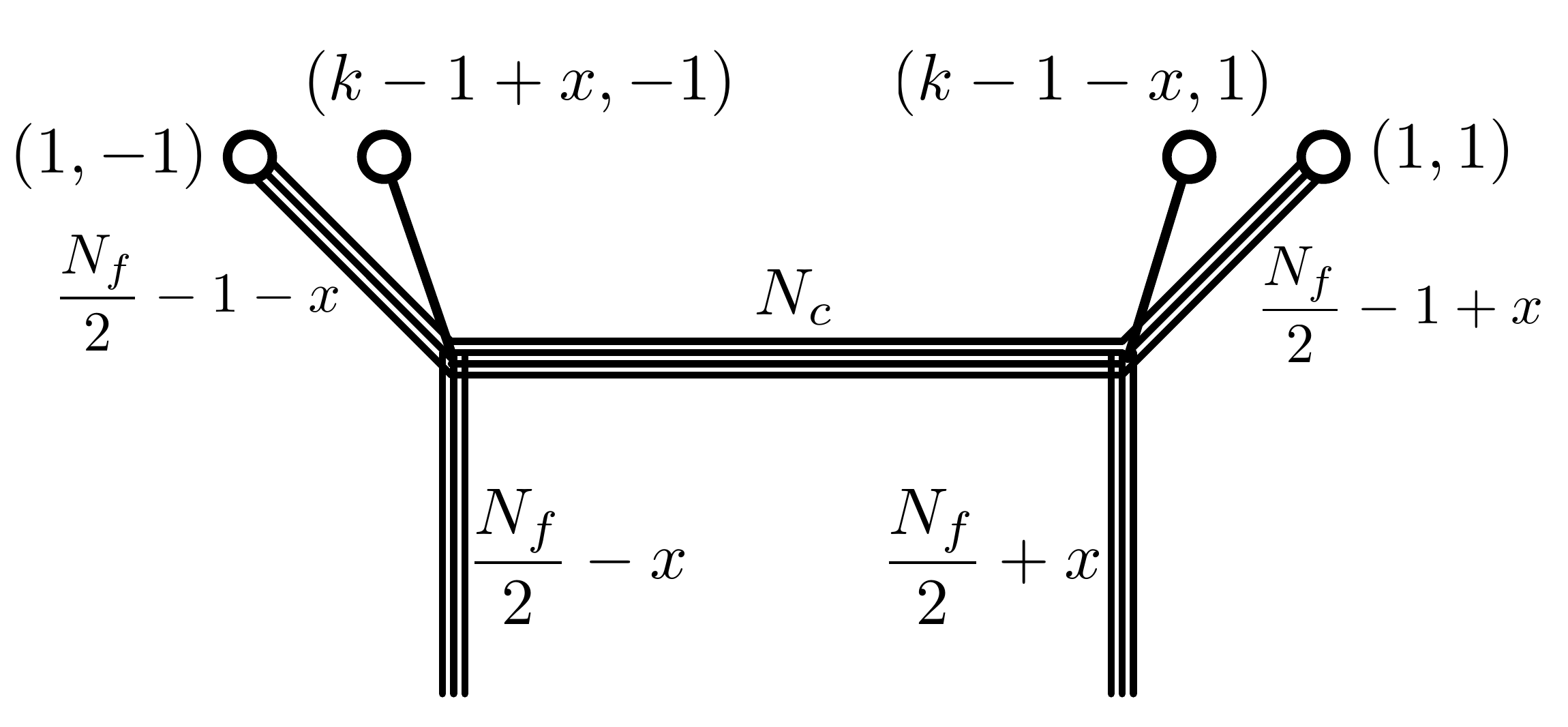}
\caption{$0 \le k=N_c - \frac{N_f}{2}+2 \le 2$. After Hanany-Witten transition.}
\label{Fig:k=Nc-Nf2+2_2_small}
\end{minipage}
\end{figure}

The brane system for the theory with $|k|=N_c - \frac{N_f}{2}+2$ at finite coupling is depicted in figure \ref{Fig:k=Nc-Nf2+2_1}. 
Performing Hanany-Witten transitions to this web diagram, we obtain figure \ref{Fig:k=Nc-Nf2+2_2_large} for $|k| > 2$ and
figure \ref{Fig:k=Nc-Nf2+2_2_small} for $|k| \le 2$.

\paragraph{General case: $|k|>2$.}

\begin{figure}
\centering
\begin{minipage}{8cm}
\includegraphics[width=7.5cm]{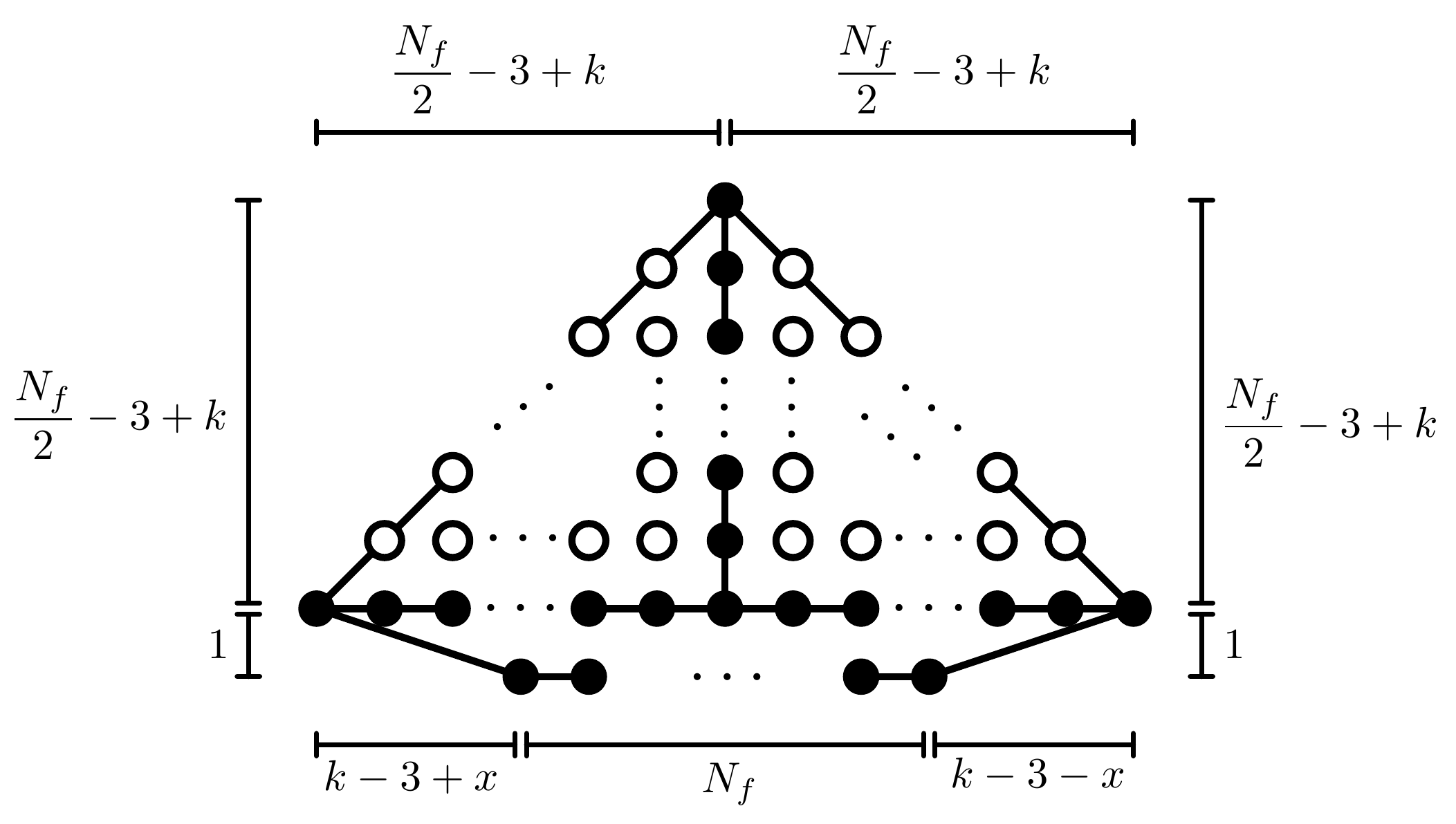}
\caption{Toric diagram for $k=N_c - \frac{N_f}{2}+2 > 2$.}
\label{Fig:toric_k=Nc-Nf2+2}
\end{minipage}
\begin{minipage}{6.5cm}
\includegraphics[width=6cm]{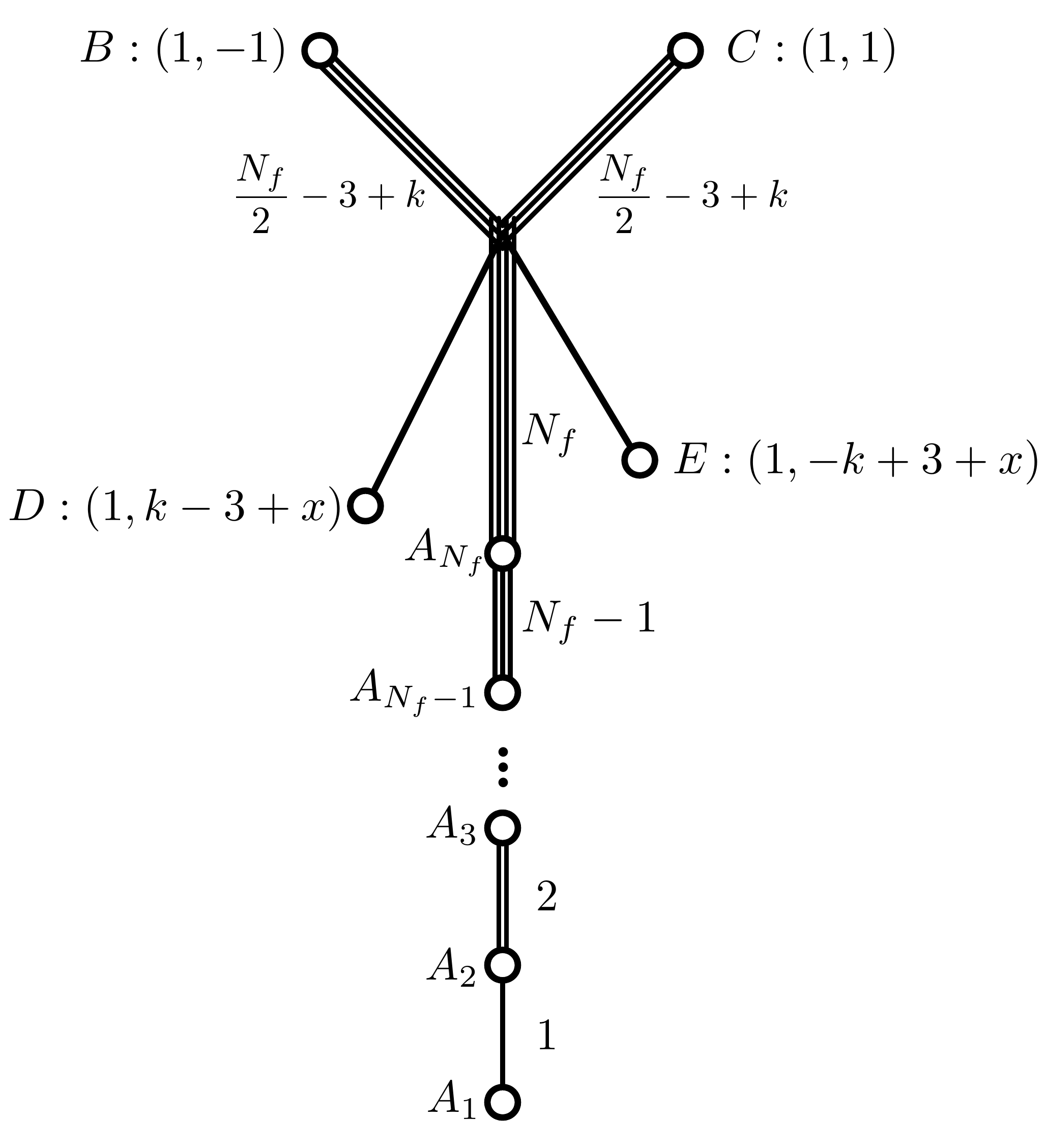}
\caption{$k=N_c - \frac{N_f}{2}+2 > 2 $. Strong coupling limit.}
\label{Fig:k=Nc-Nf2+2_3}
\end{minipage}
\end{figure}

We first concentrate on the former case.
The global symmetry is:
\begin{equation}
	SO(2N_f)\times U(1) .
\end{equation}
The corresponding toric diagram is in figure \ref{Fig:toric_k=Nc-Nf2+2}.
The final brane system after taking the gauge coupling to infinity for is depicted in figure \ref{Fig:k=Nc-Nf2+2_3}.


\begin{figure}
\centering
\includegraphics[width=14cm]{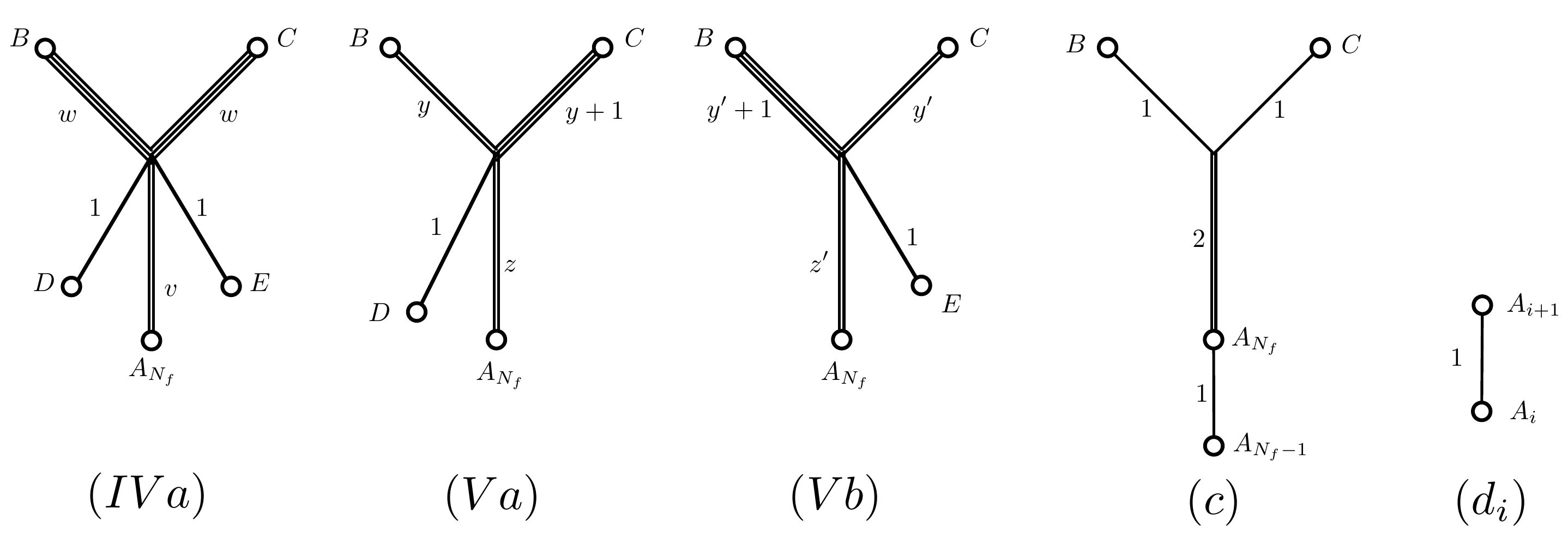}
\caption{The sub webs in the Higgs phase for $k=N_c - \frac{N_f}{2}+2 \ge 2$. 
$(IVa)$ exists only for integer $k(\ge 3)$.
For $k=\frac{5}{2}$, where $y'=-1$, $z'=0$ and $E$ is a (1,1) 7-brane, $(Vb)$ should be interpreted as a diagram with a single (1,1) 5-brane suspended between $C$ and $E$.
}
\label{Fig:k=Nc-Nf2+2_subgraph}
\end{figure}

We classify the Chern-Simons level $k$ into four classes as
\begin{align}
|k| = 2n + \alpha \quad \text{ with } \quad \alpha= 0, \frac{1}{2}, 1 , \frac{3}{2}.
\end{align}
In order to treat these four cases in a unified way, we introduce the variables $y$, $z$, $y'$ and $z'$ given as
\begin{align}
\begin{array}{l}
y=n-2, \quad  z=0, \quad y'=n-2, \quad z'=0  \qquad \text{for} \quad \alpha=0
\\
y=n-1, \quad  z=1, \quad y'=n-2, \quad z'=0  \qquad \text{for} \quad \alpha=\frac{1}{2}
\\
y=n-1, \quad  z=1, \quad y'=n-1, \quad z'=1  \qquad \text{for} \quad \alpha=1
\\
y=n-1, \quad  z=0, \quad y'=n-1, \quad z'=1  \qquad \text{for} \quad \alpha=\frac{3}{2}
\end{array}
\end{align}
while $v$ and $w$ are defined only for integer $k$ and given as 
\begin{align}
\begin{array}{l}
v=2, \quad w=2n-2  \qquad \text{for} \quad \alpha=0
\\
v=0, \quad w=2n-2  \qquad \text{for} \quad \alpha=1
\end{array}
\end{align}

The different sub webs that conform the maximal web sub divisions are in figure \ref{Fig:k=Nc-Nf2+2_subgraph}. 
There is a phase IV that only exists for $N_f (\ge 2)$ even, and a different phase V that changes from $N_f$ even and $N_f$ odd:

\begin{align}\label{eq:Maximal_div_k=Nc-Nf2+2}
\text{Phase VI: } &(IV a) \times 1 + (c) \times \left( \frac{N_f-v}{2} \right) 
+ (d_{N_f-1}) \times \left( \frac{N_f-2+v}{2} \right)
+ \sum_{i=1}^{N_f-2} (d_i) \times i
\cr
\text{Phase V: } &(Va) \times 1 + (Vb) \times 1 +  (c) \times \left( \frac{N_f-z-z'}{2}  \right) 
+ (d_{N_f-1}) \times \left( \frac{N_f +z+z'-2}{2}  \right)
\cr
& \qquad + \sum_{i=1}^{N_f-2} (d_i) \times i
\end{align}
Especially for $N_f=0$, Phase V does not exist if $k$ is odd ($\alpha=1$), while it exists as 
$(Va) \times 1 + (Vb) \times 1$ if $k$ is even ($\alpha=0$).
For $N_f=1$, Phase V always exists as $(Va) \times 1 + (Vb) \times 1$ for generic half integer $k$. 
The corresponding component of  $\mathcal{H}_\infty$ should be understood accordingly.

The different phases are depicted in table \ref{tab:region4}. Note again, that the intersection between the even components is a closure of a nilpotent orbit of height 2 of the global symmetry $SO(2N_f)$. Physically this means that the operators that are shared everywhere on the moduli space are mesons only, while baryons and other exotic objects are specific to individual components. It should be crucially noted that chiral ring relations do get corrections from instanton effects even at intersections. The mesons are no longer interpreted as the usual quark bilinears, hence the nilpotecy condition of at most $N_c$ is no longer correct, and becomes maximal or near maximal.

\begin{table}
	\centering
	\begin{tabular}{|c|c|c|}
		\hline
		Phase & Quiver & Global Symmetry \\ \hline
		IV ($N_f$ even) & 
		 $\node{}{1}-\node{}2-\dots-\node{}{N_f-3}-\node{\upnode{\frac{N_f-2}2}}{N_f-2}-\node{}{\frac{N_f}2}=\node{}{1}$ &  $SO(2N_f)$ \\ \hline
		V ($N_f$ even) & $\node{}{1}-\node{}2-\dots-\node{}{N_f - 3}-\node{\upnode{\frac{N_f-2}2}}{N_f - 2}-\overset{\overset{\displaystyle\overset 1 \circ}{\diagup~\Nwseline\rlap{\,\,$\scriptstyle N_c - \frac{N_f}2$}}}{\node{}{\frac{N_f}2}-\node{}{1}}~~~~~~~~~$ &  $SO(2N_f)\times U(1)$ \\ \hline
		V ($N_f$ odd) & $\node{}{1}-\node{}2-\dots-\node{}{N_f - 3}-\overset{\overset{\displaystyle\overset {\frac{N_f-1}2} \circ-~\overset {1} \circ}{~\diagup~~~~~~~~~\Nwseline\rlap{\,\,$\scriptstyle N_c - \frac{N_f-1}2$}}}{\node{}{N_f - 2}-\node{}{\frac{N_f - 1}2}-\node{}{1}} ~~~~~~~~~$  &  $SO(2N_f)\times U(1)$ \\ \hline
		IV $\cap$ V ($N_f$ even) &$\node{}{1}-\node{}2-\dots-\node{}{N_f-3}-\node{\overset{\displaystyle\overset {\frac{N_f-2}2} \circ~~ ~\overset {1} \circ~~}{\diagdown~~\diagup}}{N_f-2}-\node{}{\frac{N_f-2}2}$ &  $SO(2N_f)$\\ \hline
	\end{tabular}
	\caption{Components of $\mathcal{H}_\infty$ for  $2<|k|=N_c - \frac{N_f}{2}+2$.
	Component IV is present for $N_f \ge 2$ with $N_f$ even. Component V ($N_f$ even) is present for $N_f \ge 0$ if $k$ is even and is present for $N_f \ge 2$ if $k$ is odd. Component V ($N_f$ odd) is present for $N_f \ge 1$.}
	\label{tab:region4}
\end{table}

%

\paragraph{Exceptional cases: $|k|\leq 2$.}

\begin{figure}
\centering
\includegraphics[width=14cm]{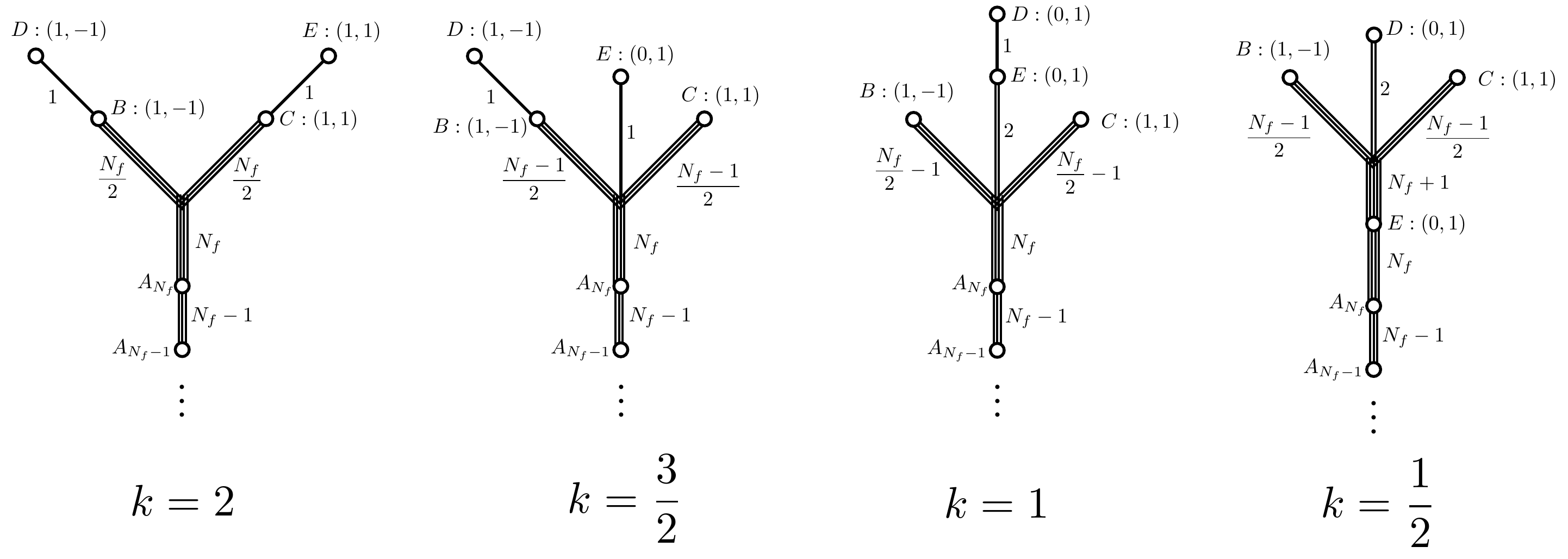}
\caption{$0 < k=N_c - \frac{N_f}{2}+2 \le 2 $. Strong coupling limit.}
\label{Fig:k=Nc-Nf2+2_exceptional}
\end{figure}

\begin{figure}
\centering
\includegraphics[width=14cm]{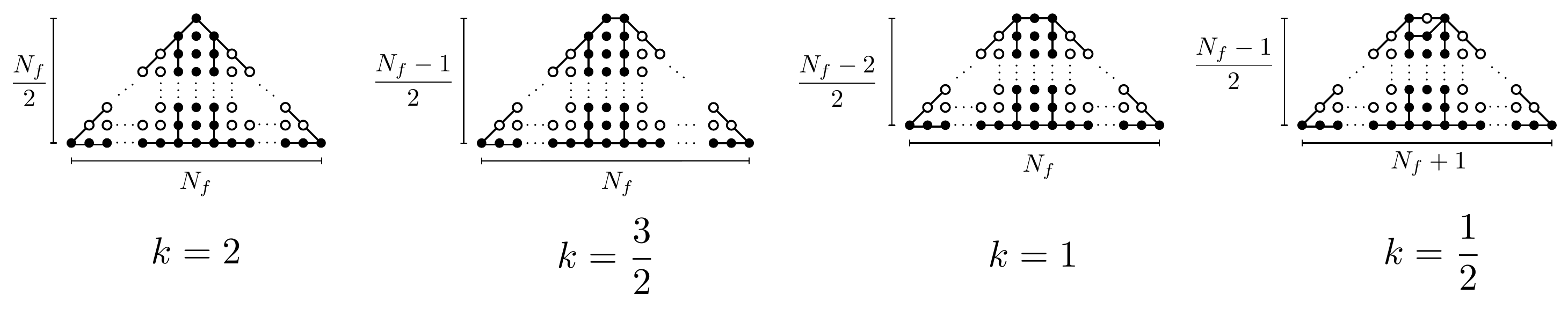}
\caption{Toric diagrams for $0 < k=N_c - \frac{N_f}{2}+2 \le 2 $. }
\label{Fig:toric_k=Nc-Nf2+2_small}
\end{figure}

\begin{figure}
\centering
\includegraphics[width=14cm]{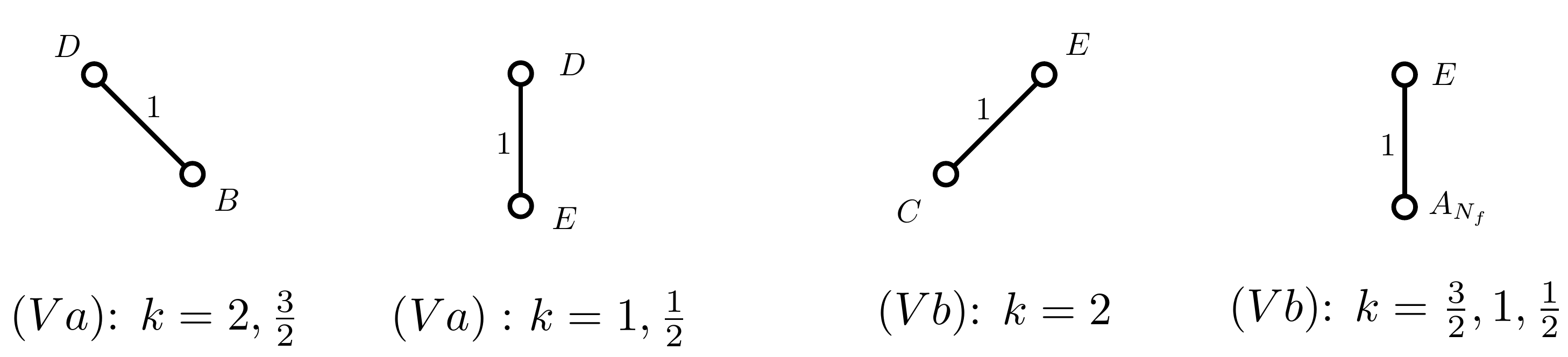}
\caption{The sub webs in the Higgs phase for $|k|=N_c - \frac{N_f}{2}+2 \le 2$.}
\label{Fig:k=Nc-Nf2+2_small_subgraph}
\end{figure}

The 5-brane web diagrams at strong coupling for $|k| \le 2$, obtained from figure \ref{Fig:k=Nc-Nf2+2_2_small},
are given in figure \ref{Fig:k=Nc-Nf2+2_exceptional}. 
The corresponding toric diagrams are given in figure \ref{Fig:toric_k=Nc-Nf2+2_small}.
The different sub webs that conform the maximal web sub divisions are given analogous to figure \ref{Fig:k=Nc-Nf2+2_subgraph}. 
However, the counterpart of $(IVa)$ does not exist in this case. For $|k|=2$, this is due to the s-rule. For $|k|=1$, it is simply because $w<0$.
The sub webs $(Va)$ and $(Vb)$ are replaced as in figure \ref{Fig:k=Nc-Nf2+2_small_subgraph} depending on the value of $k$.
The sub webs $(c)$ and $(d_i)$ are identical except that for $|k|=1/2$, $A_{N_f}$ and $A_{N_f-1}$ are replaced by $E$ and $A_{N_f}$, respectively.

Let us consider:

\begin{equation}
	|k|= 2 .
\end{equation}

The global symmetry is the same as in the general case:
\begin{equation}
	SO(2N_f)\times U(1) .
\end{equation}
There is a single phase V. 
The phase IV does not exist since $(IVa)$ is not allowed due to s-rule as stated above.
The maximal subdivision is given as in equation \ref{eq:Maximal_div_k=Nc-Nf2+2}.
The corresponding quiver is depicted in table \ref{tab:region4k2},
 which can be seen as the special case of Table \ref{tab:region4}.

\begin{table}
	\centering
	\begin{tabular}{|c|c|c|}
		\hline
		Phase & Quiver & Global Symmetry \\ \hline
		V ($N_f$ even) &   $\node{}{1}-\node{}2-\dots-\node{}{N_f-3}-\node{\upnode{\frac{N_f-2}2}}{N_f-2}-\node{\upnode{1}}{\frac{N_f}2}-\node{}{1}$ &
		  $SO(2N_f)\times U(1)$
		\\ \hline
	\end{tabular}
	\caption{The component of $\mathcal{H}_\infty$ for  $2=|k|=N_c - \frac{N_f}{2}+2$.}
	\label{tab:region4k2}
\end{table}


The next case is:
\begin{equation}
	|k|= \frac 3 2 .
\end{equation}

This case is not an exception. The global symmetry is the same as in the general case:
\begin{equation}
	SO(2N_f)\times U(1) .
\end{equation}
The maximal subdivision is given as in equation \ref{eq:Maximal_div_k=Nc-Nf2+2}
and the Higgs branch is also given by the general case, with a single component V, consistent with \cite{Ferlito:2017xdq}. It is depicted in table \ref{tab:region4k3Halves}.

\begin{table}
	\centering
	\begin{tabular}{|c|c|c|}
		\hline
		Phase & Quiver & Global Symmetry \\ \hline
		V ($N_f$ odd) &   $\node{}{1}-\node{}2-\dots-\node{}{N_f - 3}-\overset{\overset{\displaystyle\overset {\frac{N_f-1}2} \circ-~\overset {1} \circ}{~\diagup~~~~~~~~~}}{\node{}{N_f - 2}-\node{}{\frac{N_f - 1}2}-\node{}{1}} $ &
		  $SO(2N_f)\times U(1)$
		\\ \hline
	\end{tabular}
	\caption{The component of $\mathcal{H}_\infty$ for  $\frac 3 2 = |k|=N_c - \frac{N_f}{2}+2$.}
	\label{tab:region4k3Halves}
\end{table}

%

The next case is:
\begin{equation}
	|k|= 1 .
\end{equation}

The global symmetry is not the same as in the general case:
\begin{equation}
	SO(2N_f)\times SU(2) .
\end{equation}
There is a single phase V$'$,
The maximal subdivision is given as 
\begin{align}
\text{Phase V$'$: } &(Va) \times 1 + (Vb) \times 2 +  (c) \times \left( \frac{N_f-2}{2} \right)
+ (d_{N_f-1}) \times \left( \frac{N_f}{2} \right)
 + \sum_{i=1}^{N_f-2} (d_i) \times i
\end{align}
which is slightly modified from equation \ref{eq:Maximal_div_k=Nc-Nf2+2}.
The result is consistent with \cite{Ferlito:2017xdq}, see table \ref{tab:region4k1}.

\begin{table}
	\centering
	\begin{tabular}{|c|c|c|}
		\hline
		Phase & Quiver & Global Symmetry \\ \hline
		V$'$ ($N_f$ even) &   $\node{}{1}-\node{}2-\dots-\node{}{N_f-3}-\node{\upnode{\frac{N_f-2}2}}{N_f-2}-\node{}{\frac{N_f}2}-\node{}{2}-\node{}{1}$  &  $SO(2N_f)\times SU(2)$
		\\ \hline
	\end{tabular}
	\caption{The component of $\mathcal{H}_\infty$ for  $1=|k|=N_c - \frac{N_f}{2}+2$.}
	\label{tab:region4k1}
\end{table}

The next case is:
\begin{equation}
	|k|= \frac 1 2 .
\end{equation}

The global symmetry is enhanced to:
\begin{equation}
	SO(2N_f+2) .
\end{equation}
There is a single phase V$'$.
The maximal subdivision is given as 
\begin{align}
\text{Phase V$'$: } &(Va) \times 2 + (Vb) \times \left( \frac{N_f+1}{2} \right) +  (c) \times \left( \frac{N_f-1}{2} \right)
 + \sum_{i=1}^{N_f-1} (d_i) \times i
\end{align}
The quiver is depicted in table \ref{tab:region4kHalf}.

\begin{table}
	\centering
	\begin{tabular}{|c|c|c|}
		\hline
		Phase & Quiver & Global Symmetry \\ \hline
		V$'$ ($N_f$ odd) &  $\node{}{1}-\node{}2-\dots-\node{}{N_f-2}-\node{\upnode{\frac{N_f-1}2}}{N_f-1}-\node{}{\frac{N_f+1}{2}}-\node{}{2}$ & $SO(2N_f+2)$ 
		\\ \hline
	\end{tabular}
	\caption{The component of $\mathcal{H}_\infty$ for  $\frac 1 2 = |k|=N_c - \frac{N_f}{2}+2$.}
	\label{tab:region4kHalf}
\end{table}

%

The last case is:

\begin{equation}
	|k| = 0 .
\end{equation}
In this case the theory has a $6d$ fixed point.

\subsection{Excetional Case: Fifth Region with $|k|=N_c-\frac{N_f}{2}+3$ and $N_c=3$}
\begin{figure}
\centering
\includegraphics[width=8cm]{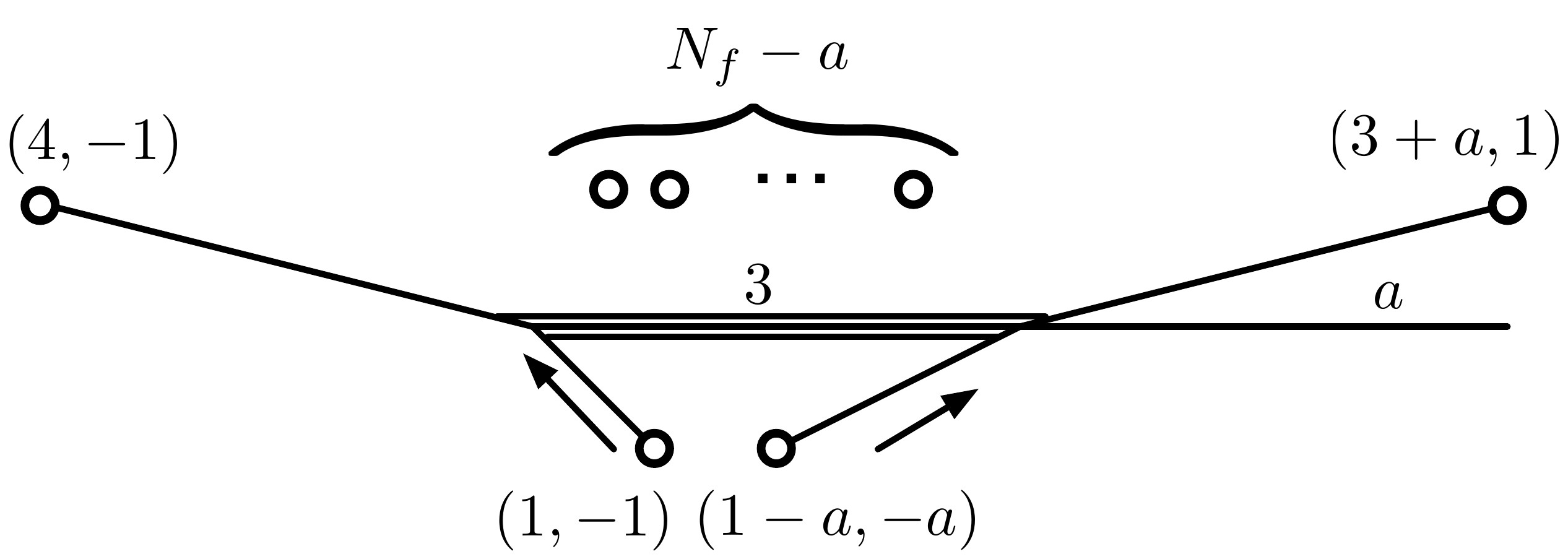}
\caption{5-brane web for $N_c=3$, $0 \le k=6 - \frac{N_f}{2}$. We define $a=0$ for $N_f=0$ and $a=1$ for $N_f \ge 1$.}
\label{Fig:SU3_k=6-Nf2}
\end{figure}

Exceptionally for $N_c = 3$, it is proposed \cite{Jefferson:2017ahm, Jefferson:2018irk} that the theory with $|k|=N_c - \frac{N_f}{2} + 3$ also has a UV fixed point. 
This case can be understood as a special case of 5d $SU(N_c)_{}$ gauge theory with one hypermultiplet in antisymmetric tensor representation and $N_f-1 (\le 8)$ hypermultiplets in fundamental representation and with $|k|=\frac{N_c-N_f+9}{2}$. This class of theories is believed to be dual to 5d $Sp(N_c-1)$ gauge theory with one antisymmetric tensor and $N_f-1$ flavors, whose UV fixed point is known to be rank $N_c-1$ SCFT with global symmetry $E_{N_f}$.
However, for $N_c =3$, the antisymmetric tensor representation is equivalent to fundamental representation, leading to $N_f$ flavor. 
The corresponding 5-brane web diagram can be depicted as in figure \ref{Fig:SU3_k=6-Nf2} with the introduction of 7-branes, where we set $a=0$ for $N_f=0$ and $a=1$ for $N_f \ge 1$. 

\begin{figure}
\centering
\begin{minipage}{6cm}
\includegraphics[width=5.5cm]{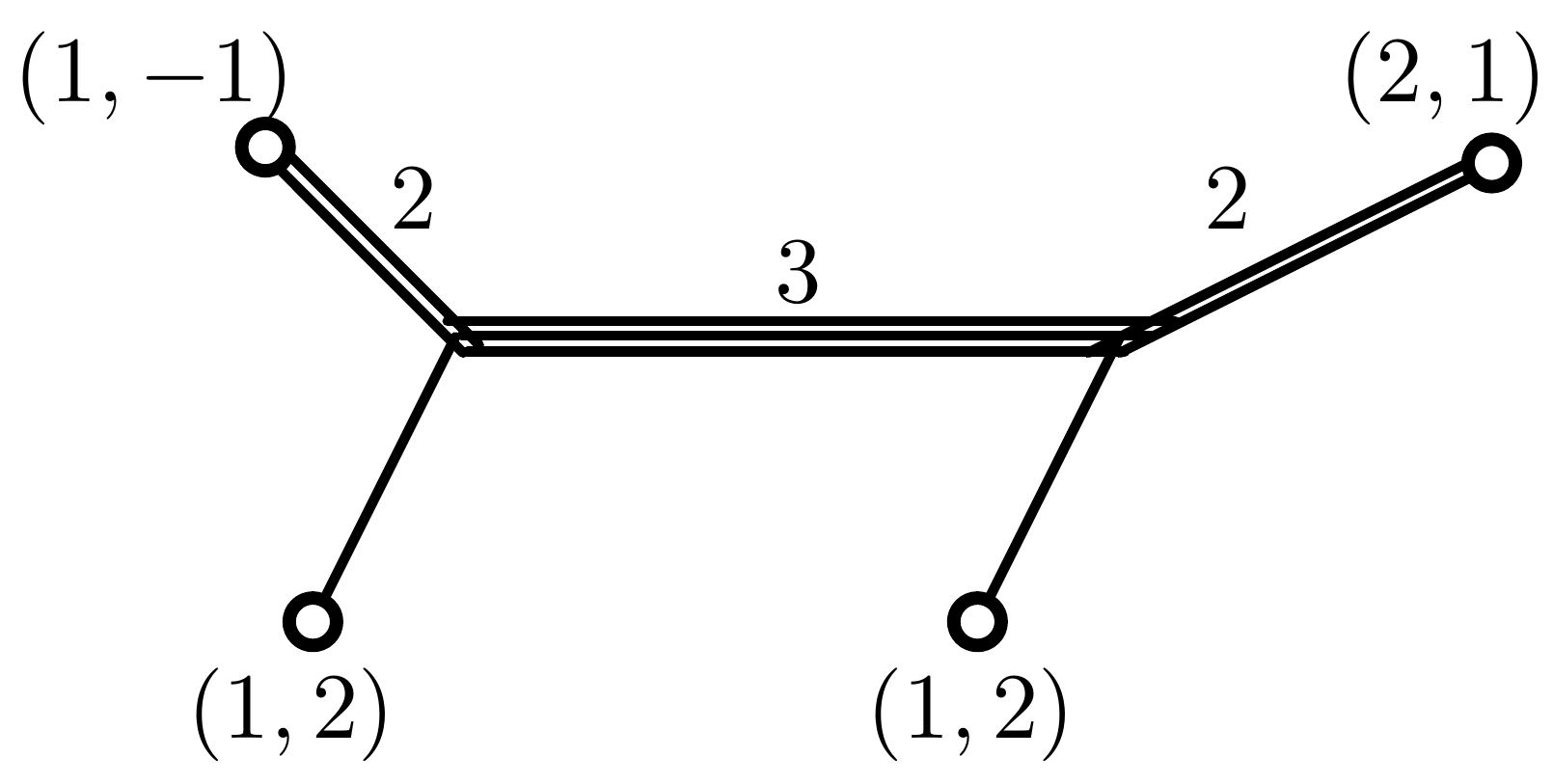}
\caption{$N_f=0, k=6$. After Hanany-Witten transition.}
\label{Fig:k=6_Nf=0}
\end{minipage}
\begin{minipage}{6cm}
\includegraphics[width=5.5cm]{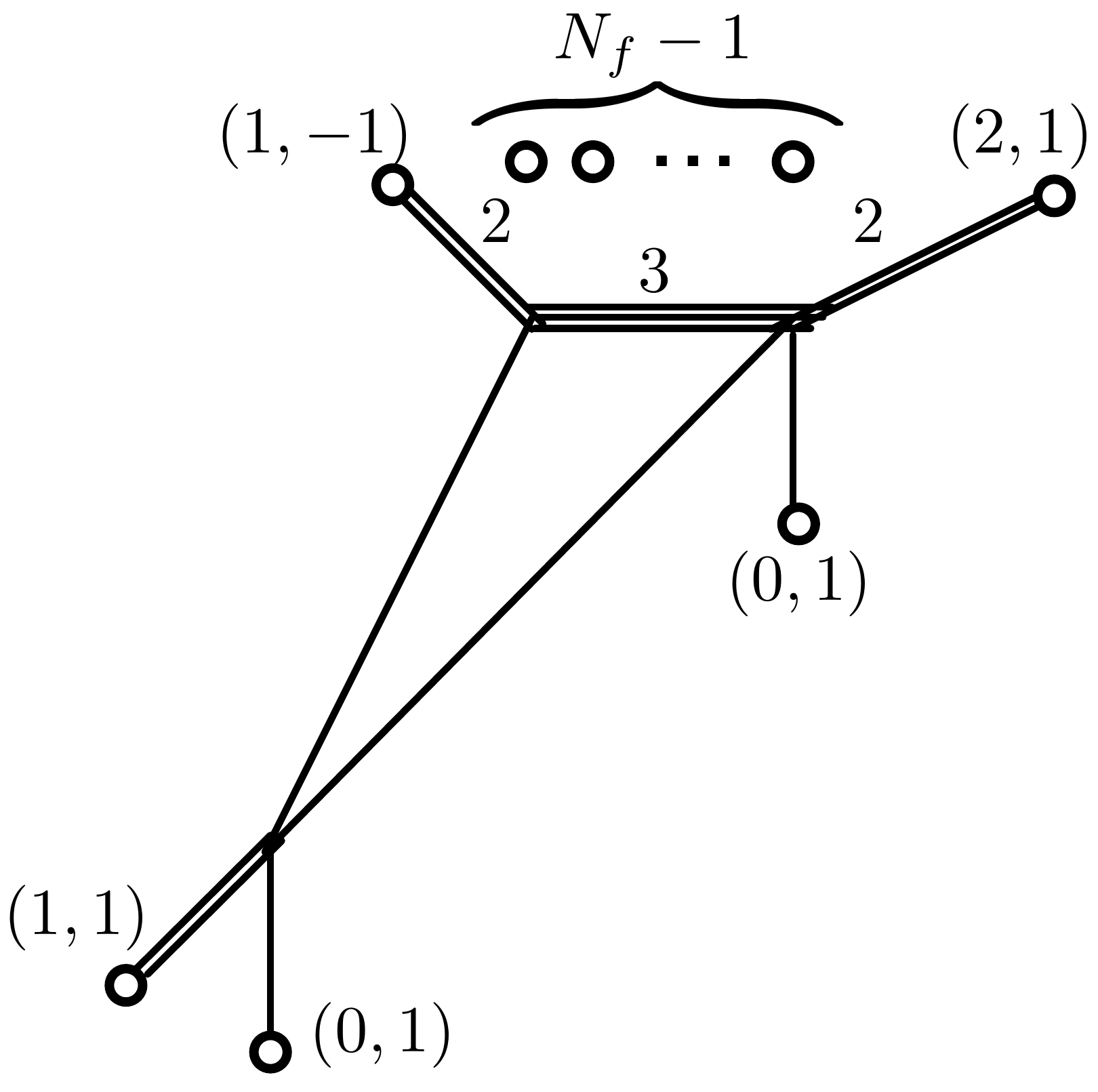}
\caption{$0 \le k=6 - \frac{N_f}{2}, N_f \ge 1$. After Hanany-Witten transition.}
\label{Fig:k=6-Nf2}
\end{minipage}
\end{figure}

By moving the 7-branes as in figure \ref{Fig:SU3_k=6-Nf2}, we obtain the 5-brane web in figure \ref{Fig:k=6_Nf=0} for $N_f=0$ and the one in figure \ref{Fig:k=6-Nf2} for $N_f \ge 1$ after Hanany-Witten transition. It is straightforward to see that the 5-brane web diagram in figure \ref{Fig:k=6-Nf2} is identical to the one suggested in \cite{Bergman:2015dpa} for 5d $Sp(2)$ gauge theory with 1 second rank antisymmetric tensor and $N_f-1$ flavors at their UV fixed point\footnote{This diagram also includes a decoupled singlet whose mass is identical to the one of the antisymmetric tensor. F.Y. thanks O.Bergman for discussion on this point.}. These theories are known to be rank 2 SCFTs with $E_{N_f}$ global symmetry.

\begin{figure}
\centering
\includegraphics[width=14cm]{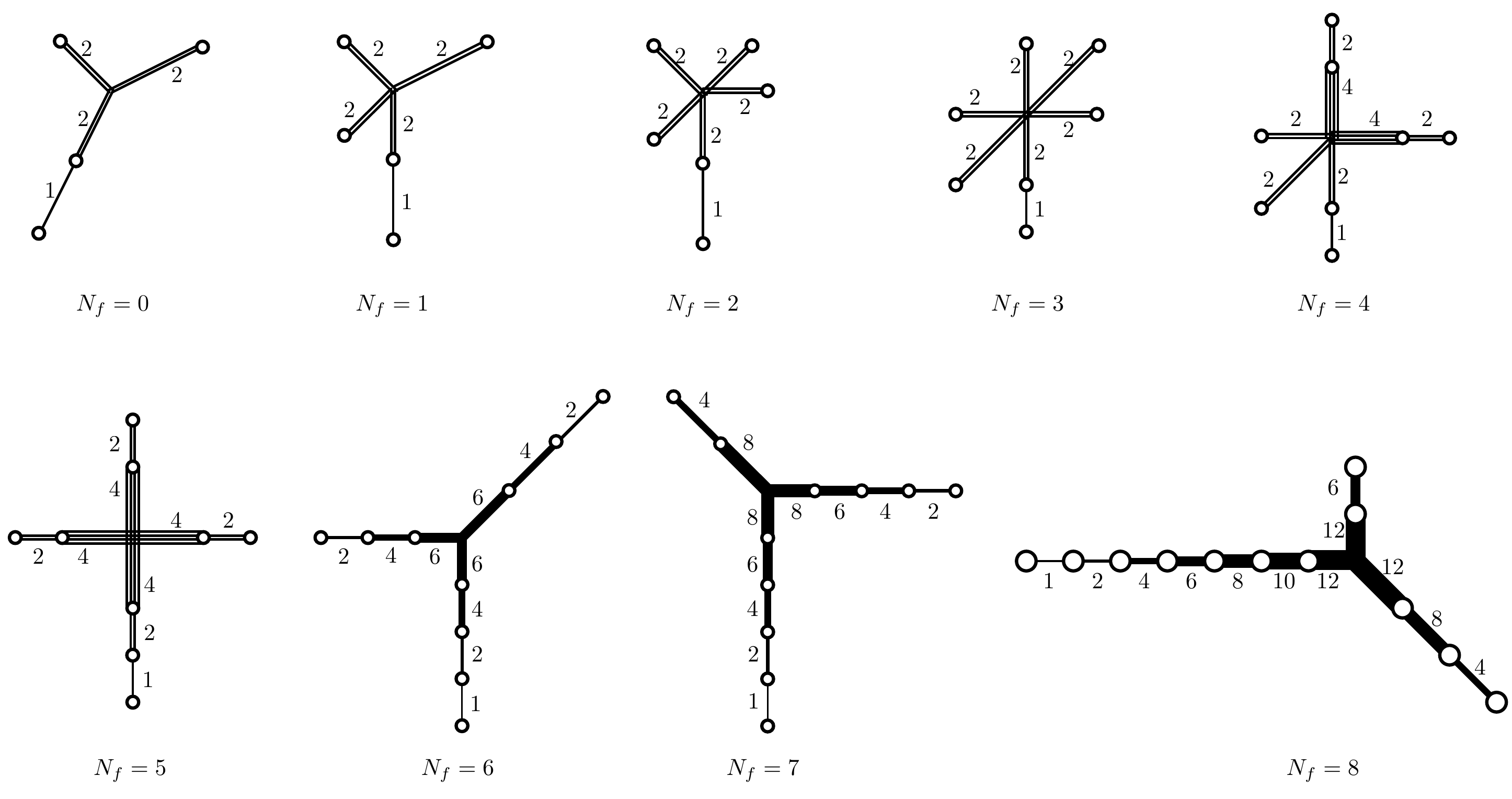}
\caption{5-brane web diagram for $N_c=3$, $|k|=6 - \frac{N_f}{2}$ at infinite coupling.}
\label{Fig:k=6-Nf2graph}
\end{figure}

\begin{figure}
\centering
\includegraphics[width=14cm]{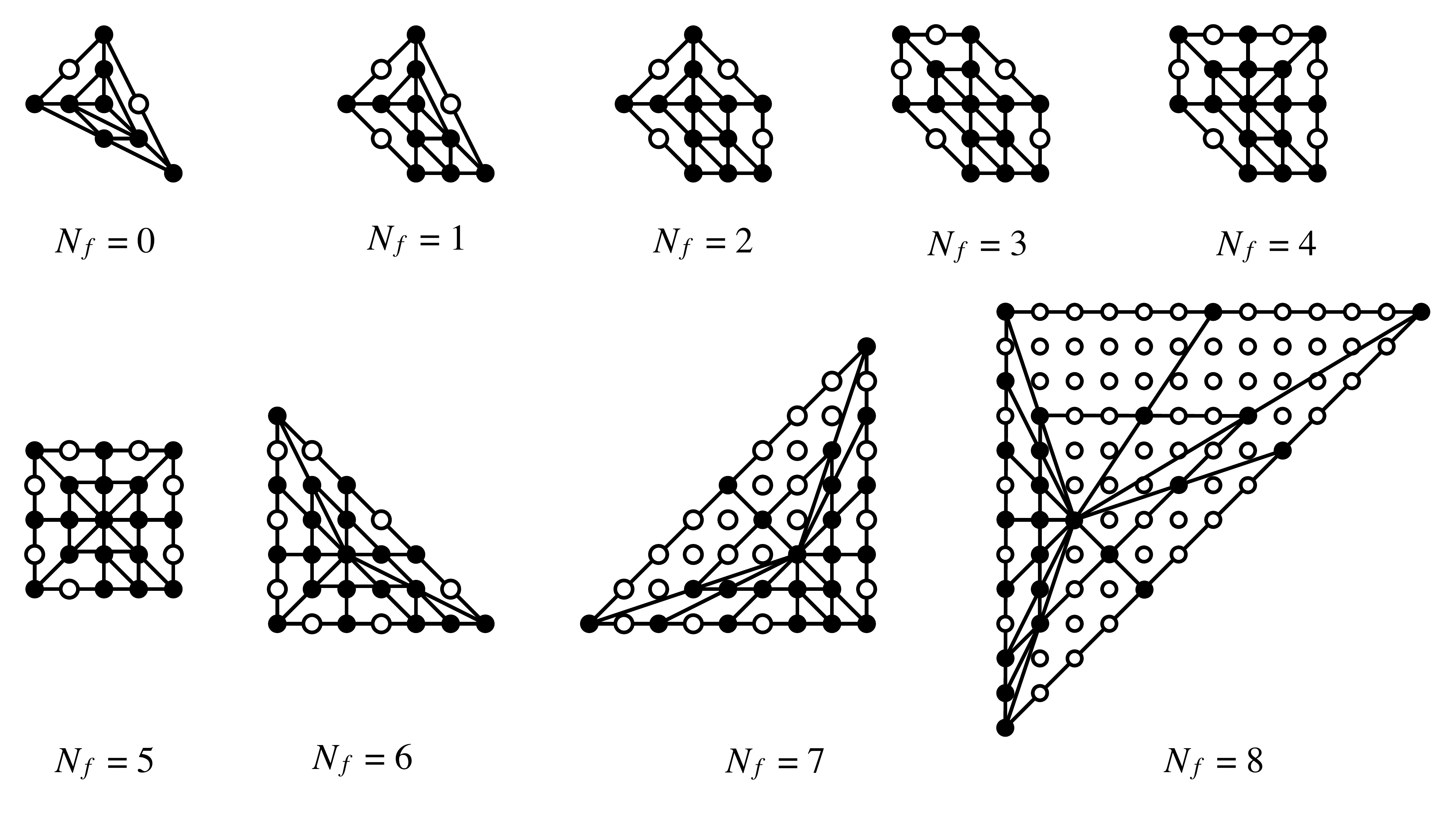}
\caption{Toric diagrams for $N_c=3$, $|k|=6 - \frac{N_f}{2}$.}
\label{Fig:toric_exceptional}
\end{figure}

Indeed, by moving 7-branes in figure \ref{Fig:k=6-Nf2}, taking into account their monodromy as well as the Hanany-Witten transitions, we see that the diagrams can be equivalently rewritten as in figure \ref{Fig:k=6-Nf2graph}. These diagrams are almost identical to the ones proposed in \cite{Benini:2009gi} as the diagram for rank 2 $E_{N_f}$ SCFTs. The only difference is the extra edge with a single 5-brane, which corresponds to the decoupled singlet. 
The corresponding toric diagrams are depicted in figure \ref{Fig:toric_exceptional}. 

%
%

From figure \ref{Fig:k=6-Nf2graph}, it is straightforward to read off the Higgs branches. They are identified as a 3d quiver with affine $E_{N_f}$ Dynkin diagram whose ranks are twice the ranks of the minimal choice for the affine $E_{N_f}$ quivers, which are discussed in section \ref{sec:knownExamples}, and with an extra $U(1)$ node attached to the null node. See tables 3 and 4 of \cite{Cremonesi:2014xha} for $k=2$ in these tables. The moduli space is, as expected, the 2 $E_{N_f}$ instanton moduli space on $\co^2$ \cite{Cremonesi:2014xha} and its Hilbert Series is computed in \cite{Hanany:2012dm}.

\section{Conclusions}\label{sec:conclusions}

The conclusion of this paper is that \emph{tropical geometry} holds the key to solve a problem in supersymmetric quantum field theory that had not been probed with the tools that were developed in the past. The problem concerns the vacuum structure of 5d theories when their gauge coupling is taken to infinity. In particular we have postulated a conjecture that is able to recover the Higgs branch of such theories. The Higgs branch at infinite gauge coupling is found to be a union of several components (hyperK\"ahler cones), and each of such components is given as a \emph{space of dressed monopole operators}. The structure of such spaces can be easily encoded in graphs (quivers), and our proposal explains how to read such quivers directly from the brane system of the 5d theory. The technique of obtaining such quivers, embodied in or Conjecture \ref{con:main}, is completely novel and we believe that its striking  simplicity makes it particularly appealing. Furthermore, the technique itself is by no means restricted to the vacuum at infinite coupling, and it can be applied to any 5d $\mathcal{N} = 1$ gauge theory that has an embedding into Type IIB superstring theory, via five brane webs ending on seven branes. It can also be used to study any hyperK\"ahler variety that exists as a subset of the full Higgs branch. In this paper for example, we have employed it to obtain the precise description of the intersections of the different components of the Higgs branch at infinite coupling.

We believe that both results, the classification of Higgs branches at infinite coupling of the three parameter family of 5d SQCD theories, and the conjecture on how to obtain the components of the Higgs branch from the brane system, are extremely relevant. The former provides an answer for a question that had been left unsolved for far too long. The latter constitutes an exceedingly simple new technique that can easily be implemented in many other analyses, and it is bound to open the door to a myriad of exciting results in the study of moduli spaces of 5d supersymmetric gauge theories. Furthermore, it gives a more robust shape to an idea that has been emerging in these types of studies during the last two years: the use of spaces of dressed monopoles to study vacuum moduli spaces is not restricted to 3d Coulomb branches, and it is in fact a property of any theory with eight supercharges in 3, 4, 5 or 6 dimensions.

\paragraph{Acknowledgements.}
The authors would like to thank Gabi Zafrir, Marcus Sperling, Erwan Brugall{\'e}, Amer Iqbal, Sebastian Franco, Oren Bergman, Kimyeong Lee, Sung-Soo Kim, and Hirotaka Hayashi for useful discussions.
We thank the Simons Center for Geometry and Physics, Stony Brook University  
for the hospitality and the 
partial support during the completion of this work at the Simons Summer workshop 2018.
S.~C.~
is supported by an EPSRC DTP studentship EP/M507878/1. A.~H.~ is supported by STFC Consolidated Grant ST/J0003533/1, and EPSRC Programme Grant EP/K034456/1.
F.~Y.~ is partly supported by Israel Science Foundation under Grant No. 352/13 and No. 1390/17, 
by NSFC grant No.~11501470 and No.~11671328, and by Recruiting Foreign Experts Program No.~T2018050 granted by SAFEA.

\appendix

\section{Example of an Explicit Computation}\label{sec:app}

\begin{figure}
	\centering
	\begin{tikzpicture}
	\draw [red] (-1,2) -- (1,2);
	\draw [red] (0.05,1) -- (0.05,2);
	\draw [red] (0.15,1) -- (0.15,2);
	\draw [blue] (-0.05,1) -- (-0.05,2);
	\draw [blue] (-0.15,1) -- (-0.15,2);
	\draw [blue] (0,0) -- (0,1);
	\draw [blue] (-0.1,2) -- (-1-0.1,3);
	\draw [blue] (-0.1,2) -- (1-0.1,3);
	\draw [red] (0.1,2) -- (-1+0.1,3);
	\draw [red] (0.1,2) -- (1+0.1,3);
	\node [7braneBig] at (0,0) {};
	\node [7braneBig, label = right: {$A_3$}] at (0,1) {}; 
	\node [7braneBig] at (-1,2) {}; 
	\node [7braneBig] at (1,2) {}; 
	\node [7braneBig, label = above: {$A_1$}] at (-1,3) {}; 
	\node [7braneBig, label = above: {$A_2$}] at (1,3) {}; 
	\end{tikzpicture}
	\caption{Example of a 5-brane web that has been maximally subdivided. The 7-branes that are shared by both sub webs are labelled $A_1$, $A_2$ and $A_3$.}
	\label{fig:appendix1}
\end{figure}
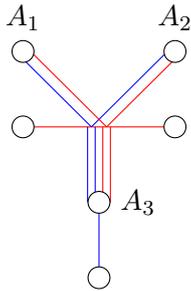

Let us illustrate the computation of the quantities $SI$, $X_i$ and $Y_i$ involved in Conjecture \ref{con:main} with an example. Let the example be the 5-brane web in figure \ref{fig:appendix1} (this brane system appears among other cases as a sub web of the system  with gauge group $SU(5)$, $N_f=8$ number of flavors, and CS level $k=3$). It can be maximally divided into two different sub webs, one depicted in color blue, the other in color red.

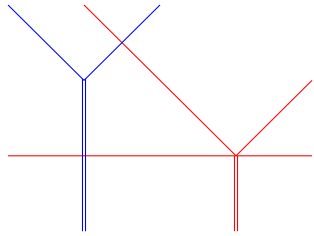
\begin{figure}
	\centering
	\begin{tikzpicture}
	\draw [red] (-1,1) -- (3,1);
	\draw [red] (2-0.02,0) -- (2-0.02,1);
	\draw [red] (2+0.02,0) -- (2+0.02,1);
	\draw [red] (2,1) -- (3,2);
	\draw [red] (2,1) -- (0,3);
	\draw [blue] (-0.02,0) -- (-0.02,2.02);
	\draw [blue] (0.02,0) -- (0.02,2.02);
	\draw [blue] (0,2) -- (-1,3);
	\draw [blue] (0,2) -- (1,3);
	\end{tikzpicture}
	\caption{Stable intersection computation. The branes that intersect in figure \ref{fig:appendix1} have been displaced with respect to each other.}
	\label{fig:appendix2}
\end{figure}

\begin{figure}
	\centering
	\begin{tikzpicture}
	\draw [red] 	(0,0) -- (0,1)
				(1,2) -- (2,3)
				(2,0) -- (2,1) -- (3,2) -- (4,1) -- (4,0) -- (2,0);
	\draw [blue] (0,0) -- (2,0)
				(0,1) -- (2,1)--(1,2)--(0,1)
				(2,3)--(3,2);
	\node[toric] at (0,0) {};
	\node[toric] at (0,1) {};
	\node[toric] at (1,0) {};
	\node[toric] at (1,1) {};
	\node[toric] at (1,2) {};
	\node[toric] at (2,0) {};
	\node[toric] at (2,1) {};
	\node[toric] at (2,2) {};
	\node[toric] at (2,3) {};
	\node[toric] at (3,0) {};
	\node[toric] at (3,1) {};
	\node[toric] at (3,2) {};
	\node[toric] at (4,0) {};
	\node[toric] at (4,1) {};
	\end{tikzpicture}
	\caption{Toric diagram dual to the configuration in figure \ref{fig:appendix2}.}
	\label{fig:appendix3}
\end{figure}
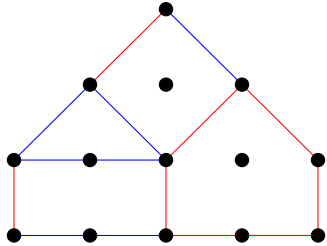

First, let us compute the \emph{stable intersection} $SI$. In order to do this we focus on the local properties of the system near the intersection point in the center of figure \ref{fig:appendix1}. We depict the neighborhood of this point in figure \ref{fig:appendix2} (note that the 7-branes are taken to be very far away), with the difference that the sub webs have been slightly displaced with respect to each other. Now it is possible to obtain a dual toric diagram of the system, represented in figure \ref{fig:appendix3}. The $SI$ is the sum of the areas of all polygons with edges of different colors. In this case there are two such polygons: one rectangle of area 2 and one square of area 2. Hence,
\begin{equation}
	SI = 2+2=4
\end{equation}

Now, let us compute the $X_i$ contributions. These come from the 7-branes that are shared by both sub webs. There are 3 such branes, denoted in figure \ref{fig:appendix1} as $A_1$, $A_2$ and $A_3$. For each brane $A_i$ we compute how many possible combinations exist of branes from different sub webs ending on it from opposite directions. This number is zero for $A_1$ and $A_2$. For $A_3$ the blue $(0,1)$ 5-brane that ends on it from below can be paired with two different red $(0,1)$ 5-branes that end on it from above, giving a total number of 2:
\begin{align}
	X_1&=0\\
	X_2&=0\\
	X_3&={\color{blue}1}\times{\color{red} 2} = 2
\end{align} 

In order to compute the $Y_i$ contributions we look again at the $A_i$ 7-branes. In each 7-brane we ask how many pairs of 5-branes from different sub webs exist such that both 5-branes end in the 7-brane coming from the same direction. For $A_1$ there is only one pair, and the same is true for $A_2$. For $A_3$ there are two blue $(0,1)$ 5-branes that end on $A_3$ from above, and there are also two red $(0,1)$ 5-branes that end on $A_3$ from above, so there is a total of four different combinations:
\begin{align}
	Y_1&={\color{blue}1}\times{\color{red} 1} = 1\\
	Y_2&={\color{blue}1}\times{\color{red} 1} = 1\\
	Y_3&={\color{blue}2}\times{\color{red} 2} = 4
\end{align} 

According to Conjecture \ref{con:main}, each sub web gives rise to a different $U(1)$ gauge node on the corresponding quiver; the number of edges $E$ between both nodes is then given by:
\begin{equation}
	\begin{aligned}
	E &= SI + \sum_{i=1}^3 X_i - \sum_{i=1}^3 Y_i\\	&= 4 + (0+0+2) - (1+1+4) \\
		&=0
	\end{aligned}
\end{equation}
Hence, the quiver corresponding to figure \ref{fig:appendix1} has two nodes of rank 1 with no hypermultiplets in between:
\begin{equation}
 \cnode{red}{}{1}~~\cnode{blue}{}{1}
\end{equation}
and the Higgs branch of the system is isomorphic to the 3d Coulomb branch (or alternatively, the space of dressed monopole operators):
\begin{equation}
 	\mathcal{C}^{3d}\left(\cnode{black}{}{1}~~\cnode{black}{}{1}\right)
\end{equation}

\bibliography{ref}
\bibliographystyle{JHEP}

\end{document}